\documentclass[aps,prd,twocolumn,superscriptaddress,amssymb,nofootinbib,floatfix]{revtex4}
\usepackage{fancyhdr}
\usepackage{fancybox}
\usepackage{graphicx}
\usepackage{color}
\usepackage{soul}
\usepackage{latexsym}

\def\cta{\cos\theta_A}
\def\sta{\sin\theta_A}

\def\tauptaum{\tau^+\tau^-}

\def\falam{F_{\alam}}
\def\fakap{F_{\akap}}
\def\fcal{F^{\cta}_{\alam}}
\def\fcak{F^{\cta}_{\akap}}

\def\cbb{C_{\rm eff}^{2b}}
\def\cbbbb{C_{\rm eff}^{4b}}
\def\hu{h_u}
\def\hd{h_d}
\def\ma{m_a}
\def\mh{m_h}
\def\hi{h_1}
\def\hii{h_2}
\def\ai{a_1}
\def\mhi{m_{h_1}}
\def\mhii{m_{h_2}}
\def\mai{m_{a_1}}
\def\mueff{\mu_{\rm eff}}
\def\mtau{m_\tau}

\def\ptl{\partial}
\def\lam{\lambda}
\def\kap{\kappa}
\def\alam{A_\lambda}

\def\akap{A_\kappa}

\def\mh{m_h}

\def\mhusq{m_{H_u}^2}
\def\mhdsq{m_{H_d}^2}
\def\mssq{m_S^2}
\def\mhu{m_{H_u}}
\def\mhd{m_{H_d}}
\def\ms{m_{S}}
\def\mqsq{m_Q^2}
\def\musq{m_U^2}
\def\mdsq{m_D^2}
\def\mlsq{m_L^2}
\def\mesq{m_E^2}

\def\h{h}
\def\mh{m_{h}}

\def\hbar{\overline h}
\def\lam{\lambda}

\def\mx{M_X}

\def\mz{m_Z}

\def\hi{h_i^0}
\def\mhi{m_{\hi}}

\def\h{h}
\def\mh{m_{\h}}

\def\lam{\lambda}

\def\nn{\nonumber}

\def\wtil{\widetilde}

\def\tauptaum{\tau^+\tau^-}

\def\lsim{\mathrel{\raise.3ex\hbox{$<$\kern-.75em\lower1ex\hbox{$\sim$}}}}
\def\gsim{\mathrel{\raise.3ex\hbox{$>$\kern-.75em\lower1ex\hbox{$\sim$}}}}
\def\ifmath#1{\relax\ifmmode #1\else $#1$\fi}
\def\half{\ifmath{{\textstyle{1 \over 2}}}}

\def\third{\ifmath{{\textstyle{1 \over 3}}}}

\def\vev#1{\langle #1 \rangle}
\def\lam{\lambda}

\def\Eq#1{Eq.~(\ref{#1})}

\def\mplanck{M_{\rm P}}

\def\mhi{m_{h_1}}

\def\gl{\wt g}
\def\mgl{m_{\gl}}

\def\stop{\wt t}
\def\stopone{\wt t_1}
\def\stoptwo{\wt t_2}

\def\mstopone{m_{\stopone}}
\def\mstoptwo{m_{\stoptwo}}

\def\mstopbar{\overline m_{\stop}}

\def\msusy{m_{\rm SUSY}}

\def\susy{{\rm SUSY}}

\def\gl{\wt g}
\def\mgl{m_{\gl}}

\def\mstopone{m_{\stopone}}

\def\hsm{h_{\rm SM}}
\def\mhsm{m_{\hsm}}
\def\hl{h}
\def\hh{H}
\def\ha{A}

\def\mha{m_{\ha}}

\def\tanb{\tan\beta}

\def\mt{m_t}
\def\mb{m_b}
\def\mz{m_Z}

\def\mgut{M_U}
\def\mx{M_X}

\def\wt{\widetilde}

\def\cpmone{\wt \chi^{\pm}_1}

\def\mcpmone{m_{\cpmone}}

\def\MPL #1 #2 #3 {{\sl Mod.~Phys.~Lett.}~{\bf#1} (#3) #2}
\def\NPB #1 #2 #3 {{\sl Nucl.~Phys.}~{\bf #1} (#3) #2}
\def\PLB #1 #2 #3 {{\sl Phys.~Lett.}~{\bf #1} (#3) #2}
\def\PR #1 #2 #3 {{\sl Phys.~Rep.}~{\bf#1} (#3) #2}
\def\PRD #1 #2 #3 {{\sl Phys.~Rev.}~{\bf #1} (#3) #2}
\def\PRL #1 #2 #3 {{\sl Phys.~Rev.~Lett.}~{\bf#1} (#3) #2}
\def\RMP #1 #2 #3 {{\sl Rev.~Mod.~Phys.}~{\bf#1} (#3) #2}
\def\ZPC #1 #2 #3 {{\sl Z.~Phys.}~{\bf #1} (#3) #2}
\def\IJMP #1 #2 #3 {{\sl Int.~J.~Mod.~Phys.}~{\bf#1} (#3) #2}
\def\NIM #1 #2 #3 {{\sl Nucl.~Inst.~and~Meth.}~{\bf#1} {#3} #2}

\def\lam{\lambda}
\def\br{B}
\def\tauptaum{\tau^+\tau^-}

\def\gam{\gamma}

\def\anti{\overline}
\def\epem{e^+e^-}

\def\rts{\sqrt s}
\def\ie{{\it i.e.}}

\def\anti{\overline}

\def\ai{a_1}
\def\aii{a_2}
\def\mai{m_{\ai}}
\def\maii{m_{\aii}}

\def\gev{~{\rm GeV}}
\def\tev{~{\rm TeV}}
\def\mt{m_t}
\def\mb{m_b}

\def\hi{\h_1}
\def\hii{\h_2}
\def\hiii{\h_3}
\def\mhi{m_{\hi}}
\def\mhii{m_{\hii}}
\def\mhiii{m_{\hiii}}

\newcommand{\nc}{\newcommand}
\nc{\beq}{\begin{equation}}   \nc{\eeq}{\end{equation}}
\nc{\bea}{\begin{eqnarray}}   \nc{\eea}{\end{eqnarray}}
\nc{\baa}{\begin{array}}      \nc{\eaa}{\end{array}}
\nc{\bit}{\begin{itemize}}    \nc{\eit}{\end{itemize}}
\nc{\ben}{\begin{enumerate}}  \nc{\een}{\end{enumerate}}
\nc{\bce}{\begin{center}}     \nc{\ece}{\end{center}}
\def\beqa{\begin{eqnarray}}
\def\eeqa{\end{eqnarray}}
\def\bed{\begin{description}}
\def\eed{\end{description}}

\def\mhi{m_{h_1}}

\def\gl{\wt g}
\def\mgl{m_{\gl}}

\def\half{\frac{1}{2}\,}
\def\third{\frac{1}{3}\,}
\def\quart{\frac{1}{4}\,}

\def\tanb{\tan\beta}

\def\Eq#1{Eq.~(\ref{#1})}

\def\simle{
    \mathrel{\rlap{\raise 0.511ex 
        \hbox{$<$}}{\lower 0.511ex \hbox{$\sim$}}}}

\def\slashchar#1{\setbox0=\hbox{$#1$}           
   \dimen0=\wd0                                 
   \setbox1=\hbox{/} \dimen1=\wd1               
   \ifdim\dimen0>\dimen1                        
      \rlap{\hbox to \dimen0{\hfil/\hfil}}      
      #1                                        
   \else                                        
      \rlap{\hbox to \dimen1{\hfil$#1$\hfil}}   
      /                                         
   \fi}

\def\lam{\lambda}

\def\ls#1{\ifmath{_{\lower1.5pt\hbox{$\scriptstyle #1$}}}}
\def\lss#1{\ifmath{^{\,\lower2.5pt\hbox{$\scriptstyle #1$}}}}

\def\nn{\nonumber}
\tighten \preprint{UCD-HEP-???}
\begin{document}

\title{\Large\bf The NMSSM Solution to the Fine-Tuning Problem,
  Precision Electroweak Constraints and the Largest LEP Higgs Event Excess
}
\author{
Radovan Derm\' \i \v sek$^*$ and John F. Gunion$^\dagger$}
\address{
$^*$School of Natural Sciences, Institute for Advanced Study, Princeton,
NJ 08540 \break
$^\dagger$Department of Physics, University of California at Davis, Davis, CA 95616} 
\begin{abstract}
  We present an extended study of how the Next to Minimal
  Supersymmetric Model easily avoids fine-tuning in electroweak
  symmetry breaking for a SM-like light Higgs with mass in the
  vicinity of $100\gev$, as beautifully consistent with precision
  electroweak data, while escaping LEP constraints due to the
  dominance of $h\to aa$ decays with $m_a<2m_b$ so that $a\to
  \tauptaum$ or jets.  The residual $\sim 10\%$ branching ratio for
  $h\to b\anti b$ explains perfectly the well-known LEP excess at
  $\mh\sim 100\gev$. Details of model parameter correlations and
  requirements are discussed as a function $\tanb$. Comparisons of
  fine-tuning in the NMSSM to that in the MSSM are presented. We also
  discuss fine-tuning associated with scenarios in which the $a$ is
  essentially pure singlet, has mass $m_a>30\gev$, and decays
  primarily to $\gam\gam$ leading to an $h\to aa\to 4\gam$ Higgs
  signal.
\end{abstract}
\maketitle
\thispagestyle{empty}

\section{Introduction}

In the Standard Model (SM), electroweak symmetry breaking, whereby the
$W$ and $Z$ bosons and the quarks and leptons acquire mass, gives rise
to a Higgs boson, $\hsm$.  However, the value of $\mhsm$ is
quadratically sensitive to the cutoff scale of the theory, $\Lambda$,
especially through top quark loops which give a one-loop correction of
\beq
\delta\mhsm^2=-{3\over 4\pi^2}{\mt^2\over v^2}\Lambda^2\,,
\eeq
where $\Lambda$ is the high energy cutoff and $v=176\gev$.
For $\Lambda$ of order the GUT scale, $\mgut$, or the Planck scale,
$\mplanck$, an extreme cancellation between the one-loop 
contribution(s) and the bare Higgs mass is required in order that the
physical Higgs mass be below a $\tev$, as required in order for the
scattering of longitudinally polarized $W$ bosons to obey unitarity in
a perturbative fashion.

Supersymmetric (SUSY) models, such as the
Minimal Supersymmetric Model (MSSM), cure this naturalness / hierarchy
problem associated with the quadratically divergent 1-loop corrections
via the introduction of superpartners for each SM particle.
Because the spin of the superpartners differs by 1/2 unit from that of
the corresponding SM particle, the 1-loop correction from the
superpartner will cancel that of the SM particle once the energy scale
being integrated over in the loop is above the mass
of the (presumed to be heavier) superpartner.
So long as the superpartners have mass somewhat below $1\tev$ (say
$\sim 500\gev$),
the cancellation is not particularly extreme and the hierarchy /
naturalness problem associated with the quadratic divergences is
ameliorated.  However, there remains the question of how finely the
GUT-scale parameters must be adjusted in order to get appropriate
electroweak symmetry breaking, that is to say correctly predict the
observed value of $\mz$.  It is here that LEP limits on a
SM-like Higgs boson play a crucial role.

Supersymmetric models most naturally predict that the lightest Higgs
boson, generically $h$, is SM-like and that it has a mass closely
correlated to $\mz$, typically lying in the range $\lsim 105\gev$ for
stop masses $\lsim 500\gev$, with an upper bound, for example, of
$\lsim 135\gev$ in the MSSM for stop masses $\sim 1\tev$ and large
stop mixing.  If the stop masses are large, the predicted value of
$\mz$ is very sensitive to the GUT scale parameters. Such sensitivity
is termed 'fine tuning'.  Models with minimal fine tuning provide a
much more natural explanation of the $Z$ mass than those with a high
level of fine tuning. The degree of fine tuning required is thus quite
closely related to the constraints on a SM-like $h$, and these in turn
depend on how it decays.

\begin{figure}[h!]
  \centerline{\includegraphics[width=2.4in,angle=90]{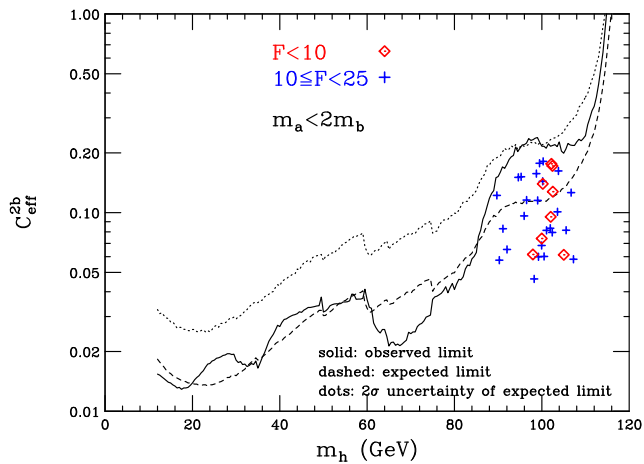}}
  \vspace*{-.1in}
\caption{Expected and observed 95\% CL limits on
  $\cbb$ from
  Ref.~\cite{oldleplimits} are shown vs. $\mh$.  Also plotted are the
  predictions for the NMSSM parameter cases discussed in
  \cite{Dermisek:2005gg} having 
  fixed $\tanb=10$,
  $M_{1,2,3}(\mz)=100,200,300\gev$ that give
  fine-tuning measure $F<25$ and $\mai<2\mb$ and that are consistent
  with Higgs constraints obtained using the preliminary LHWG analysis code \cite{bechtle}.}
\label{zbblimits}
\vspace*{-.1in}
\end{figure}

The SM and the MSSM predict that $h\to b\anti b$ decays are dominant
and LEP has placed strong constraints on $Zh\to Z b\anti b$.  The
limits on
\beq
\cbb\equiv [g_{ZZh}^2/g_{ZZ\hsm}^2]\br(h\to b\anti b)
\eeq 
are shown in Fig.~\ref{zbblimits} (from
Ref.~\cite{oldleplimits}). From this plot, one concludes that
$\mh<114\gev$ is excluded for a SM-like $h$ that decays primarily to
$b\anti b$.  In fact, because of the manner in which the analysis is
done, at a first level of approximation this limit applies for an $h$
that decays to any combination of $2b$ and $4b$. For $\br(h\to b\anti
b)\sim 0.15$ and $\br(h\to b\anti b b\anti b)\sim 0.8$ (with $\tau$
channels making up the rest) $\mh\lsim 110\gev$ is excluded. This will
be important later.  In the case of the CP-conserving MSSM, one always
obtains $\br(h\to b\anti b)\gsim 0.88$. For $\msusy\lsim 1\tev$, most
of parameter space will yield $\mh<114\gev$ and thus be ruled out by
the SM-like Higgs LEP limit .  The remaining part of MSSM parameter
space either has at least one very large parameter, most typically a
soft-SUSY-breaking stop mass close to a TeV at scale
$\mz$, or else large mixing in the stop sector.  In the former case,
one always finds that to predict the observed $\mz$ requires
very careful adjustment, \ie\ fine-tuning, of the GUT-scale parameters
with accuracies better than 1\%. In the latter case, fine-tuning can
be reduced to the 3\% level. To achieve small fine-tuning, let us
say no worse than 10\%, the soft-SUSY-breaking parameters that affect
the Higgs sector should be well below a TeV, in which case the
lightest CP-even MSSM Higgs boson would have mass $\sim 100\gev$.

As suggested in \cite{Dermisek:2005ar}, the simplest way to allow a
Higgs mass of order $100\gev$, thus making possible a light SUSY
spectrum and low fine-tuning, is to modify Higgs decays so that the
$b\anti b$ branching ratio is small and primary decays are to
channel(s) to which LEP is less sensitive. This is very natural in
models in which the Higgs sector is extended and Higgs to Higgs decays
are kinematically allowed. The decay widths for Higgs to Higgs decays
can easily exceed the very small width for the $b\anti b$ channel.
The simplest supersymmetric model that gives rise to this possibility
is the Next-to-Minimal Supersymmetric Model (NMSSM). The NMSSM yields
a preferred value of $\mh\sim 100\gev$ purely on the basis of
minimizing fine-tuning.  A Higgs mass near $100\gev$ is also strongly
preferred by precision electroweak measurements. Further, there is a
well-known $2.3\sigma$ excess in the $\epem\to Z+b's$ channel 
in the LEP data for $M_{b's}\sim 100\gev$ when a final state that
contains two or more $b's$ is assumed to contain exactly 2 $b's$.
If the Higgs decays only to $b\anti b$ then this
excess and limits on the $Z+b's$ final state 
would apply to $\cbb$ defined by 
\beq
\cbb=[g_{ZZh}^2/g_{ZZ\hsm}^2]\br(h\to b\anti b)\,.
\label{cbbdef}
\eeq
The excess is apparent in the higher observed vs. expected $\cbb$
limits for a test Higgs mass of $\mh\sim 100\gev$ shown in
Fig.~\ref{zbblimits}.  This excess is particularly apparent in the
$1-CL_b$ result (Fig.~7 of \cite{oldleplimits}) obtained after
combining all four LEP experiments.

In a previous paper~\cite{Dermisek:2005gg}, we have shown that the
above excess is consistent with a scenario in which the Higgs boson
has SM-like $ZZh$ coupling, but has reduced $\br(h\to b\anti b)$ by
virtue of the presence of $h$ decays to a pair of lighter Higgs
bosons, $h\to aa$, where $\br(a\to b\anti b)$ is small, as is
automatic if $\ma<2\mb$ so that $a\to \tauptaum$ or light quarks and
gluons.~\footnote{ If $a\to b\anti b$ is dominant, as occurs for
  $\ma>2\mb$, then, as noted earlier in the text, $\mh\gsim 110\gev$
  is required by LEP data~\cite{bechtle}.}  (The importance of such
decays was first emphasized in \cite{Gunion:1996fb}, and later in
\cite{Dobrescu:2000jt}, followed by extensive work in
\cite{Ellwanger:2001iw,higsec3,Ellwanger:2005uu,Dermisek:2006wr}.)
For example, if the $ZZh$ coupling is full SM strength, then $\mh\sim
100\gev$ with $\br(h\to b\anti b)\sim 0.08$ and $\br(h\to aa)\sim 0.9$
fits the observed $Z2b$ excess nicely.  Meanwhile, there are no
current limits on the $Zh\to Zaa\to Z\tauptaum\tauptaum$ final state
for $\mh\gsim 87\gev$ \cite{newleplimits}. And limits in the case of
$a\to jets$ run out at slightly lower $\mh$.  As already stressed and
as described below in more detail, we are particularly led to the
above interpretation of LEP data since fine-tuning within the NMSSM is
absent for model parameters that yield precisely this kind of scenario
\cite{Dermisek:2005ar,Dermisek:2005gg}.  While various alternative
interpretations of this excess in terms of a non-SM Higgs sector have
been suggested~\cite{Drees:2005jg,newleplimits}, the
NMSSM scenario has the lowest fine tuning of any such scenario and
has particularly strong theoretical motivation.

The NMSSM is an extremely attractive model \cite{allg}.
First, it provides a very elegant solution to the $\mu$ problem of the
MSSM via the introduction of a singlet superfield $\widehat{S}$.  For
the simplest possible scale invariant form of the superpotential, the
scalar component of $\widehat{S}$ naturally acquires a vacuum
expectation value of the order of the \susy\ breaking scale, giving
rise to a value of $\mu$ of order the electroweak scale. The NMSSM is
the simplest supersymmetric extension of the standard model in which
the electroweak scale originates from the \susy\ breaking scale only.
Hence, the NMSSM deserves very serious consideration.

Apart from the usual quark and lepton Yukawa couplings, the scale
invariant superpotential of the NMSSM is
$
W=\lambda \widehat{S} \widehat{H}_u \widehat{H}_d + \third\kappa
\widehat{S}^3
$ 
depending on two dimensionless couplings $\lambda$, $\kappa$ beyond
the MSSM.  [Hatted (unhatted) capital letters denote superfields
(scalar superfield components).]  The associated trilinear soft terms
are
%
$
\lambda A_{\lambda} S H_u H_d + \third\kappa A_\kappa S^3 \,. 
$
%
The final two input parameters are 
%
$
\tan \beta = h_u/h_d$ and $\mu_\mathrm{eff} = \lambda
s \,, 
$
%
where $h_u\equiv
\vev {H_u}$, $h_d\equiv \vev{H_d}$ and $s\equiv \vev S$.
The Higgs sector of the NMSSM is thus described by the six parameters
$\lambda\ , \ \kappa\ , \ A_{\lambda} \ , \ A_{\kappa}, \ \tan \beta\ ,
\ \mu_\mathrm{eff}\ .$
In addition, values must be input for the gaugino masses 
and for the soft terms related to the (third generation)
squarks and sleptons that contribute to the
radiative corrections in the Higgs sector and to the Higgs decay
widths. 

The particle content of the NMSSM differs from
the MSSM by the addition of one CP-even and one CP-odd state in the
neutral Higgs sector (assuming CP conservation), and one additional
neutralino.  The result is three CP-even Higgs bosons ($h_{1,2,3}$)
two CP-odd Higgs bosons ($a_{1,2}$) and a total of
five neutralinos $\wtil\chi^0_{1,2,3,4,5}$. It will be convenient to
denote the CP-even and CP-odd neutral Higgs bosons of the MSSM as $h,H$
and $A$, respectively, while those of the NMSSM will be denoted by
$\hi,\hii,\hii$ and $\ai,\aii$, respectively. In the latter case, our
focus will be on the lightest states $\hi$ and $\ai$. 
 The NMHDECAY program \cite{Ellwanger:2004xm}, which
includes most LEP constraints, allows easy exploration
of Higgs phenomenology in the NMSSM.

\begin{figure}[ht!]
  \centerline{\includegraphics[width=2.4in,angle=90]{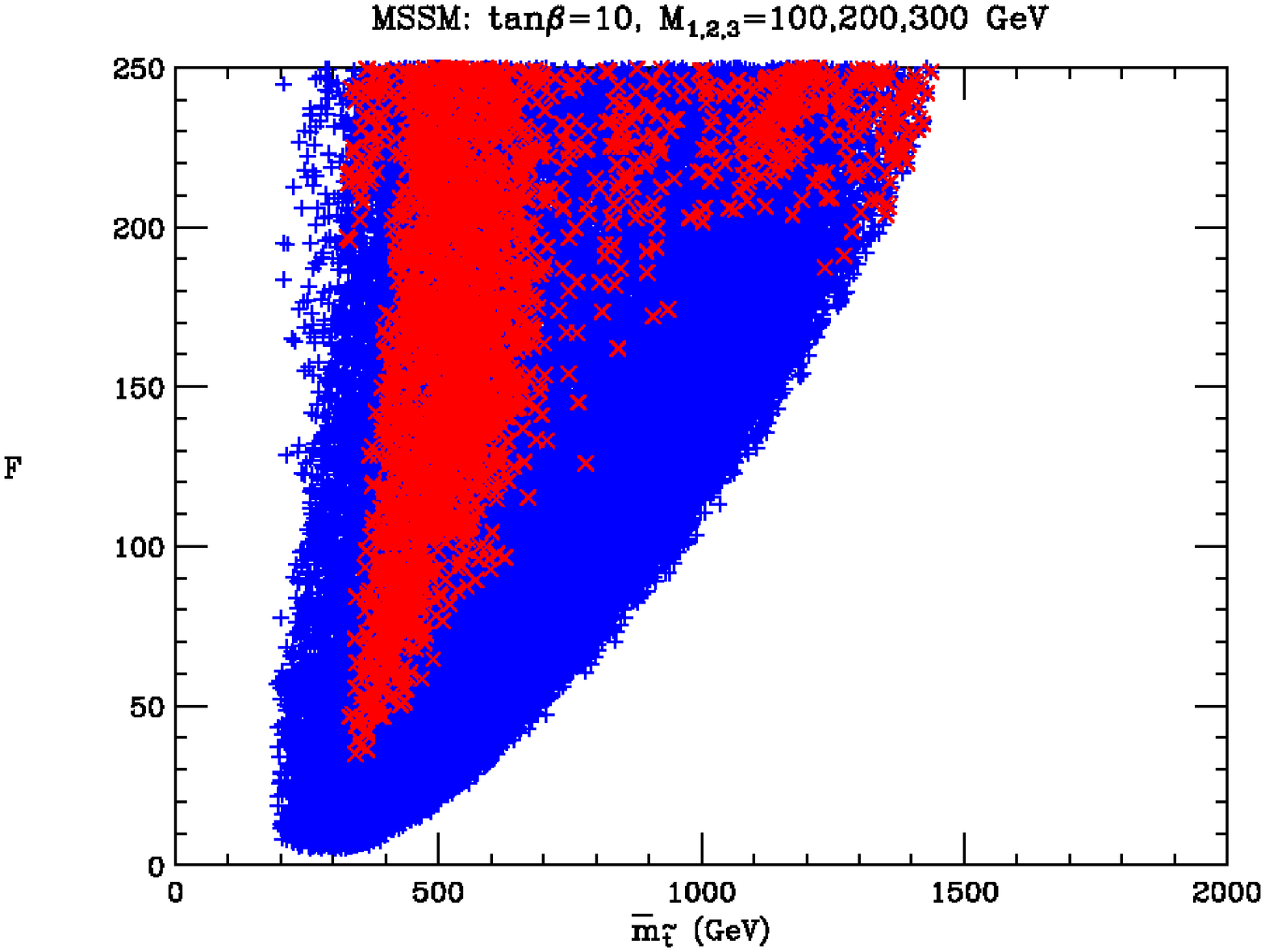}}
\vspace*{.05in}
  \centerline{\includegraphics[width=2.4in,angle=90]{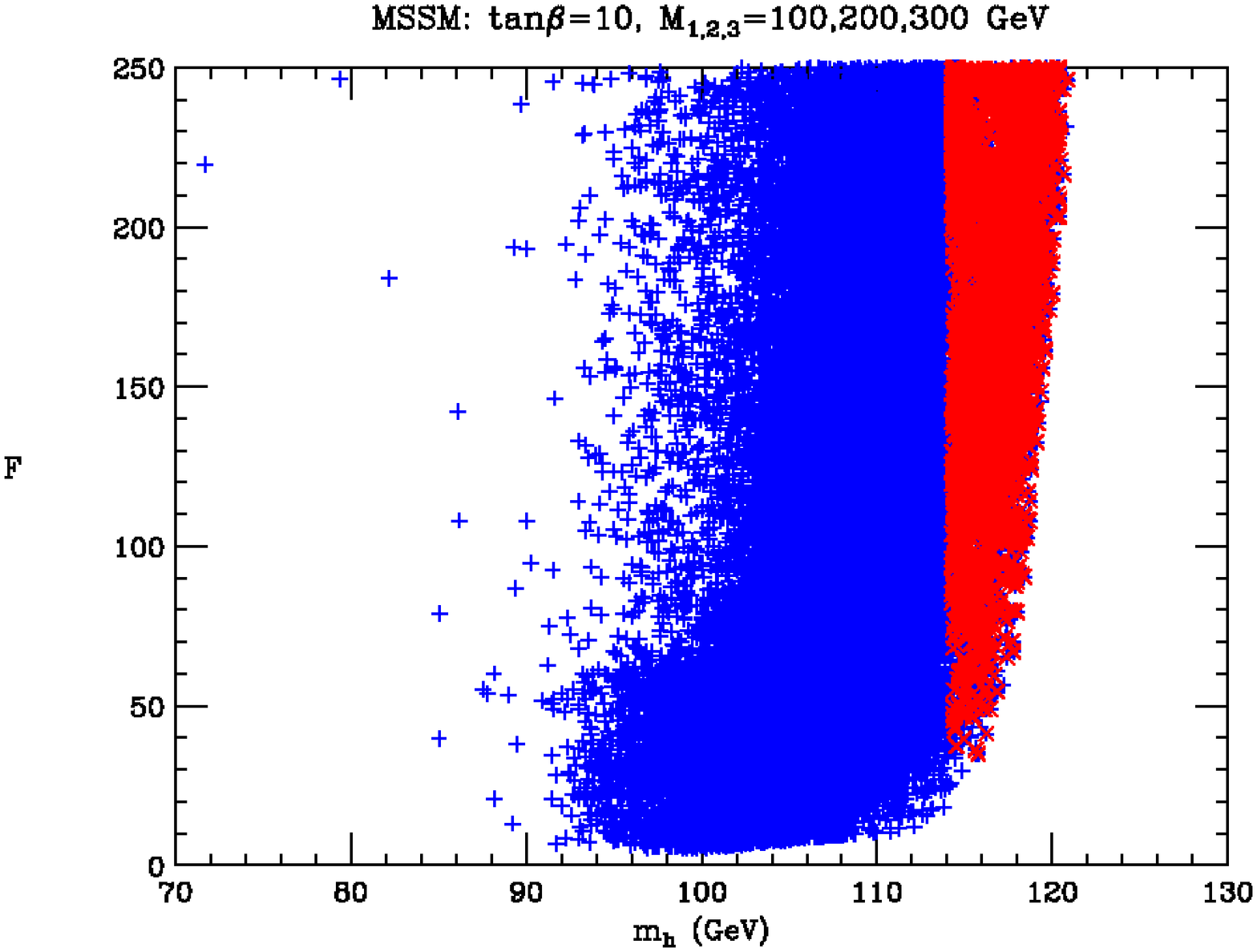}}
\vspace*{.05in}
  \centerline{\includegraphics[width=2.4in,angle=90]{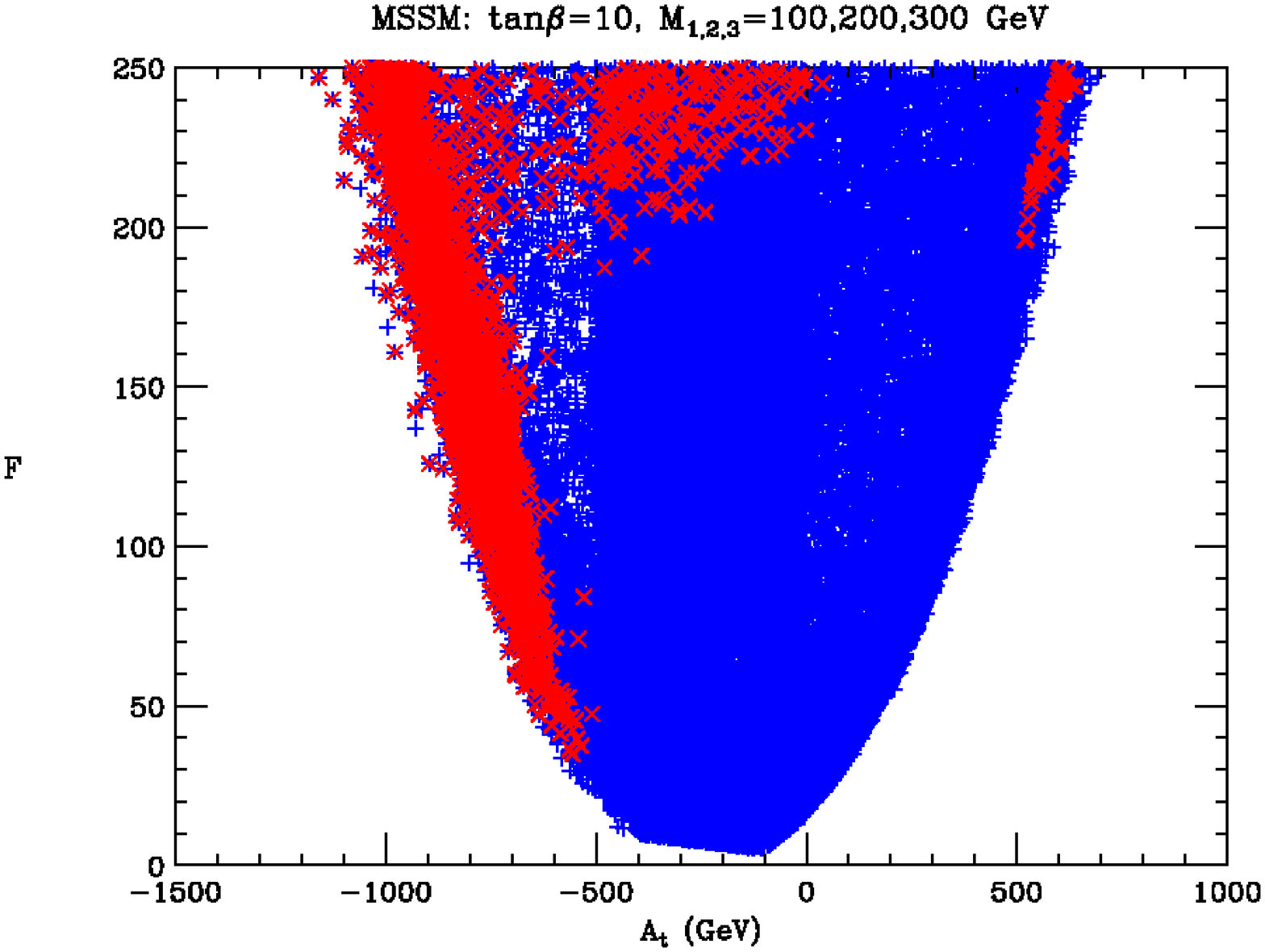}}
  \vspace*{-.1in}
\caption{ Fine tuning vs. $\mstopbar$ (top), $\mh$ (middle)
and $A_t$ (bottom) for randomly generated MSSM parameter choices 
  with $\tanb=10$ and  $M_{1,2,3}(\mz)=100,200,300\gev$.  Blue pluses
  correspond to parameter choices such that $\mh<114\gev$.  Red
  crosses are points with $\mh>114\gev$.}
\label{fvsmhmssm}
\vspace*{-.1in}
\end{figure}

In \cite{Dermisek:2005ar,Dermisek:2005gg}, we presented a first study
of the fine-tuning issues for the NMSSM vs. the MSSM.  We define the
fine-tuning measure to be
\beq F={\rm Max}_p F_p \equiv {\rm Max}_p\left|{d\log \mz\over d\log
    p}\right|\,, \eeq
where the parameters $p$ comprise all GUT-scale soft-SUSY-breaking
parameters.

\section{Comparison of the MSSM with the NMSSM}
\label{sec:compare}

In this section, we will consider scenarios associated with minimal
fine tuning in the MSSM and the NMSSM.  In the following section,
we will give a broader overview of all types of NMSSM scenarios
and will show how it is that one is lead to the NMSSM scenarios
considered in this section.

We discuss fine tuning for the MSSM first.  In this case, the GUT
scale parameters comprise: : $M_{1,2,3}$, $\mu$, $B_\mu$, $\mqsq$,
$\musq$, $\mdsq$, $\mlsq$, $\mesq$, $\mhusq$, $\mhdsq$, $A_t$, $A_b$,
and $A_\tau$. In principle, soft masses squared for the first two
generations should be included above, but they have negligible effect
upon $\mz$.  In our approach, we choose $\mz$-scale values for all the
squark and slepton soft masses squared at scale $\mz$, for the gaugino
masses, $M_{1,2,3}(\mz)$, and for $A_t(\mz)$, $A_b(\mz)$ and
$A_\tau(\mz)$ (with no requirement of universality at the GUT scale).
We also choose $\mz$-scale values for $\tanb$, $\mu$ and $\mha$; these
uniquely determine $B_\mu(\mz)$.  The vevs $h_u$ and $h_d$ at scale
$\mz$ are fixed by $\tanb$ and $\mz$ via $\mz^2=\anti
g^2(h_u^2+h_d^2)$ (where $\anti g^2=g^2+g^{\prime\,2}$). Finally,
$\mhusq(\mz)$ and $\mhdsq(\mz)$ are determined from the two potential
minimization conditions. [From here on, all parameters displayed
without an explicit argument are $\mz$-scale values, although we
sometimes give them an explicit $(\mz)$ argument for emphasis.  All
GUT-scale parameters will be specifically indicated using an explicit
argument $(\mgut)$.]  We then evolve all parameters to the MSSM GUT
scale (including $\mu$ and $B_\mu$).  Next, we shift each of the
GUT-scale parameters in turn, evolve back down to scale $\mz$, and
re-minimize the Higgs potential using the shifted values of $\mu$,
$B_\mu$, $\mhusq$ and $\mhdsq$. This gives new values for $h_u$ and
$h_d$ from which we compute a new value for $\mz$ (and $\tanb$).

It is not difficult to understand why fine tuning is typically large
in the MSSM given LEP constraints. Minimization of the Higgs potential
gives (at scale $\mz$)
\beq
\half\mz^2=-\mu^2+{\mhdsq-\tan^2\beta \mhusq\over \tan^2\beta-1}\,.
\label{mzsquared}
\eeq

The $\mz$-scale $\mu,\mhusq,\mhdsq$ parameters can be determined from
the GUT-scale values of all SUSY-breaking parameters via the
renormalization group equations.  The result
for $\tanb=10$ (similar to the $\tanb=2.5$ results in
Refs.~\cite{Bastero-Gil:1999gu,Kane:1998im}) is
\bea
\mz^2&\sim&-2.0\mu^2(\mgut)+5.9M_3^2(\mgut)+0.8\mqsq(\mgut)\nn\\
&&+0.6\musq(\mgut) -1.2\mhusq(\mgut)\nn\\
&&-0.7M_3(\mgut)A_t(\mgut)+0.2 A_t^2(\mgut)+\ldots
\label{mzfromgut}
\eea
All of the above terms aside from $-2\mu^2(\mgut)$ and $-2\mhusq(\mgut)$
arise from the RGE evolution result for $2\mhusq(\mgut)-2\mhusq(\mz^2)$.
Similarly, one can expand $\mz$-scale values for soft-SUSY-breaking
parameters in terms of GUT-scale parameters.  In particular, one finds
(at $\tanb=10$)
\bea
A_t(\mz)&\sim& -2.3 M_3(\mgut)+0.24 A_t(\mgut)\label{atmz}\\
M_3(\mz)&\sim& 3 M_3(\mgut)\label{m3mz}\\
\mstopbar^2(\mz)&\sim&
5.0 M_3^2(\mgut)+0.6 \mstopbar^2(\mgut)\nn\\
&&+0.2 A_t(\mgut)M_3(\mgut)\label{mstopbarmz}
\,.
\label{rgeatm3}
\eea  
In the above,
\beq
\mstopbar\equiv \left[\half(\mstopone^2+\mstoptwo^2)\right]^{1/2}\,.
\label{mstopbardef}
\eeq 
Unless there are large cancellations (fine-tuning), one would expect that 
\beq
\mz \sim \mgl,\mstopbar,m_{\wtil H^\pm},
\eeq
where $m_{\wtil H^\pm}$ is similar in size to $\mu$.
We would need a very light gluino, and a rather light stop, to avoid
fine-tuning. More precisely, if $A_t(\mgut)=0$, then it is clear from
Eq.~(\ref{mzfromgut}) that the minimum of $F$ is determined by the
$5.9M_3^2(\mgut)$ term, which would give $F\sim 6$ for
$M_3(\mgut)=300\gev$. Allowing for small
positive $A_t(\mgut)$ reduces this minimum $F$ somewhat, as will be
illustrated below.    Of
course, in specific models you can also have correlations among the
GUT-scale parameters that would reduce $F$.

The problem is that the small $A_t$ value required for minimal $F$ does not
yield a MSSM Higgs mass $\mh$ above the $114\gev$ LEP limit unless
$\mstopbar$ is very large (which causes a high level of fine-tuning,
$F> 175$). To maximize $\mh$  at moderate $\mstopbar$, one should
consider parameters corresponding to $|A_t/\mstopbar|\sim \sqrt 6$,
termed an '$\mh$-max scenario'. For such choices it is also possible to
obtain $\mh>114\gev$. To
simultaneously minimize $F$, the sign of $A_t$ must be chosen
negative.  To understand this, we first note that $\mh>114\gev$ can be
achieved with $-A_t\gsim 500\gev$ or $A_t\gsim 500 \gev$ and
$\mstopbar\sim 300\gev$. Given Eq.~(\ref{rgeatm3}), this translates to
$-A_t(\mgut)\gsim 1\tev$ or $A_t(\mgut)\gsim 3\tev$, respectively. In
both cases, $F$ will be determined by the $0.2 A^2_t(\mgut)$ term in
Eq.~(\ref{rgeatm3}), yielding
\beq
F\sim 0.2 {A^2_t(\mgut)\over \mz^2}\,. 
\eeq
Obviously the case of $A_t(\mgut)\gsim 3\tev$ case will correspond to
very large fine tuning, roughly $F\gsim 180$.  For the
$-A_t(\mgut)\gsim 1\tev$ case, $F\gsim 30$ is obtained. Similar values
of $F$ can be obtained for much smaller $|A_t(\mgut)|$ values provided
one allows for moderate $\mstopbar^2(\mgut)<0$~\cite{Dermisek:2006ey}.

The above generic features are apparent in the numerical results
presented in Fig.~\ref{fvsmhmssm} for the case of $\tanb(\mz)=10$ and
$M_{1,2,3}(\mz)=100,200,300\gev$.  We scan randomly over $A_t(\mz)$,
$A_b(\mz)$, $A_\tau(\mz)$ and 3rd generation squark and slepton soft
masses-squared above $(200\gev)^2$, as well as over $|\mu(\mz)|\geq
100\gev$, ${\rm sign}(\mu)=\pm $ and over $\mha>120\gev$. For such
values of $\mha$, the $h$ is quite SM-like and only allowed by LEP
data if $\mh\gsim 114\gev$.  If lower values of $\mha$ are allowed, in
particular $\mha\sim 100\gev$, lower values of $F\sim 16$ can be obtained for
experimentally allowed scenarios.  In these latter scenarios, the
$\hh$ is typically fairly SM-like but will have mass above $114\gev$
while the $\hl$ can have mass below $114\gev$ by virtue having weak
$ZZ$ coupling.  These scenarios are characterized by mixing among the
Higgs bosons. Analogous mixed-Higgs scenarios are also possible in the
NMSSM.  The MSSM and NMSSM mixed-Higgs scenarios will be considered in
a separate paper~\cite{mixedscenarios}.  The fine tuning in  NMSSM mixed-Higgs
scenarios have also been discussed in~\cite{BasteroGil:2000bw}. The
main drawback of mixed-Higgs scenarios is that they require
adjustments in other parameters besides those necessary for correct
electroweak symmetry breaking.

Returning to Fig.~\ref{fvsmhmssm}, the top plot gives $F$ as a
function of $\mstopbar$.  The latter enters into the computation of
the radiative correction to the SM-like light Higgs mass $\mh$.  In
the middle plot, we display $F$ as a function of $\mh$.  And, in the
bottom plot we display $F$ as a function of $A_t$.  We first of all
note that the very smallest values of $F$ are achieved for
$\mstopbar\in[ 300\gev,400\gev]$, $\mh \sim 90-105\gev$ and
$A_t\in[-400\gev,0]$.  As stated above, for $\mha>120\gev$, as
considered here, the $h$ is fairly SM-like in all its couplings to SM
particles.  Thus, points with $\mh<114\gev$, plotted as (blue) $+$'s,
are excluded by LEP data, whereas those with $\mh>114\gev$, plotted as
(red) $\times$'s, are not excluded by LEP.  Although very modest
values of $F$ (of order $F\sim 5$) are possible for $\mh<114\gev$, the
smallest $F$ value found for $\mh\geq114\gev$ is of order $F\sim 34$,
as explained earlier.  The increase of the smallest achievable $F$
with $\mh$ is illustrated in the middle plot.  The modest $F\gsim
34$ values are achieved for special parameter choices, namely, $\mstopbar\sim 300\gev$ and
$A_t\sim -500\gev$ corresponding to a large ratio of $A_t/\mstopbar$;
see earlier discussion.  A value of $F\sim 34$ corresponds to roughly
$3\%$ fine tuning.  Generally speaking, however,  
it would obviously be nicer if
the $\mh\sim 100\gev$ points with $F\sim 5$ were not excluded by LEP.

\begin{figure}[ht!]
  \centerline{\includegraphics[width=2.4in,angle=90]{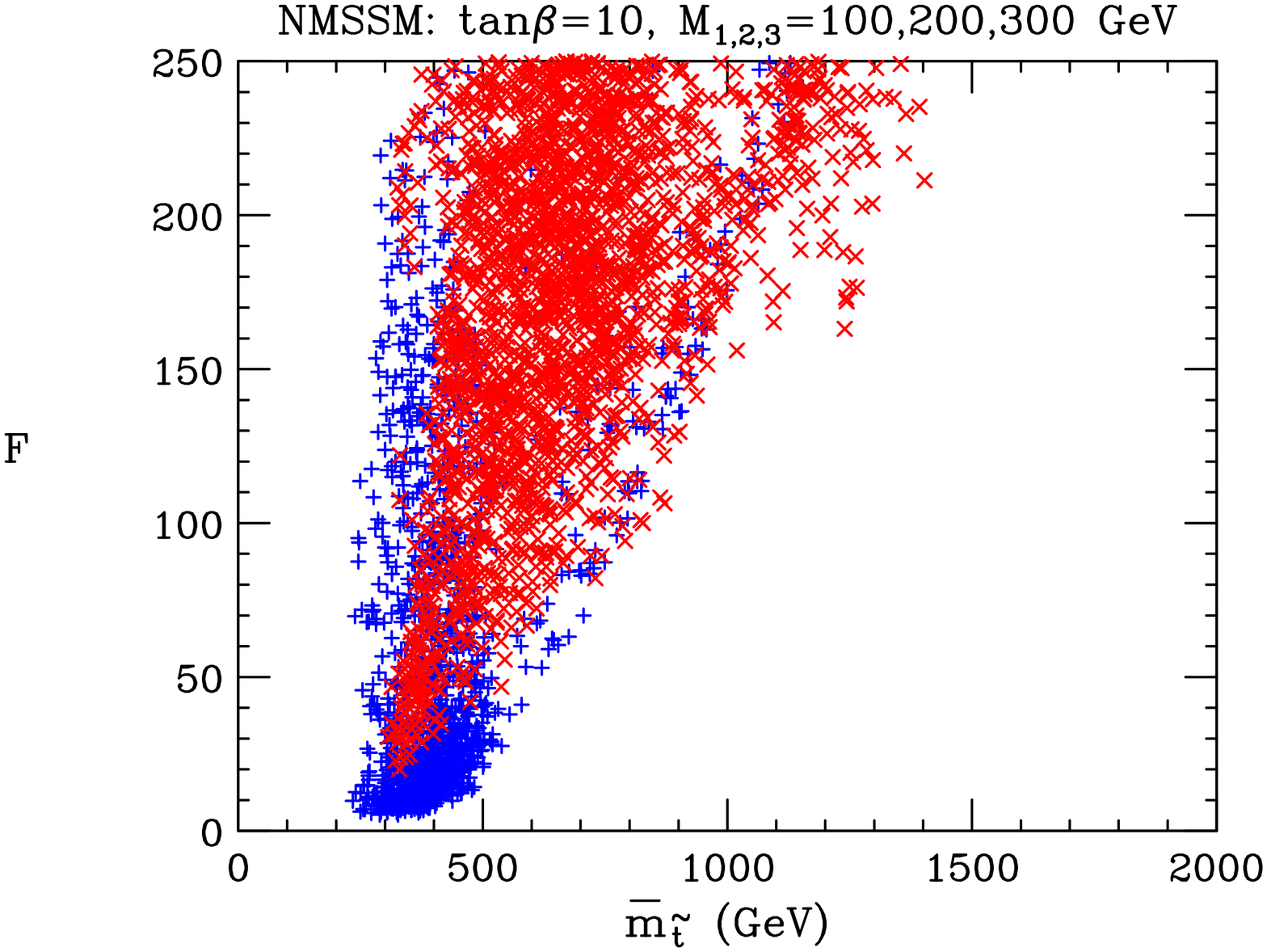}}
\centerline{\includegraphics[width=2.4in,angle=90]{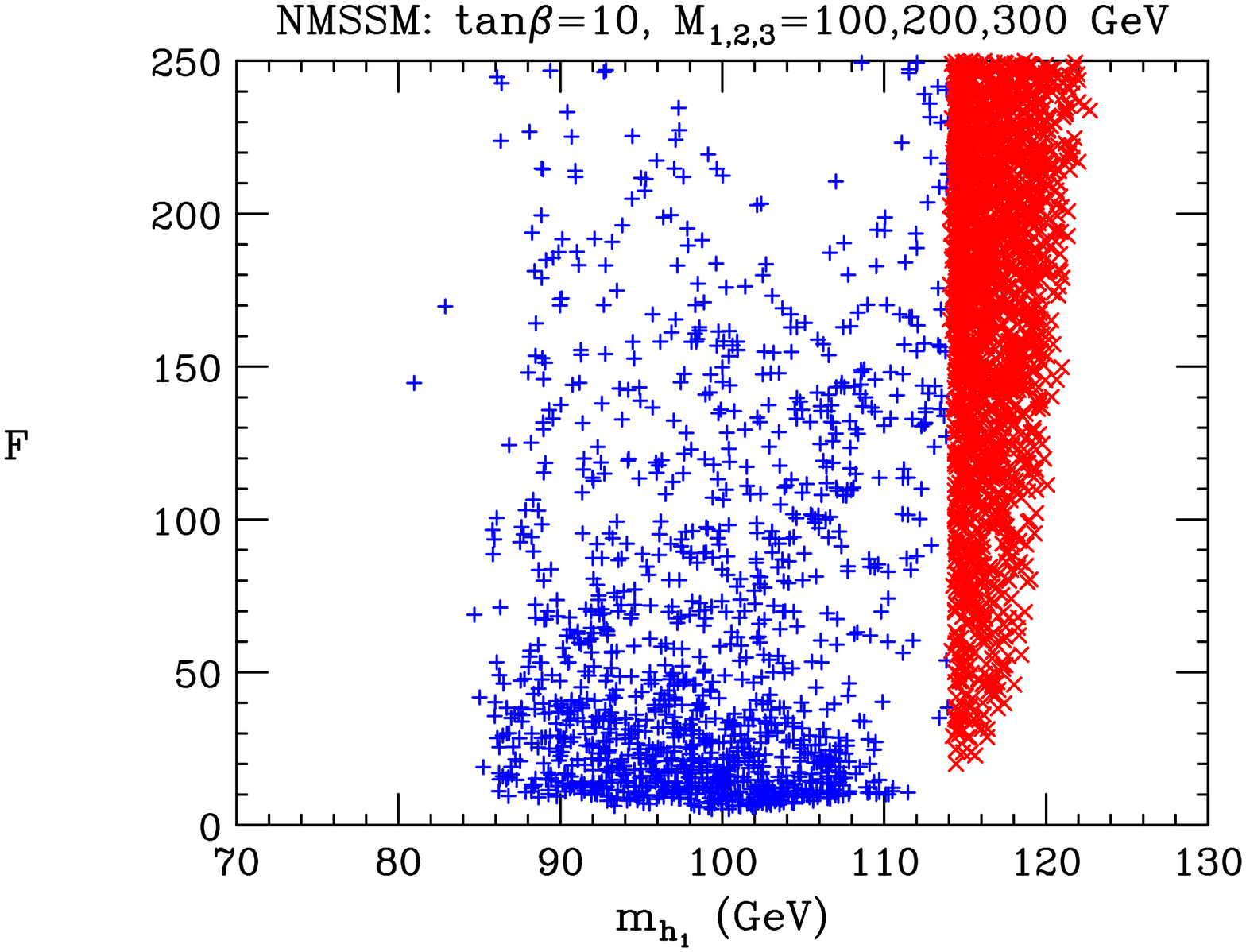}}
\centerline{\includegraphics[width=2.4in,angle=90]{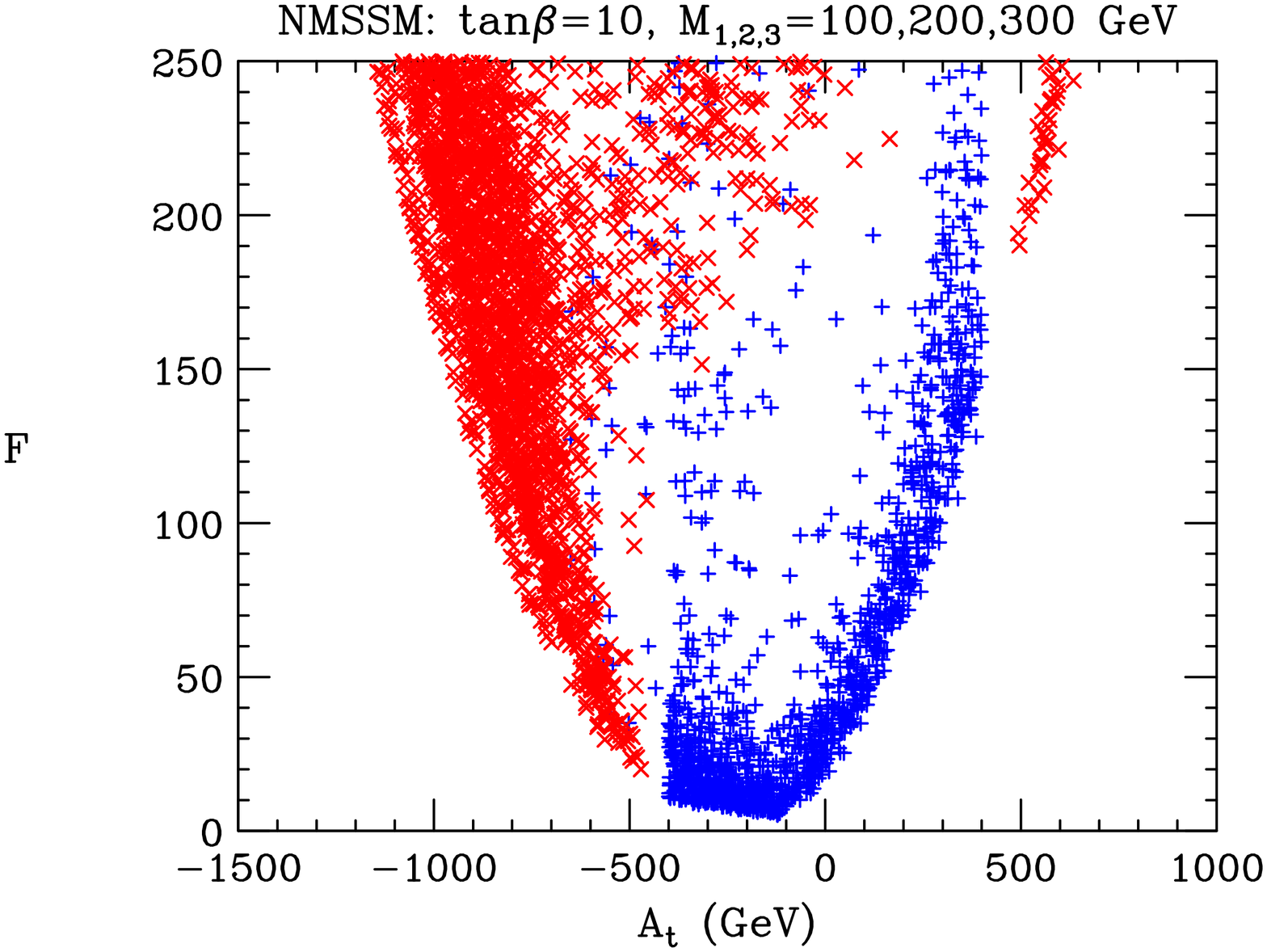}}
  \vspace*{-.1in}
\caption{ For the NMSSM, we plot the fine-tuning measure $F$
  vs. $\mstopbar$, $\mhi$ and $A_t$ for NMHDECAY-accepted scenarios
with $\tanb=10$ and $M_{1,2,3}(\mz)=100,200,300\gev$. 
Points marked by blue '$+$'s are consistent with LEP limits on the
$Z+2b$ channel and the $Z+4b$ channel \cite{newleplimits}, considered separately, {\it but not necessarily with LEP limits on the
  combined $Z+2b$ and $Z+4b$ channels).}
Points marked by red '$\times$'s escape LEP limits due to $\mhi>114\gev$. }
\label{fvsmh1nmssm1}
\vspace*{-.1in}
\end{figure}

We now compare these results to what is found in the NMSSM. Plots
analogous to those for the MSSM appear in Fig.~\ref{fvsmh1nmssm1}.
Let us first first define our conventions (we follow those of Ref.~
\cite{Ellwanger:2004xm}) and discuss a few theoretical points.  
The superpotential for the Higgs fields is 
\beq
W = \lam \widehat{S}
\widehat{H}_u \cdot \widehat{H}_d + \frac{1}{3} \kappa \widehat{S}^3 \ ,
\eeq 
where
\beq
\widehat{H}_u \cdot \widehat{H}_d = \widehat{H}_u^+ \widehat{H}_d^- 
- \widehat{H}_u^0 \widehat{H}_d^0\ .
\eeq
For the soft \susy\ breaking terms we take
\bea
V_\mathrm{soft}&=&
m_\mathrm{H_u}^2 | H_u |^2 + m_\mathrm{H_d}^2 | H_d |^2 + m_\mathrm{S}^2 | S |^2
\nn \\
&+&\left(\lam A_\lam\ H_u \cdot H_d S + \third \kap A_\kappa\ S^3 + \mathrm{h.c.}
\right)\,. 
\eea
(Above, we have not written the usual terms involving Higgs fields and quark/squark
 fields.) Assuming that the parameters of the potential are real, $W$ and
$V_\mathrm{soft}$ together yield a
full potential for the neutral components of the $H_u$, $H_d$ and $S$
scalar fields of the form
\bea
V & = & \lam^2 (h_u^2 s^2 + h_d^2 s^2 + h_u^2 h_d^2) + \kappa^2 s^4
+ \quart \anti g^2 (h_u^2 - h_d^2)^2 \nn\\
&&\quad - 2 \lam\kappa h_u h_d s^2 - 2\lambda A_\lam\ h_u h_d s 
+ \frac{2}{3}\,\kappa A_\kappa  s^3 \nn\\
&&\quad + m_{H_u}^2 h_u^2 + m_{H_d}^2 h_d^2 + m_S^2 s^2
\ .
\label{vhiggs}
\eea

There are now three minimization conditions
\beq
{\ptl V\over \ptl h_u}=0,\quad {\ptl V\over \ptl h_d}=0,\quad {\ptl
  V\over \ptl s}=0
\eeq
which are to be solved for $m_{H_u}^2$, $m_{H_d}^2$ and $m_S^2$ in
terms of the vevs and other parameters appearing in \Eq{vhiggs}.  One
combination of the minimization equations yields the MSSM-like
expression for $\mz^2$ in terms of $\mu^2$, $\tanb$, $\mhusq$ and
$\mhdsq$ with $\mu$ replaced by $\mueff$.  However, a second
combination gives an expression for $\mueff$ in terms of $\mz$ and
other Higgs potential parameters:
\bea
&& \kappa \lambda \left( \frac{1}{\tan \beta} m_{H_d}^2 - m_{H_u}^2
  \tan \beta \right) - \lambda^2 \left( m_{H_d}^2 - m_{H_u}^2  
\right) \nn\\
&&
= \frac{1}{2} \mz^2 \frac{\tan^2\beta - 1}{\tan^2\beta +1}
\left[\kappa \lambda \left(\frac{1}{\tan \beta} + \tan \beta \right) 
- 2 \lambda^2 + \frac{2}{\anti g\,^2} \lambda^4  \right]\nn\\
&& \qquad\qquad+ \mu_{eff} A_\lambda \lambda^2 \left(\frac{1}{\tan \beta} - \tan 
\beta \right) \label{mzform_s}\,.
\eea
Eliminating $\mueff$, we arrive at an equation of the form $ \mz^4 + B
\mz^2 + C = 0$, with solution $\mz^2=-B\pm\sqrt{B^2-4C}$, where $B$
and $C$ are given in terms of the soft SUSY breaking parameters,
$\lam$, $\kappa$ and $\tanb$.  Only one of the solutions to the
quadratic equation applies for any given set of parameter choices.

To explore fine tuning numerically, we proceed analogously to the
manner described for the MSSM. At scale $\mz$, we fixed $\tanb$ and scanned
over all allowed values of $\lam$ ($\lam>0$ by convention and 
$\lam\lsim 0.7$ is required for
perturbativity up to the GUT scale) and $\kap$, and over $100\gev\leq |\mueff|\leq 1.5\tev$, ${\rm
  sign}(\mueff)=\pm$.  We also choose $\mz$-scale values for the
soft-SUSY-breaking parameters $ A_\lam$, $A_\kappa$, $ A_t=A_b$,
$M_1$, $M_2$, $M_3$, $\mqsq$, $\musq$, $\mdsq$, $\mlsq$, and $\mesq$,
all of which enter into the evolution equations. We process each such
choice through NMHDECAY to check that the scenario satisfies all
theoretical and experimental constraints, with the exception that we
plot some points that are consistent with LEP limits on the $Z+2b$ and
$Z+4b$ channels considered separately as in \cite{newleplimits}, but
inconsistent with the LEP constraints on the $Z+b's$ final states,
where $b's=2b+4b$. We shall return to this point shortly.  For
accepted cases, we then evolve to determine the GUT-scale values of
all the above parameters.  The fine-tuning derivative for each
parameter is determined by shifting the GUT-scale value for that
parameter by a small amount, evolving all parameters back down to
$\mz$, redetermining the potential minimum (which gives new values
$\hu^\prime$ and $\hd^\prime$) and finally computing a new value for
$\mz^2$ using $\mz^{\prime\,2}=\anti g^{\,2}(
\hu^{\prime\,2}+\hd^{\prime\,\,2})$.

\begin{figure}[ht!]
\centerline{\includegraphics[width=2.4in,angle=90]{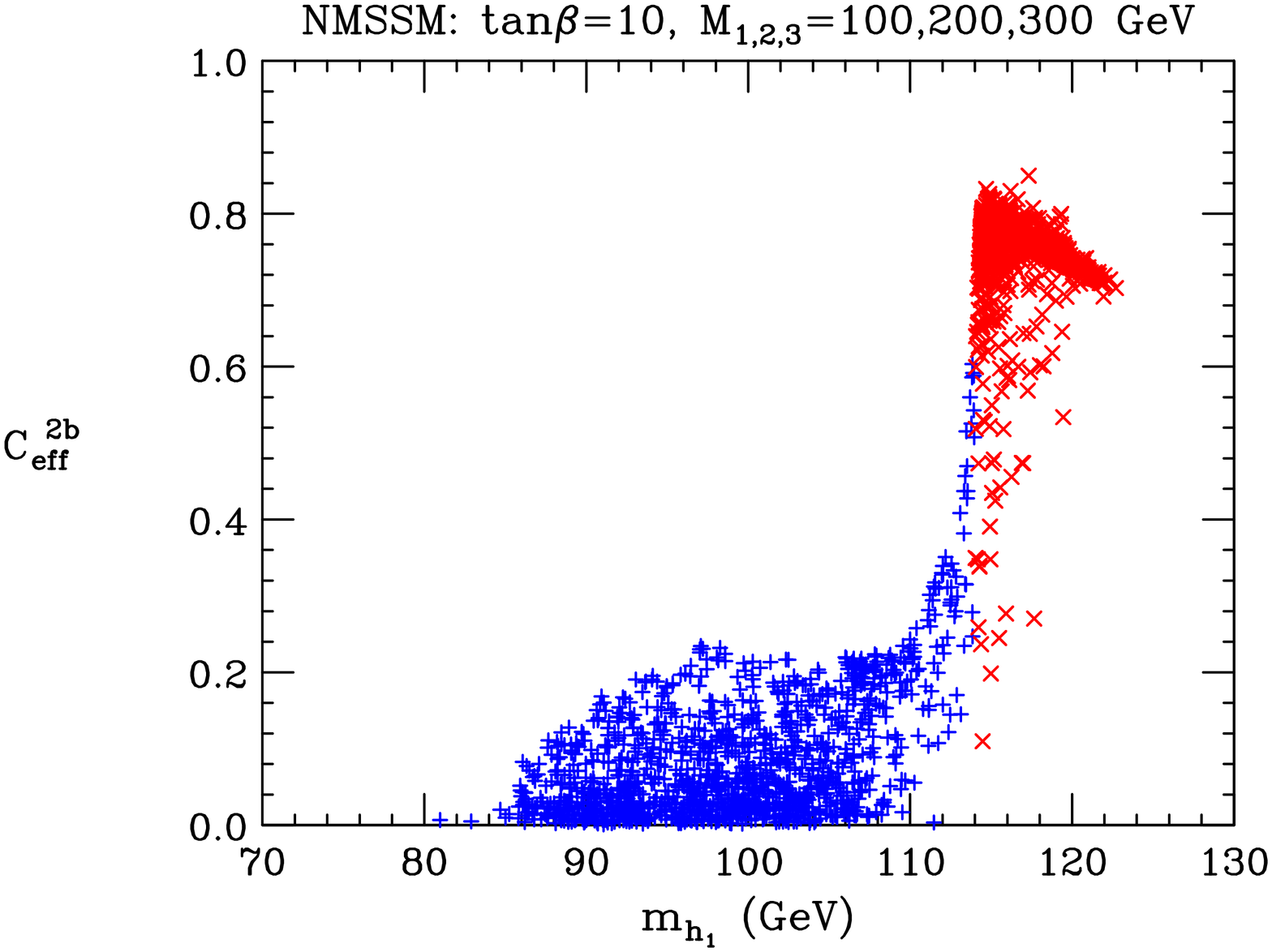}}
\centerline{\includegraphics[width=2.4in,angle=90]{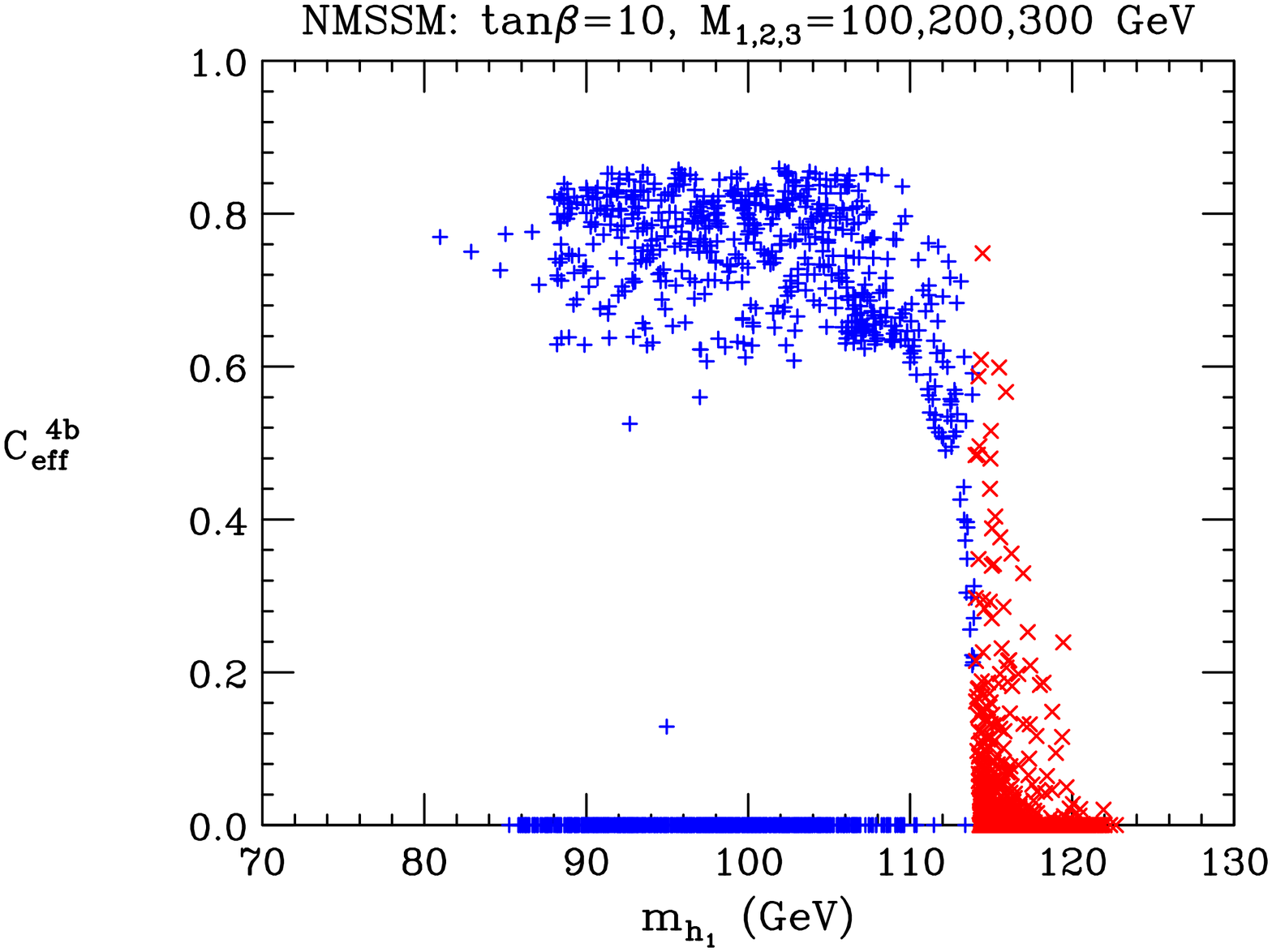}}
\caption{For the NMSSM, we plot   $\cbb$ 
and $\cbbbb$ as a functions of $\mhi$ for NMHDECAY-accepted scenarios
with $\tanb=10$ and $M_{1,2,3}(\mz)=100,200,300\gev$.
 Point notation as in Fig.~\ref{fvsmh1nmssm1}.}
\label{nmssmceff}
\end{figure}

\begin{figure}[ht!]

  \centerline{\includegraphics[width=2.4in,angle=90]{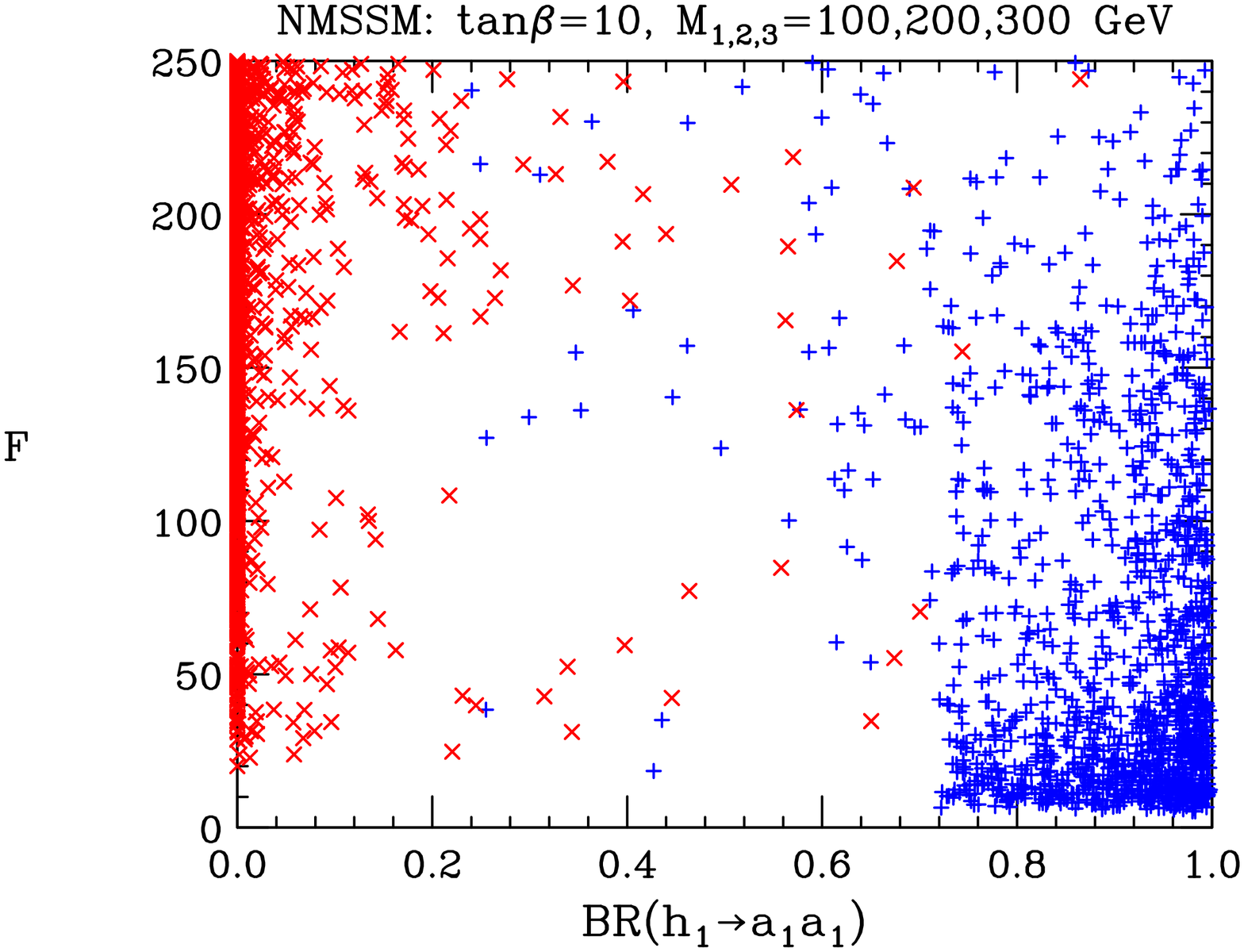}}
  \centerline{\includegraphics[width=2.4in,angle=90]{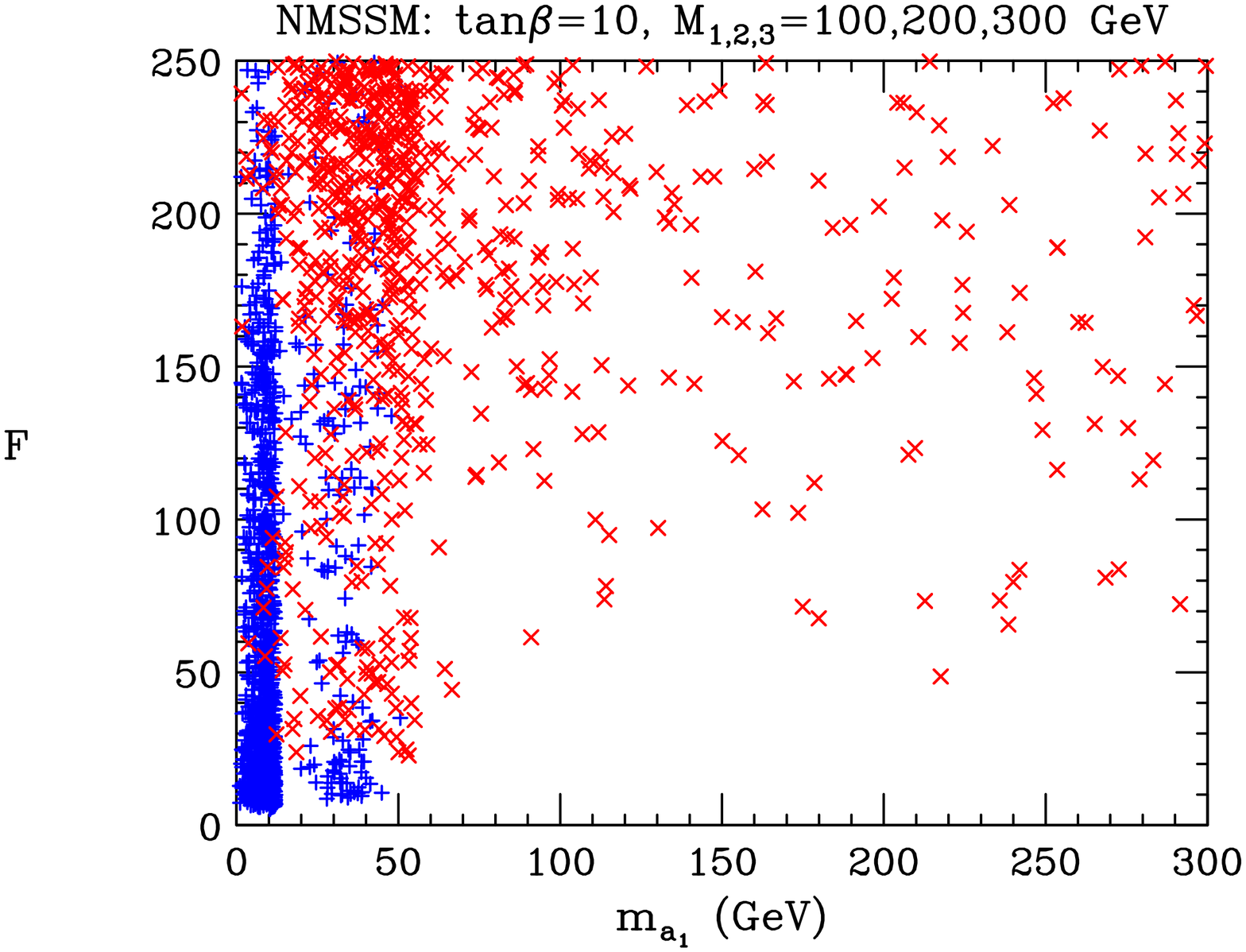}}
\caption{For the NMSSM, we plot the fine-tuning measure $F$
  vs. $BR(\hi\to\ai\ai)$ and vs. $\mai$ for NMHDECAY-accepted scenarios
with $\tanb=10$ and $M_{1,2,3}(\mz)=100,200,300\gev$.
 Point notation as in Fig.~\ref{fvsmh1nmssm1}.}
\label{nmssmbrhaa}
\end{figure}

Our basic results are displayed in Fig.~\ref{fvsmh1nmssm1}. The
density of points of a given type should not be taken as having any
significance --- it simply reflects the nature of the scanning
procedures employed for the various cases. In particular, the scans
used to obtain results presented in this section were designed to
focus on parameter regions with $F<250$. Further, we focused a
lot of our scans on keeping only points with $\mai<20\gev$. 

In Fig.~\ref{fvsmh1nmssm1}, one sees
a lot of similarity between the NMSSM
 plots and those for the MSSM, with differences
to be noted below. Again, one finds that $F<10$ (\ie\ 
no worse than 10\% fine-tuning of GUT scale parameters) is easily
achieved in the NMSSM for the present modest gluino mass of $300\gev$
if the mean stop mass is of order $300-400\gev$ which yields $\mhi$
$\sim 100\gev$ (for the case of $\tanb=10$ --- variation with $\tanb$
will be noted later). The associated $A_t$ values are of very modest
size, lying in the range $[-400\gev,-100\gev]$.  Further, as described
in more detail later, the low-$F$ scenarios are once again such that
the $\hi$ is quite SM-like as regards its couplings to $WW$, $ZZ$ and
$f\anti f$.  The difference between these plots and the earlier MSSM
plots is that many (but not all, as we shall explain) of the
$\mhi<114\gev$ points plotted escape all LEP limits.  Of course,
points with $\mhi>114\gev$ escape LEP limits simply by being above the
maximum LEP-excluded mass. 

Let us next discuss in more detail the points with $\mhi<114\gev$.
They are plotted in Fig.~\ref{fvsmh1nmssm1} provided that they are
consistent with LEP limits on the $Z+2b$ and $Z+4b$ channels,
considered separately as plotted and tabulated in \cite{newleplimits},
but not necessarily the combined $Z+b's$ limit.
 All the plotted $\mhi<114\gev$ points pass the $Z+2b$ and $Z+4b$
 separate channel
limits by virtue of the fact that $\br(\hi\to\ai\ai)$ is large enough
that $\br(\hi\to b\anti b)$ is sufficiently suppressed that $Z+\hi\to
Z+b\anti b$ lies below the LEP $Z+2b$ limit, while simultaneously
$Z+\hi\to Z+\ai\ai\to b\anti b b\anti b$ lies below the $Z+4b$ limit.
The values of $\cbb$ and $\cbbbb$ are plotted in Fig.~\ref{nmssmceff} as
functions of $\mhi$.  In the $\cbbbb$ plot one finds two classes of
points with $\mhi<114\gev$. The first class has large $\cbbbb$ 
(but still below LEP limits for this individual channel) by
virtue of the fact that $\br(\hi\to\ai\ai)$ is large {\it and}
$\mai>2\mb$. The second class of points has zero $\cbbbb$ since
$\mai<2\mb$.  $F$ as a function of $\br(\hi\to\ai\ai)$ is shown in the
top plot of Fig.~\ref{nmssmbrhaa}.

However, the plotted $\mhi<114\gev$ points with non-zero $\cbbbb$ are
mostly not consistent with the LEP data.  NMHDECAY allows these points
because it does not take into account the need to combine $Z+2b$ and
$Z+4b$ final states in confronting the LEP limits, which are
effectively (at least roughly) on the sum of these two final states.
For those $\mhi<114\gev$ points with non-zero $\cbbbb$, the sum of
$\cbb$ and $\cbbbb$ is typically large and one expects such points to
have too large a net $Z+b's$ rate, where $b's=2b+4b$.  Indeed, for the
limited number of points that were analyzed using the full LEP Higgs
working group code one indeed finds \cite{bechtle} that those points
with $\mhi\lsim 110\gev$ that have large $\br(\hi\to\ai\ai)$ {\it and} $\mai>2\mb$ (implying $\ai\to
b\anti b$ predominantly) are inconsistent with LEP limits on the net
$Z+b's$ rate. Without analyzing every one of our $\mai>2\mb$ points
using the full code, we cannot be sure that this same statement
applies to all of them.

With this proviso, we thus find that when $\mhi<114\gev$, one needs
large $\br(\hi\to\ai\ai)$ {\it and} $\mai<2\mb$ to evade LEP limits.
The bottom plot of Fig.~\ref{nmssmbrhaa} shows that it is easy to
obtain very low $F$ points that satisfy both criteria.  (The less
frequent occurrence of $\mai>20\gev$ points in this plot is purely an
artifact of our scan procedure.) We will turn to a discussion of this
in more detail shortly.  

As regards $\mhi\geq 114\gev$ points (which
are not subject to LEP limits), returning to Fig.~\ref{fvsmh1nmssm1},
we find that $F$ values as low as $\sim 20$ (\ie\ only $5\%$ tuning of
GUT-scale parameters) can be achieved for the special choices of
$\mstopbar\sim 300\gev$ and $A_t\sim -500\gev$. This is the same
region of stop parameter space that yields a minimum $F\sim 34$ with
$\mhi>114\gev$ in the MSSM.

We have also found another type of point with low $F$ and $\mhi\sim
100\gev$ that escapes published LEP limits as follows.  First, for
these points $\br(\hi\to\ai\ai)$ is large, $\gsim 0.75$, so that
$\br(\hi\to b\anti b)\lsim 0.2$, implying a perfectly acceptable LEP
rate in the $Zh\to Z2b$ channel.  Second, the $\ai$ is highly singlet
and decays mainly into two photons, $\br(\ai\to\gam\gam)\gsim 0.9$.
Thus, there is negligible contribution to the $Zh\to Z4b$ channel.
Thirdly, $\mai$ is typically fairly substantial for these points,
$\mai\sim 30-45\gev$.  However, these points are highly fine-tuned in
the sense that the highly singlet nature of the $\ai$ required for
large $\br(\ai\to \gam\gam)$ is very sensitive to GUT scale
parameters. This is why they do not appear in the random scans
discussed above. Locating such points requires an extremely fine scan
over a carefully chosen part of parameter space.
We will give more details regarding these points later.

We now briefly describe a third class of points that manage to have
relatively low fine-tuning. Generically, in the NMSSM it is easy to
have $\mhi<114\gev$ without violating LEP limits simply by choosing
parameters so that the $\hi$ has substantial singlet $S$ component.
In this way, the $ZZ\hi$ coupling is suppressed and the $\epem\to
Z^*\to Z\hi$ production rate is reduced to an allowed level even if $\hi\to
b\anti b$ decays are dominant. In such scenarios, it is typically the
$\hii$ that is the most SM-like CP-even Higgs boson, but
$\mhii>114\gev$ and LEP constraints do not apply to the $\hii$.
We have performed a broad scan over NMSSM parameter space to look for
and investigate the fine tuning associated with
scenarios of this type.  We find that not
all the points of this type found in our scans
are highly fine-tuned.  There is a specific parameter region that
produces points of this type 
that are only moderately fine-tuned and for which the $\hi$ escapes
LEP limits by virtue of small $ZZ\hi$ coupling. The lowest $F$ value
that we have found for such points is $F\sim 16$.  In a separate paper
\cite{mixedscenarios}, we will describe these scenarios and their
fine-tuning in detail and compare to similar MSSM scenarios that are
found when $\mha\sim 100\gev$ points are included in the MSSM parameter scans.

\begin{figure}[h!]
  \centerline{\includegraphics[width=2.4in,angle=90]{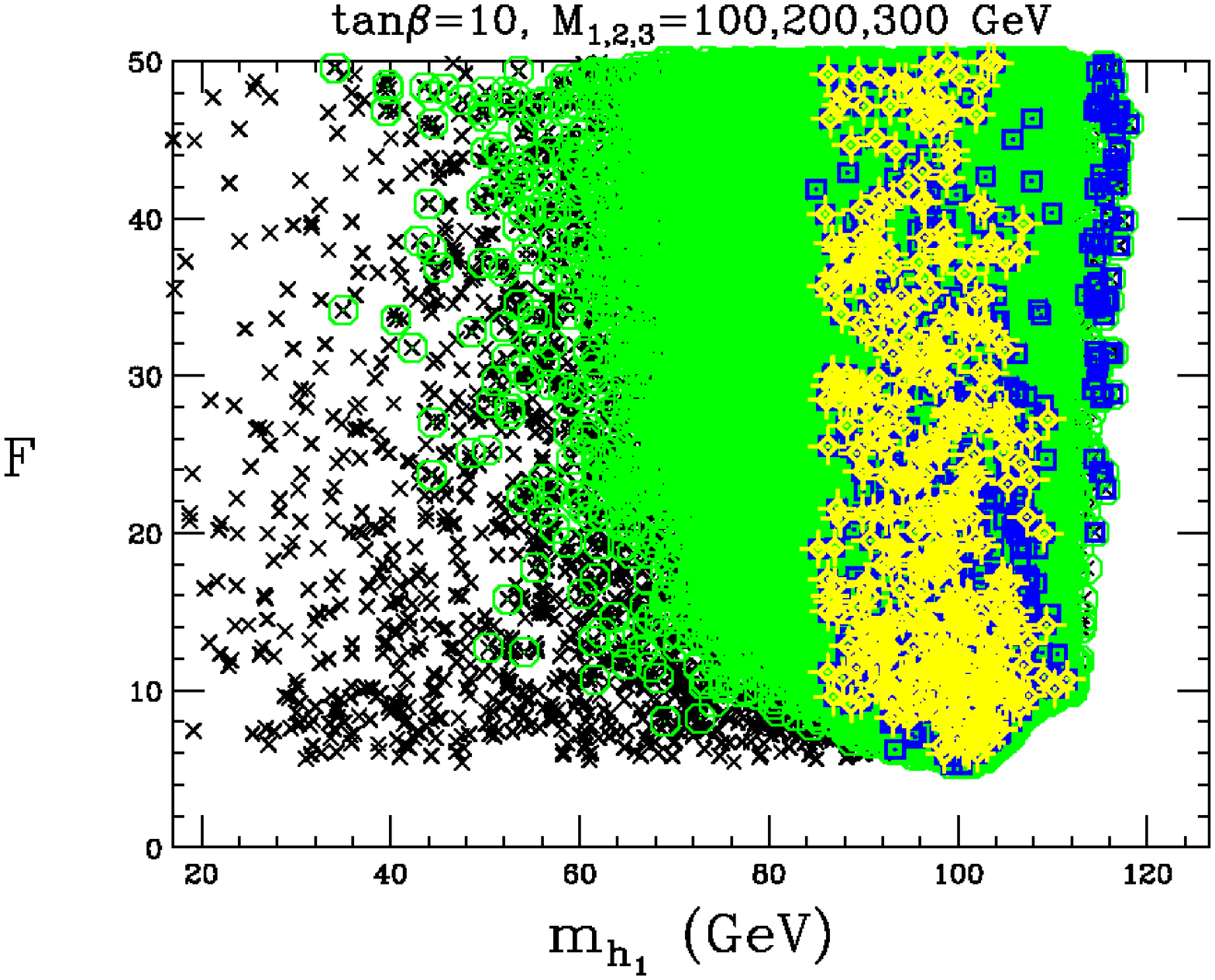}}
  \centerline{\includegraphics[width=2.4in,angle=90]{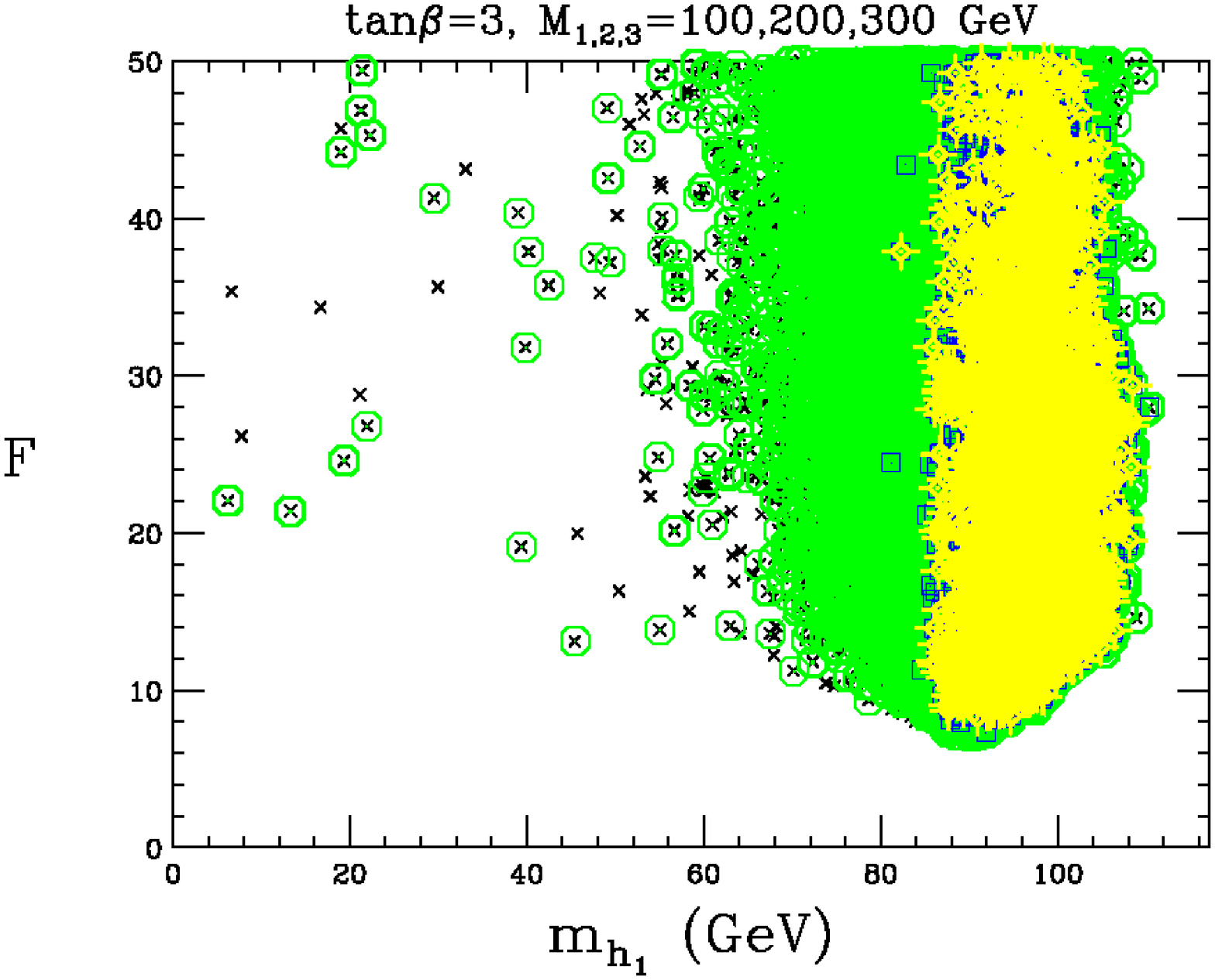}}
  \centerline{\includegraphics[width=2.4in,angle=90]{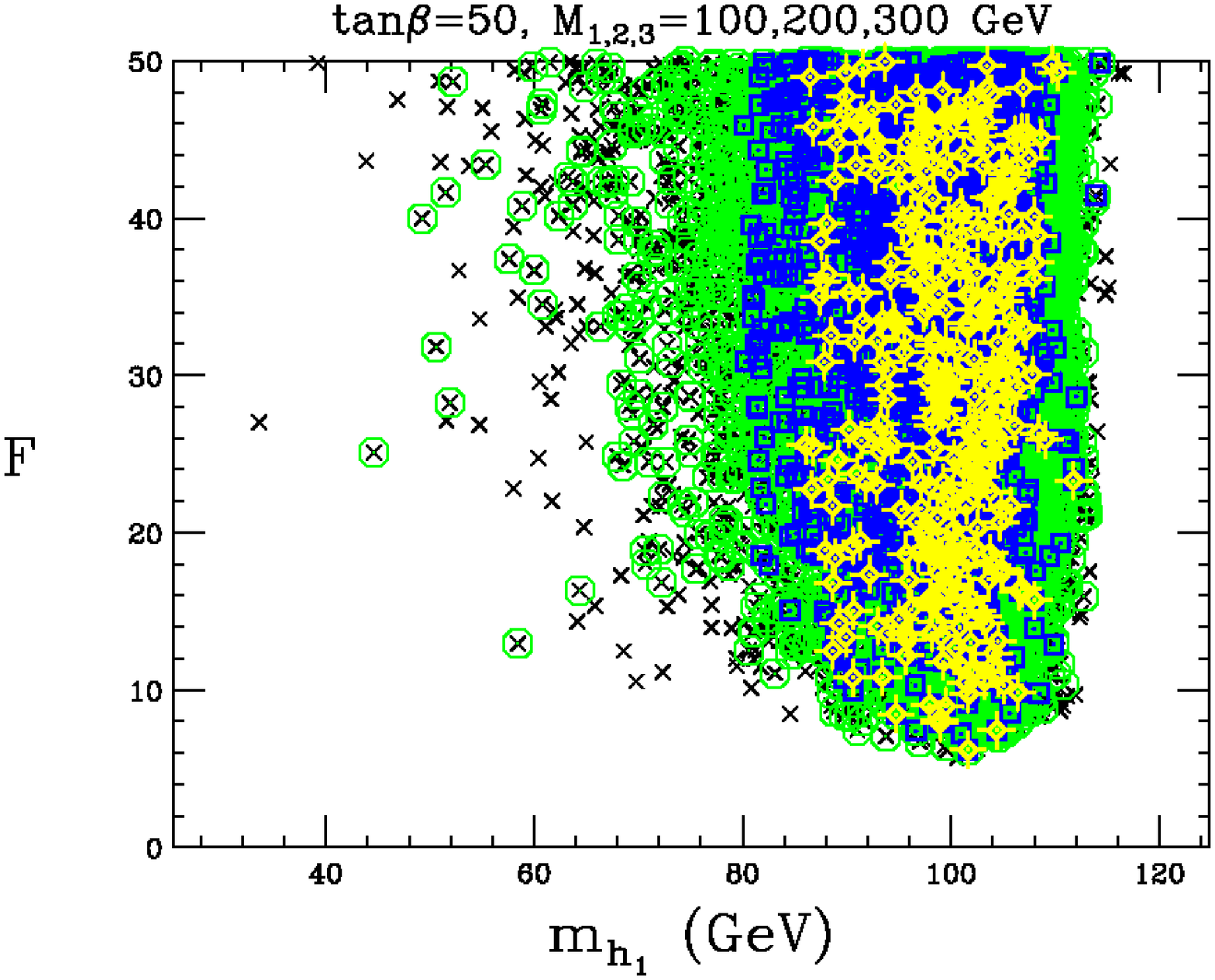}}
  \vspace*{-.14in}
\caption{ Fine tuning vs. $\mhi$ for $\tanb=10$, $\tanb=3$  and 
$\tanb=50$  for
  points with $F<50$,  taking $M_{1,2,3}(\mz)=100,200,300\gev$. Small
black $\times$ points are those obtained  after requiring a global and local minimum, no Landau pole before $\mgut$ and
a neutralino LSP. The green O's are those that 
in addition satisfy experimental limits on stops
and charginos, but not necessarily Higgs limits. The blue
  $\square$'s are the points that remain after imposing 
all LEP single channel Higgs limits, in particular limits
\cite{newleplimits} on the $Z+2b$
and $Z+4b$ channels considered separately. The 
yellow fancy crosses are the blue square points that remain 
after requiring $\mai<2\mb$, 
so that LEP limits on $Z+b's$, where $b's=2b+4b$, are not violated. }
\label{fvsmh1}
\vspace*{-.1in}
\end{figure}

\section{The Low-Fine-Tuning Region}
\label{sec:lowftregion}

Armed with this overview, we now return to the parameter region of the
NMSSM that allows for the lowest possible fine tuning, as studied
earlier in Section~\ref{sec:compare} for $\tanb=10$. Here, we consider
also $\tanb=3$ and $\tanb=50$.  These scans are focused very
much on parameter choices that can yield the  lowest $F$ values.
The relevant plots are presented in Fig.~\ref{fvsmh1}.
We present our results in a somewhat different manner than in
Sec.~\ref{sec:compare} so as to stress the remarkable preference
for $\mhi\sim 100\gev$ in order to achieve the very lowest $F$ values
at $\tanb=10$, with corresponding preferences for $\mhi\sim 101\gev$
at $\tanb=50$ and $\mhi\sim 90\gev$ at $\tanb=3$. First, we stress
that the above $\mhi$ values are the largest ones consistent with low
$F$ in an unbiased (\ie\ before applying experimental constraints of
any kind) scan over the part of parameter space that is simply
theoretically consistent (see below).  Once one imposes lower bounds
on the stop (and chargino) masses, $F$ shows a distinct minimum at the
above $\mhi$ values.  The preference for these values of $\mhi$ to
achieve low $F$ becomes progressively more apparent as one imposes in
addition: (a) LEP constraints on Higgs bosons, including the important
$Z+2b$ channel and $Z+4b$ channels, considered separately (as plotted
and tabulated in \cite{newleplimits}), but not the
combined $Z+2b$ and $Z+4b$ channels; and (b) LEP constraints on the
combined $Z+2b$ and $Z+4b$ channels.

Let us first focus on the $\tanb=10$ case. Four different
types of points are displayed.   The black crosses show
$F$ as a function of $\mhi$ after requiring only that the scenario be
theoretically consistent, but before any experimental constraints
whatsoever are imposed.  The most important components of the
theoretical consistency are: (1) that the vacuum corresponds to a
proper electroweak symmetry breaking vacuum at a true minimum of the
potential; and (2) that the couplings remain perturbative during
evolution up to the GUT scale. The black crosses already single out
$\mhi\sim 100\gev$ as the point above which $F$ rises rapidly.
Black points with low $\mhi$ typically have a rather low value for
$\mstopone$ that is clearly inconsistent with LEP and Tevatron
limits. The minimum $F$ for these low-$\mhi$ black cross points
is fairly independent of $\mhi$. 
The (green) circles correspond to the
black crosses that survive after imposing experimental limits on
$\mstopone$ and $\mcpmone$ and similar non-Higgs constraints.  
We immediately see a striking preference for $\mhi\sim 100\gev$ in
order to achieve minimum $F$. The
(blue) squares indicate the points that survive after requiring {\it
  in addition} that the scenario be consistent with LEP Higgs limits,
including the $Z+2b$ and  $Z+4b$ final state limits considered
separately \cite{newleplimits}, {\it but before imposing a
  limit on the combined $Z+b's$ ($b's=2b+4b$) final state}. In the case of
$\tanb=10$, these blue squares are the union of the $F<50$ red
$\times$'s and blue $+$'s of the middle plot of
Fig.~\ref{fvsmh1nmssm1}. 
The blue-square points now show a very strong preference for $\mhi\sim
100\gev$, even before, but especially after, focusing on minimal $F$.
The final large (yellow) crosses are the
$\mhi<114\gev$ points among the (blue) square points that have
$\mai<2\mb$ so that there is no contribution to the $Z+4b$ channel
from the $\hi\to\ai\ai$ decay that, in turn, has a sufficiently large
branching ratio to allow these points to escape the $Z+2b$ channel LEP
limit. Now $\mhi\sim 100\gev$ is clearly singled out.

In the plots for the $\tanb=3$ and $\tanb=50$ cases, 
we did not bother to generate points with low $\mstopone$. So the
black cross points in these cases simply indicate the presence of a
few scenarios with $\mstopone$ above experimental limits but with
$\mcpmone$ below existing limits or some other non-Higgs experimental
inconsistency. The green-circle, blue-square and yellow-cross are
as described above.

The large number of blue-square points with very low $F$ indicate that
a significant fraction of the very lowest $F$ scenarios are such that
$\hi$ decays primarily into a pair of the lightest CP-odd Higgs bosons
of the model, $\hi\to\ai\ai$.  The yellow crosses show that low-$F$
points with large $\br(\hi\to\ai\ai)$ {\it and} $\mai<2\mb$ are often
found.  For such points, $\ai\to \tauptaum$ (or $q\anti q+gg$ if
$\mai<2m_\tau$) thereby allowing consistency with LEP constraints on
the $Z+b's$ channel and, in many cases, the LEP excess in the $\hi\to
b\anti b$ channel for Higgs mass of order $100\gev$.

\subsection{Is small $\mai$ natural?}

Given that low $F$ can be easily 
achieved without violating LEP constraints if $\mai<2\mb$, an
important issue is whether obtaining small $\mai$ requires fine-tuning
of GUT-scale parameters. In fact, 
a light $a_1$ is natural in the NMSSM in the
$\akap,\alam\to 0$ limit. This can be understood as a
consequence of a global $U(1)_R$ symmetry of the scalar potential (in
the limit $\akap, \alam \to 0$) which is spontaneously broken
by the vevs, resulting in a Nambu-Goldstone boson in the
spectrum~\cite{Dobrescu:2000jt,Hiller:2004ii,Dermisek:2006wr}.\footnote{The
  alternative for getting a light $\ai$ is to have a slightly broken
  Peccei-Quinn symmetry.  However, the models with low $F$ are not
  close to the Peccei-Quinn symmetry limit.}
This symmetry is explicitly broken
by the trilinear soft terms so that for small $\akap, \alam$
the lightest CP odd Higgs boson is naturally much lighter than other
Higgs bosons. In fact, as discussed in depth
in~\cite{Dermisek:2006wr}, the values of $\akap, \alam$ needed 
to have $\mai<2\mb$, $\br(\hi\to\ai\ai)>0.7$ and low $F$ are quite
natural in the context of the NMSSM starting from small or zero values
of $\akap(\mgut)$ and $\alam(\mgut)$. In particular, large
$\br(\hi\to\ai\ai)$ is essentially automatic for typical RGE generated
values of $\alam$ and $\akap$  ($|\akap(\mz)|<10\gev$ and
$|\alam(\mz)|<200\gev$) and it is only a question of whether the
requirement $\mai<2\mb$ is naturally
achieved. In~\cite{Dermisek:2006wr}, we found that essentially no tuning
of $\alam$ and $\akap$ is required in many model contexts. For
example, in the case of $\tanb=10$ tuning of the GUT-scale parameters
needed to achieve appropriate $\alam$ and $\akap$ is likely to be minimal 
for scenarios in which the  $\ai$ is about 10\% non-singlet at the
state-mixing, amplitude level, \ie\ 1\% at the probability level.

More precisely, let us define
\beq
\ai=\cta A_{MSSM}+\sta A_S\,,
\eeq
where $A_{MSSM}$ is the usual two-doublet CP-odd state and $A_S$ is
the CP-odd state coming from the $S$ field. Then,
 the mass of the lightest CP-odd Higgs boson in the simplest
 approximation is given by:
\begin{equation}
m_{a_1}^2  \simeq 3s \left( \frac{3 \lambda A_\lambda \cos^2
\theta_A }{2 \sin 2 \beta}
 - \kappa  A_\kappa \sin^2 \theta_A \right)\,. \label{eq:ma1_first}
\end{equation}
We see that the $\alam$ contribution to $\mai$ is suppressed
relative to the $\akap$ contribution for small $\cta$ and large
$\tanb$ and an appropriate balance between the
contributions is naturally achieved. 
In \cite{Dermisek:2006wr}, we defined a measure called $G$ that
encapsulates the amount of tuning at the GUT scale that is likely to
be needed to achieve small $\mai$. $G$ is defined using
\bea
\falam\equiv {\alam \over \mai^2}{d\mai^2\over
  d\alam},\quad \fakap \equiv {\akap \over \mai^2}{d\mai^2\over
  d\akap}
\eea
as
\beq
G\equiv {\rm Min}\left\{ {\rm Max}\left[|\falam|,|\fakap|\right],|\falam+\fakap|\right\}\,.
\eeq
As shown in \cite{Dermisek:2006wr}, small $G$ implies it
is quite natural to get small $\mai$ even for fairly general
$\mgut$-scale boundary conditions. 
For example, if Eq.~(\ref{eq:ma1_first}) is approximately correct, so
that $\mai^2$ is linear in $\alam$ and $\akap$, and if $\alam$ and
$\akap$ are primarily sensitive to a single GUT-scale parameter $p$,
then, if $|\falam+\fakap|$ is small, sensitivity of $\mai^2$ to $p$ is
guaranteed to cancel. Nonetheless, the measure $G$
should not be overemphasized since specific boundary condition choices
can give small $\mai$ even when $G$ is large.

\begin{figure}
\begin{center}
\includegraphics[width=2.4in,angle=90]{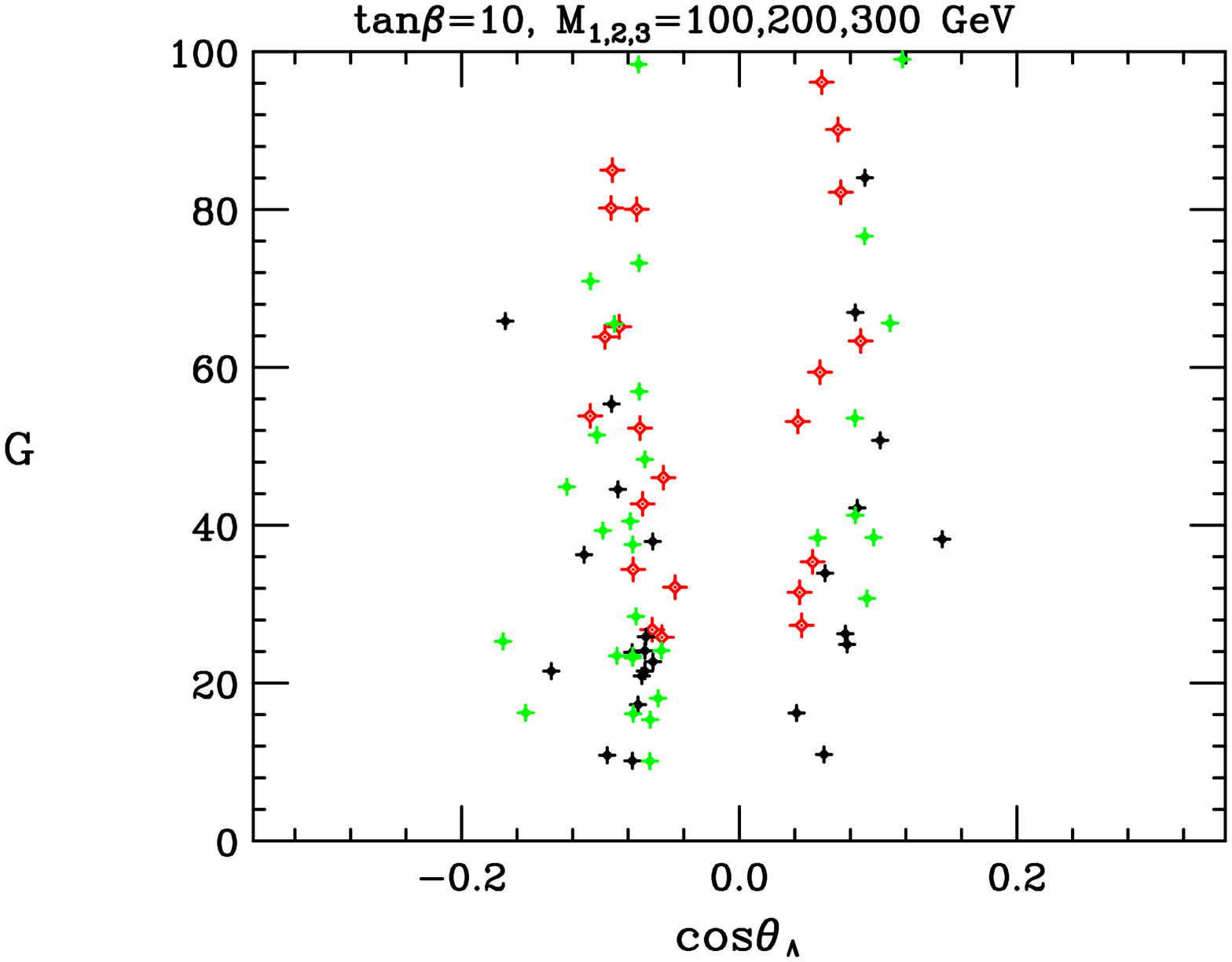}
\includegraphics[width=2.4in,angle=90]{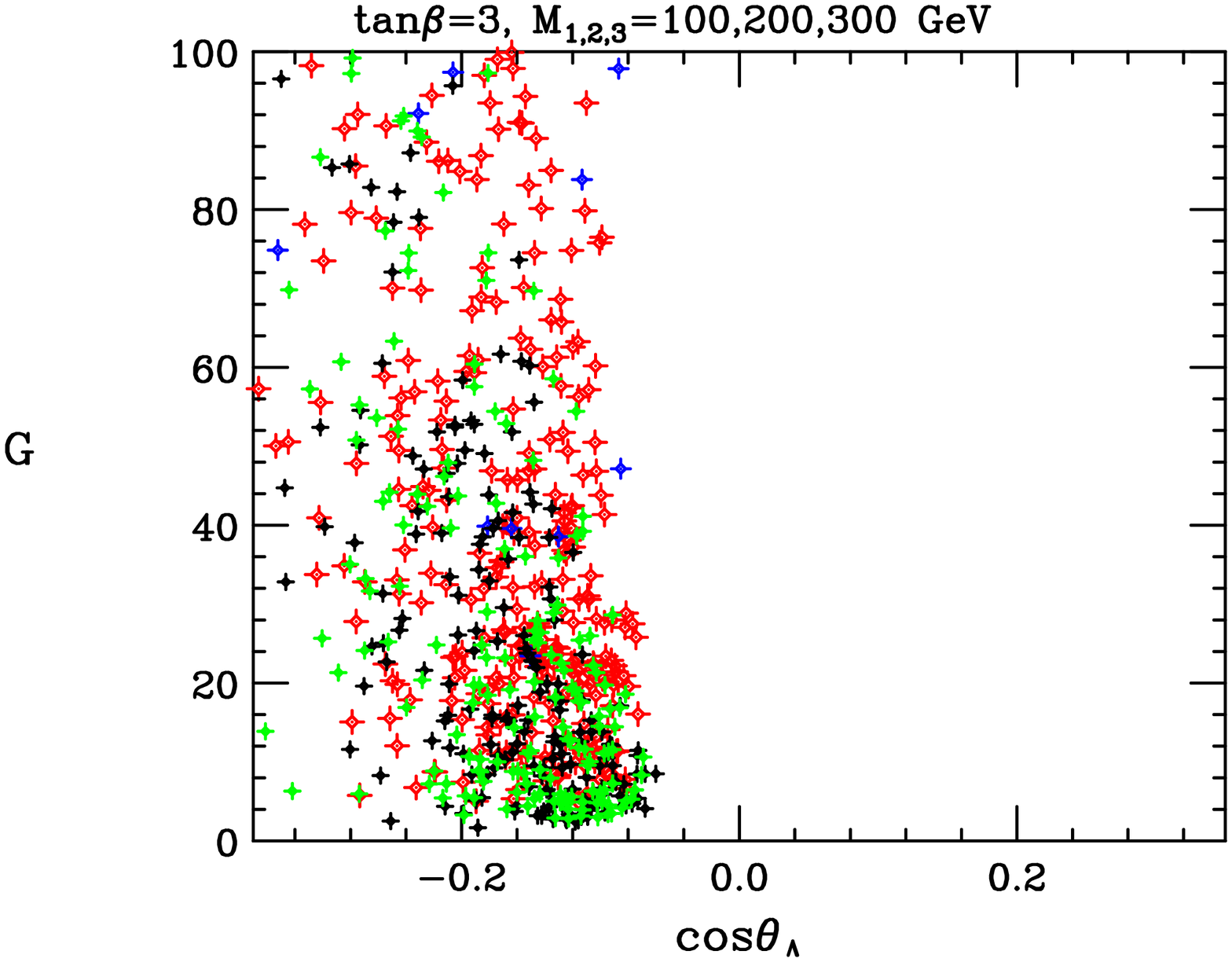}
\includegraphics[width=2.4in,angle=90]{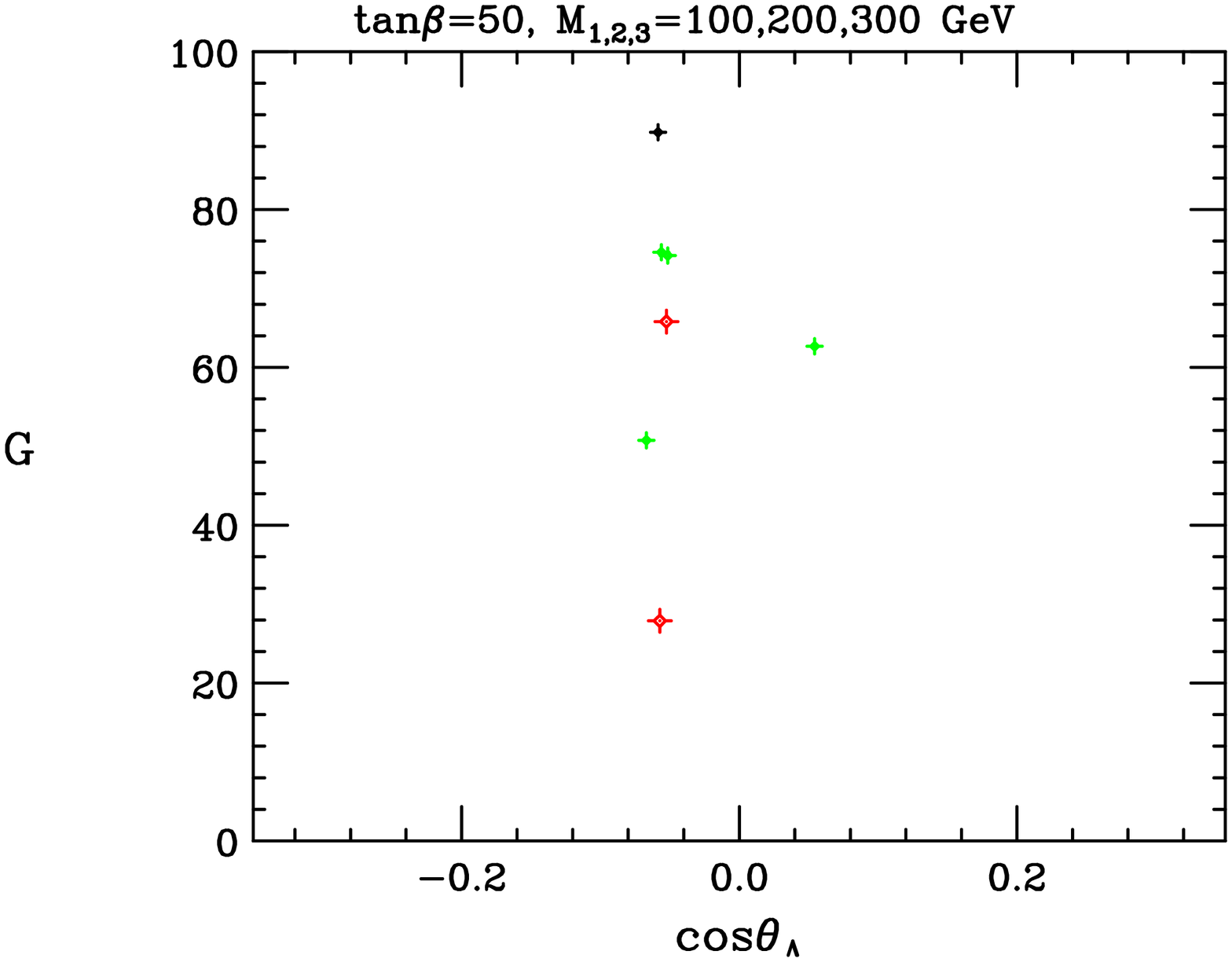}
\end{center}
\caption{ For the $F<15$ scenarios that are fully consistent with all
  LEP constraints, we plot $G$ vs. $\cta$ taking
$M_{1,2,3}=100,200,300\gev$ and $\tanb=10$ (top), $3$ (middle) and
$50$ (bottom). 
The point coding is: black $= 8.8\gev<\mai<\mhi/2$; 
dark grey (red) $=2\mtau<\mai<7.5\gev$;  
light grey (green) $=7.5\gev<\mai<8.8\gev$; and darkest grey (blue) = $\mai<2\mtau$.}
\label{gvscta}
\end{figure}

In Fig.~\ref{gvscta}, we plot $G$ as a function of $\cta$ for
parameters choices that yield $F<15$, taking 
$\tanb=10$, 3, and 50. The results are displayed using different
shadings (colors) for different ranges of $\mai$, as delineated in the
figure caption. For $\tanb=10$, we see that $G$ is minimal for
$|\cta|\sim 0.08$, not much above the rough lower bound of $|\cta|\sim
0.06$. This lower bound is a direct consequence of the {\it combined} requirements
that: 1) $\br(\hi\to\ai\ai)$ be so large that $\br(\hi\to b\anti b)$
is small enough to escape the LEP limit on the
$Z+2b$ channel; and 2) that $\mai<2\mb$ so that $\ai$ does not decay 
to $b\anti b$ and there is no $\hi\to \ai\ai\to b\anti b b\anti b$
decay contribution to the net $Z+b's$ channel.  
We also note that small $G$
is only achieved for $\mai>2\mtau$, with $\mai\gsim 7\gev$ preferred.

For $\tanb=3$, there is again a preference for larger $\mai$ in order to achieve
small $G$.  However, small $G$ can be achieved for a
much larger range of $\cta$. Of course, one should also notice that
all the low-$F$ solutions have $\cta<0$, with lower bound of
$|\cta|\gsim 0.06$.

The bottom of the three plots shows that if
$\tanb=50$ then it is much more difficult to find solutions with low $F$
that also have low $G$.  The lower bound on $|\cta|$ 
needed to achieve large $\br(\hi\to\ai\ai)$ shifts downwards slightly
to about $0.05$.

One should note that the coupling of the $\ai$ to $b\anti b$
is proportional to $\tanb\cta$ times the usual SM-like $\gamma_5$
coupling strength.  The lower limits on $|\cta|$  at $\tanb=10$, $3$, $50$ are
such that $|\tanb\cta|\sim 1$, $\sim 0.6$, $\sim 2.5$. This means that the
$b\anti b$ coupling is not particularly suppressed, and can even be enhanced with
respect to the SM-like $\gam_5$ value. (Of course, the $t\anti t$
coupling of the $\ai$, proportional to $\cot\beta\cta$,  is very
suppressed.) The fact that the $b\anti b$ coupling is always
significant implies that there is always a significant branching ratio
for $\Upsilon\to\gam\ai$ (where the $\Upsilon$ can be the $1S$, $2S$
or $3S$ state) so long as there is adequate phase space for the decay.
The predictions for $\br(\Upsilon_{1S}\to\gam\ai)$ 
and further discussion appear in \cite{Dermisek:2006py}.

\begin{figure}[ht!]

  \centerline{\includegraphics[width=2.4in,angle=90]{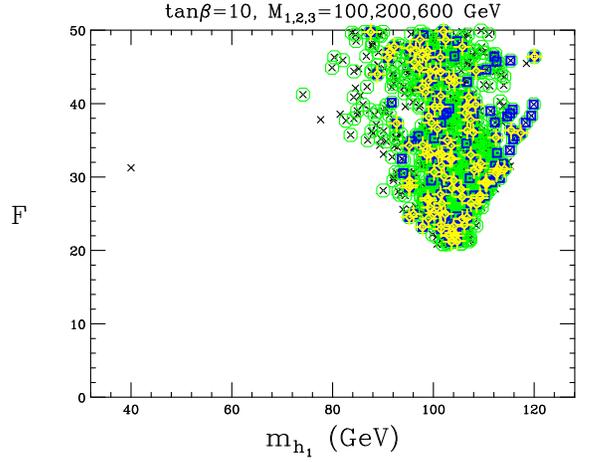}}
  \vspace*{-.1in}
\caption{ Fine tuning vs. the Higgs mass for randomly generated NMSSM parameter choices 
  except for fixed $M_{1,2,3}(\mz)=100,200,600\gev$
and a fixed value of $\tanb=10$. Notation as in
Fig.~\ref{fvsmh1}.}
\label{fvsmh1meq126}
\vspace*{-.1in}
\end{figure}

\subsection{Dependence of $F$ on the gluino mass}

The minimum value of $F$ that can be achieved is, of course, dependent
upon $M_3$ (and is essentially independent of $M_1$ and $M_2$).
Indeed, the largest GUT-scale parameter derivative is very frequently
that with respect to $M_3(\mgut)$. To explore this sensitivity, we
have also performed a (somewhat less dense) parameter scan for the
case of $M_{1,2,3}=100,200,600\gev$ at $\tanb=10$.  The results for
$F$ as a function of $\mhi$ are presented in Fig.~\ref{fvsmh1meq126}.
We find a minimum value of $F\sim 20$ at $\mhi\sim 104\gev$, the
latter being somewhat higher than the $\mhi\sim 100\gev$ location in
the corresponding $\tanb=10$, $M_{1,2,3}=100,200,300\gev$ case.  A
SM-like $\hi$ with $\mhi\sim 104\gev$ is only consistent with LEP
limits if $\br(\hi\to\ai\ai)$ is large and $\mai<2\mb$ (the large
(yellow) crosses). The $\mhi\sim 104\gev$ location of the minimal $F$
is less consistent with the $M_{2b}\sim 100\gev$ excess in the LEP
data.  However, to have $\mhi\sim 100\gev$ in the
$M_{1,2,3}=100,200,600\gev$ case is possible for $F\sim 22$, which is
barely different from the $F\sim 20$ minimum value.

One other point of interest is that $\mai<2\mb$ points are more
easily achieved at larger $\mhi\gsim 114\gev$ when $M_3(\mz)=600\gev$
than for $M_3(\mz)=300\gev$. This can be understood by considering the
special case of $\lambda = 0.2$, $\kappa =
\pm 0.2$ and $\tan \beta = 10$.  In this case, we find
\begin{eqnarray}
A_\lambda (\mz) &\sim & -0.03 A_\kappa(\mgut) + 0.93 A_\lambda(\mgut) - 0.35 A_t(\mgut) \nn\\
&& \hspace*{-.6in} 
-   0.03 M_1(\mgut) - 0.37 M_2(\mgut) + 0.66 M_3(\mgut), \label{alrun}\\
A_\kappa (\mz) &\sim & 0.90 A_\kappa(\mgut) - 0.11 A_\lambda(\mgut) + 0.02 A_t(\mgut)
\nn\\
&& \hspace*{-.6in}  
+ 0.003 M_1(\mgut) + 0.025 M_2(\mgut) - 0.017 M_3(\mgut)\label{akrun}\,.
\end{eqnarray}
Let us consider the $\kap>0$ case (by convention, $\lam>0$), for which
it can be shown that $\akap<0$ is required~\cite{Dermisek:2006wr} to
get $\mai^2>0$. From Eqs.~(\ref{akrun}), (\ref{atmz}) and
(\ref{m3mz}), one finds that $\akap(\mz)\sim 0.06 M_3(\mz)+0.1
A_t(\mz)$,
implying that increasingly negative $A_t(\mz)$ is
required to achieve $\akap<0$ as $M_3(\mz)$ increases.  From
Eq.~(\ref{eq:ma1_first}), a small value of $\mai^2$ will be easily
achieved in the present case of $\kap\akap<0$ 
if $\alam(\mz)<0$ so that the $\kap\akap(\mz)$
and $\lam\alam(\mz)$ terms tend to cancel. 
Eqs.~(\ref{alrun}), (\ref{atmz}) and
(\ref{m3mz}) imply $\alam(\mz)\sim -0.3M_3(\mz)-1.5A_t(\mz)$ 
which gives $\alam(\mz)<0$ for increasingly negative $A_t(\mz)$ as
$M_3(\mz)$ increases.  In short, the larger $M_3(\mz)$ is the more
negative $A_t(\mz)$ can be while requiring small $\mai^2$.  The more
negative $A_t(\mz)$, the larger stop mixing is at fixed $\mstopbar$
and therefore the larger $\mhi$.

\subsection{Low-$F$ scenarios and the LEP excess}

We will now discuss in more detail other properties of
the low-$F$ scenarios with $\mhi\sim 100\gev$ and $\mai<2\mb$,
focusing first on the case of $\tanb=10$ and
$M_{1,2,3}(\mz)=100,200,300\gev$.
 First, we recall our earlier results from
\cite{Dermisek:2005gg}. There, we studied in detail the $F<25$ points
from our earliest $\tanb=10$ scans as plotted in
Fig.~\ref{zbblimits}. The plot shows the $\cbb$ predictions for all parameter
choices in our scan that had $F<25$ and $\mai<2\mb$ and that are
consistent with the experimental and theoretical constraints built
into NMHDECAY as well as with limits from the preliminary LHWG full
analysis code \cite{bechtle}, which in particular incorporates
limits on the $Z+b's$ combined channel.  Eight $F<10$ points are singled
out. As we have emphasized, these latter points cluster
near $\mhi\sim 98\div105\gev$. In \cite{Dermisek:2005gg}, we found
the remarkable result that not only are these $F<10$ $\mai<2\mb$
points consistent with LEP limits, but also most are such that $\mhi$
and $\br(\hi\to b\anti b)$ are appropriate for explaining the 
$\cbb$ excess. We wish to emphasize that in our scan there are many, many points that
satisfy all constraints and have $\mai<2\mb$. The remarkable result is
that those with $F<10$ have a substantial probability that they
predict the Higgs boson properties that would imply a LEP $Zh\to
Z+b$'s excess of the sort seen. We stress again that the $F<10$ points with $\mai$
substantially above $2\mb$ all predict a net $Z+b$'s signal that is
ruled out at better than $99\%$ CL by LEP data. Indeed, all such
$F<25$ points have a net $\hi\to b$'s branching ratio, $\br(\hi\to
b\anti b)+\br(\hi\to \ai\ai\to b\anti b b\anti b)\gsim 0.85$, which is
too large for LEP consistency. In our larger scans, as represented by
the $\cbb$ results of Fig.~\ref{nmssmceff}, we see a huge number of
$\mai<2\mb$ points with approximately the correct $\cbb$ to
explain the LEP $100\gev$ excess.

For $\tanb=50$, $M_{1,2,3}(\mz)=100,200,300\gev$, the preference for
$\mhi\sim 101\gev$ to achieve low $F$ will again imply that many of
the lowest $F$ scenarios will provide a natural explanation of the
$Z+2b$ LEP excess.  At $\tanb=3$, the very lowest $F$ values, $F\sim
7-8$, consistent with LEP limits are achieved for $\mhi\sim 95\gev$,
as shown in the middle plot of Fig.~\ref{fvsmh1}.  Such an $\hi$ mass
is too low to provide a natural explanation of the $Z+2b$ excess.
However, this same plot shows that the very slightly higher value of
$F\sim 10$ is possible for $\mhi\sim 100\gev$.  Thus, the LEP $Z+2b$
excess is fully consistent with low fine-tuning scenarios that pass
all LEP Higgs limits for all $\tanb\geq 3$. (We have not explored
still lower values.) Of course, it is equally true that at $\tanb=10$
and $\tanb=50$, only a very modest increase in $F$ would be needed for
$\mhi$ to take on a value that is not perfectly correlated with the
location at $M_{2b}\sim 100\gev$ of the $Z+2b$ LEP excess.

\subsection{Properties of the heavier Higgs bosons for low-$F$ scenarios}

\begin{figure}[h!]
  \centerline{\includegraphics[width=2.4in,angle=90]{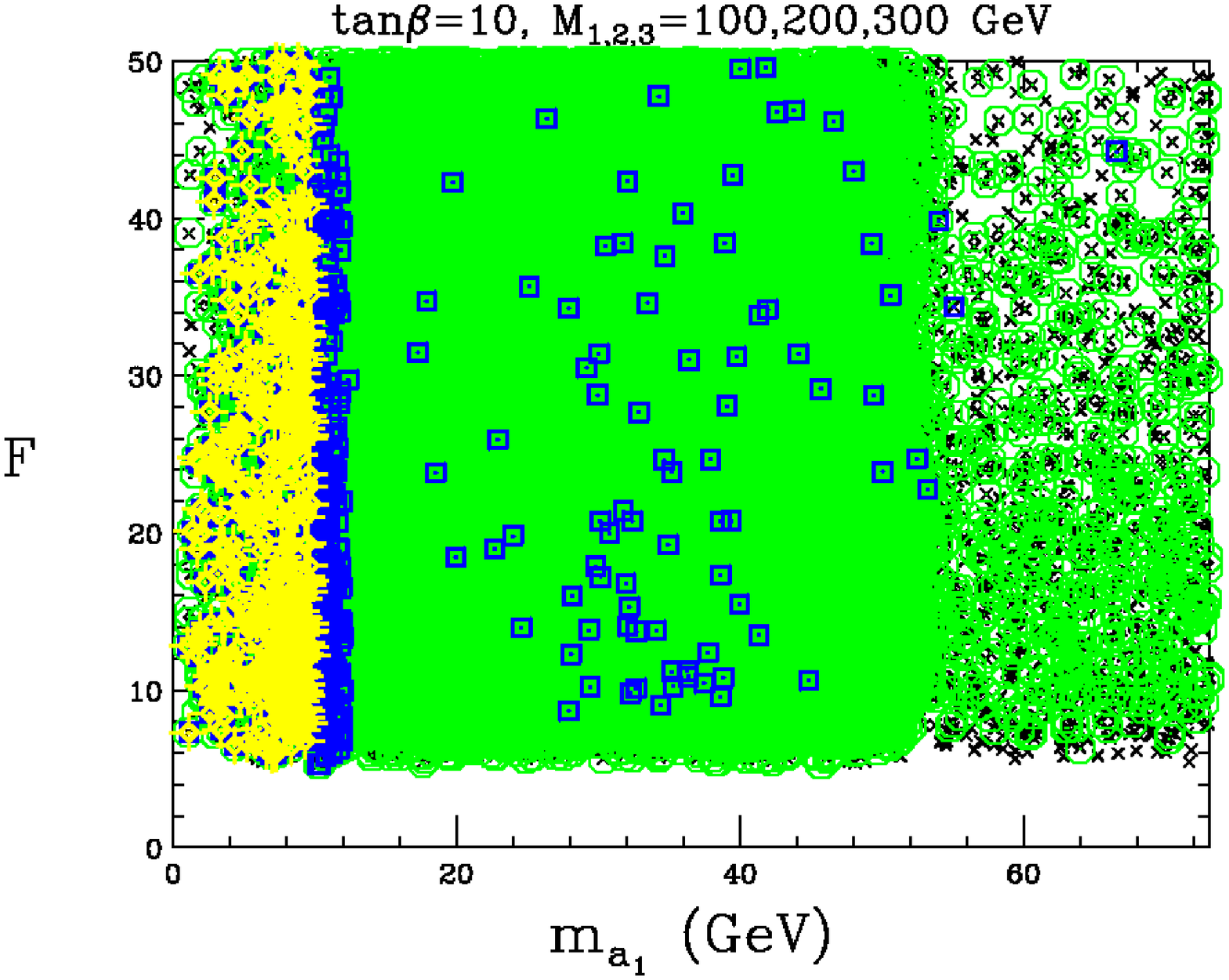}}
  \centerline{\includegraphics[width=2.4in,angle=90]{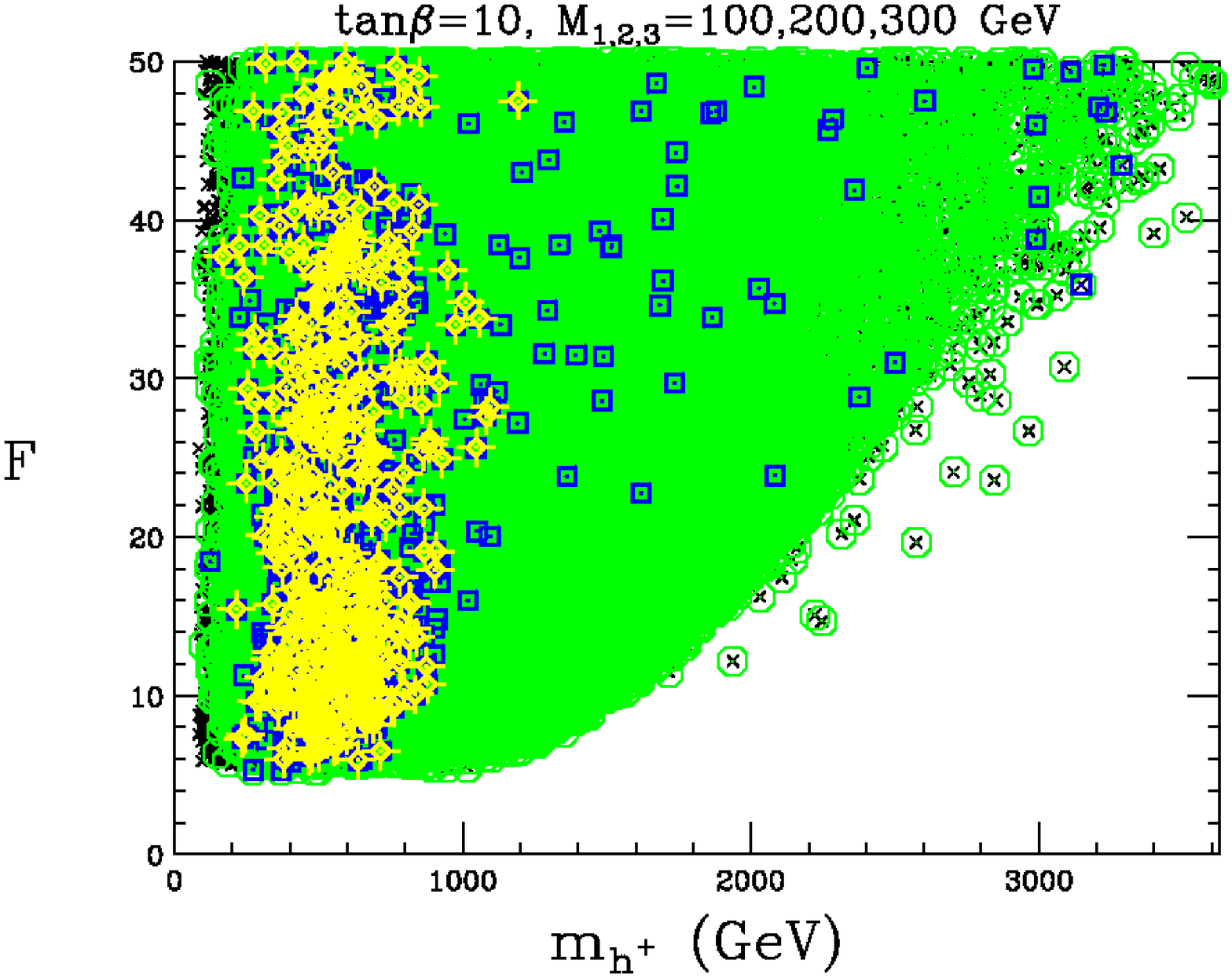}}
  \vspace*{-.1in}
\caption{ Fine tuning vs. $\mai$ (top) and vs. $m_{h^+}$ 
  (bottom) for
  points with $F<50$  taking $M_{1,2,3}(\mz)=100,200,300\gev$
and $\tanb=10$. Point notation as in Fig.~\ref{fvsmh1}.}
\label{fvsma1andmhplus}
\vspace*{-.1in}
\end{figure}

\begin{figure}[h!]
  \centerline{\includegraphics[width=2.4in,angle=90]{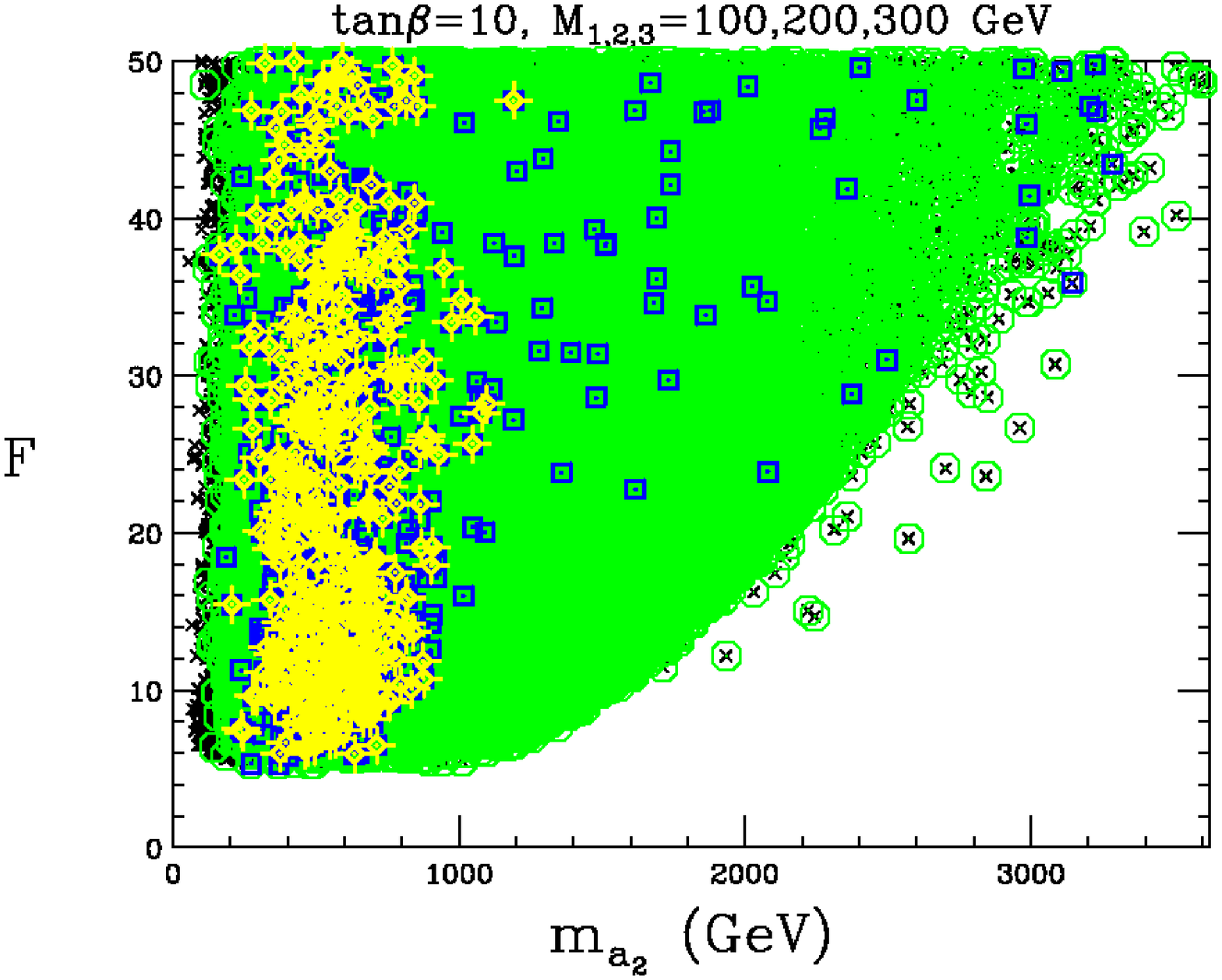}}
  \centerline{\includegraphics[width=2.4in,angle=90]{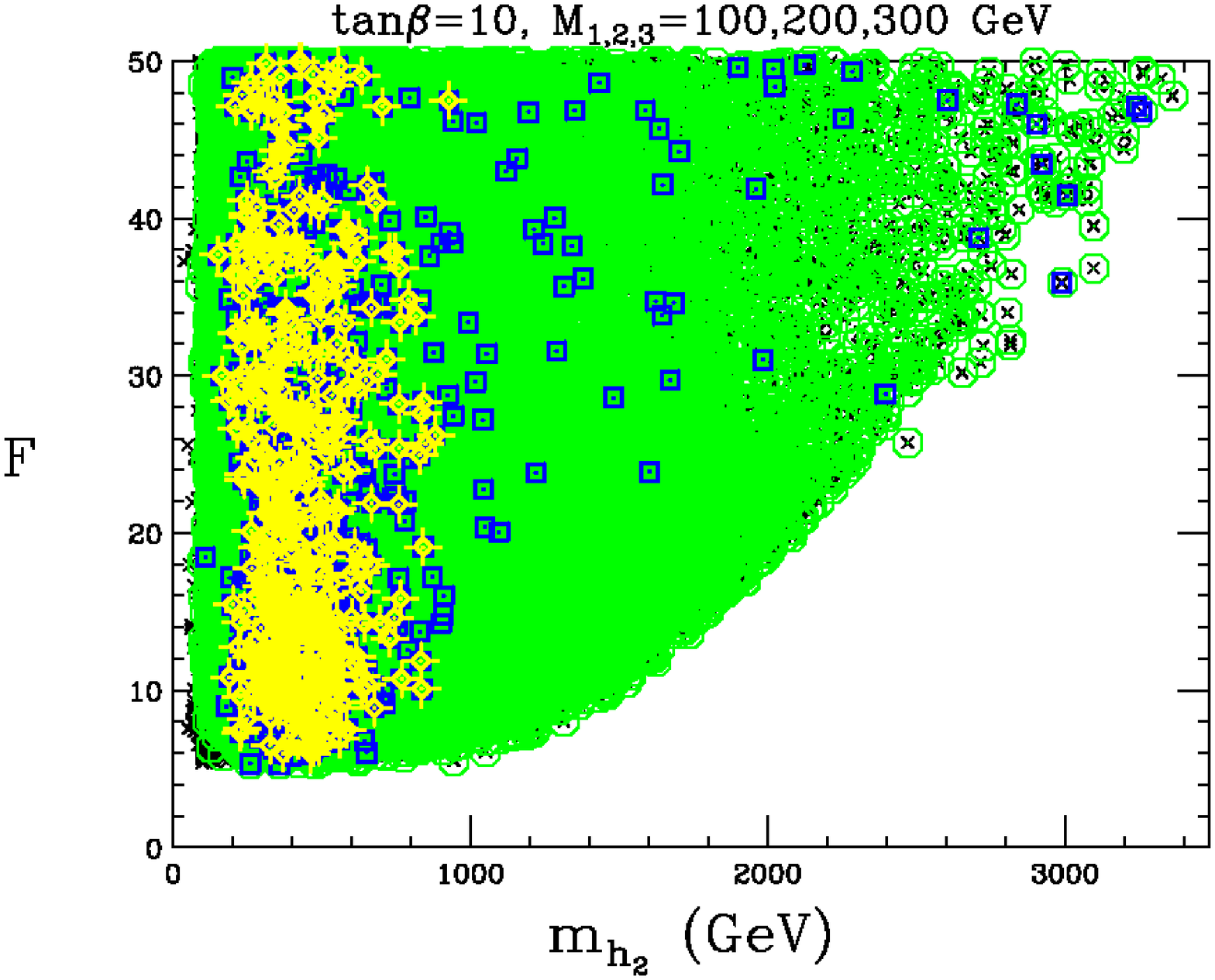}}
  \centerline{\includegraphics[width=2.4in,angle=90]{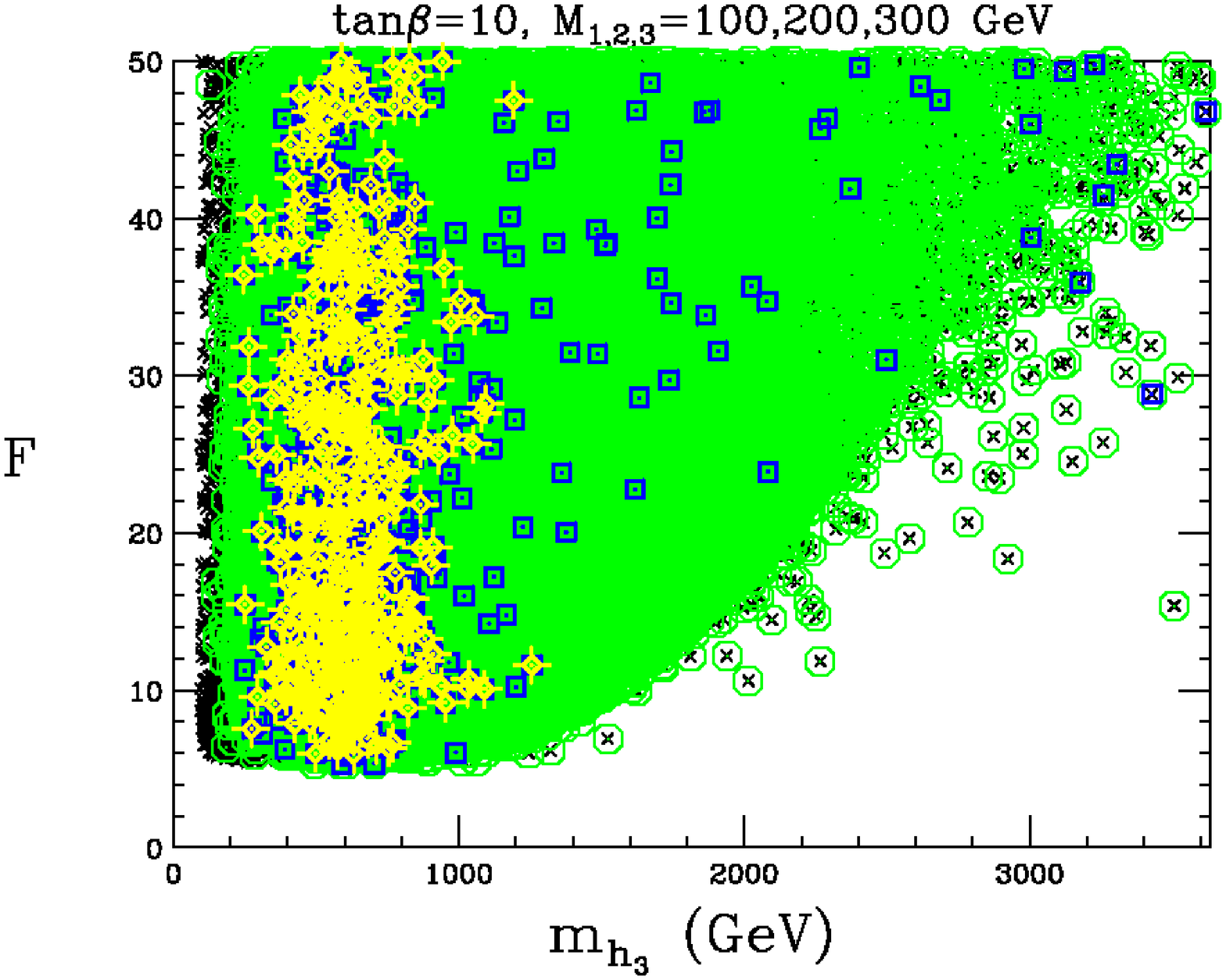}}
  \vspace*{-.1in}
\caption{ Fine tuning vs. $\maii$ (top), $\mhii$ (middle) and 
 $\mhiii$ 
  (bottom) for
  points with $F<50$  taking $M_{1,2,3}(\mz)=100,200,300\gev$
and $\tanb=10$. Point notation as in Fig.~\ref{fvsmh1}.}
\label{fvsma2mh2mh3}
\vspace*{-.1in}
\end{figure}

\begin{figure}[h!]
  \centerline{\includegraphics[width=2.4in,angle=90]{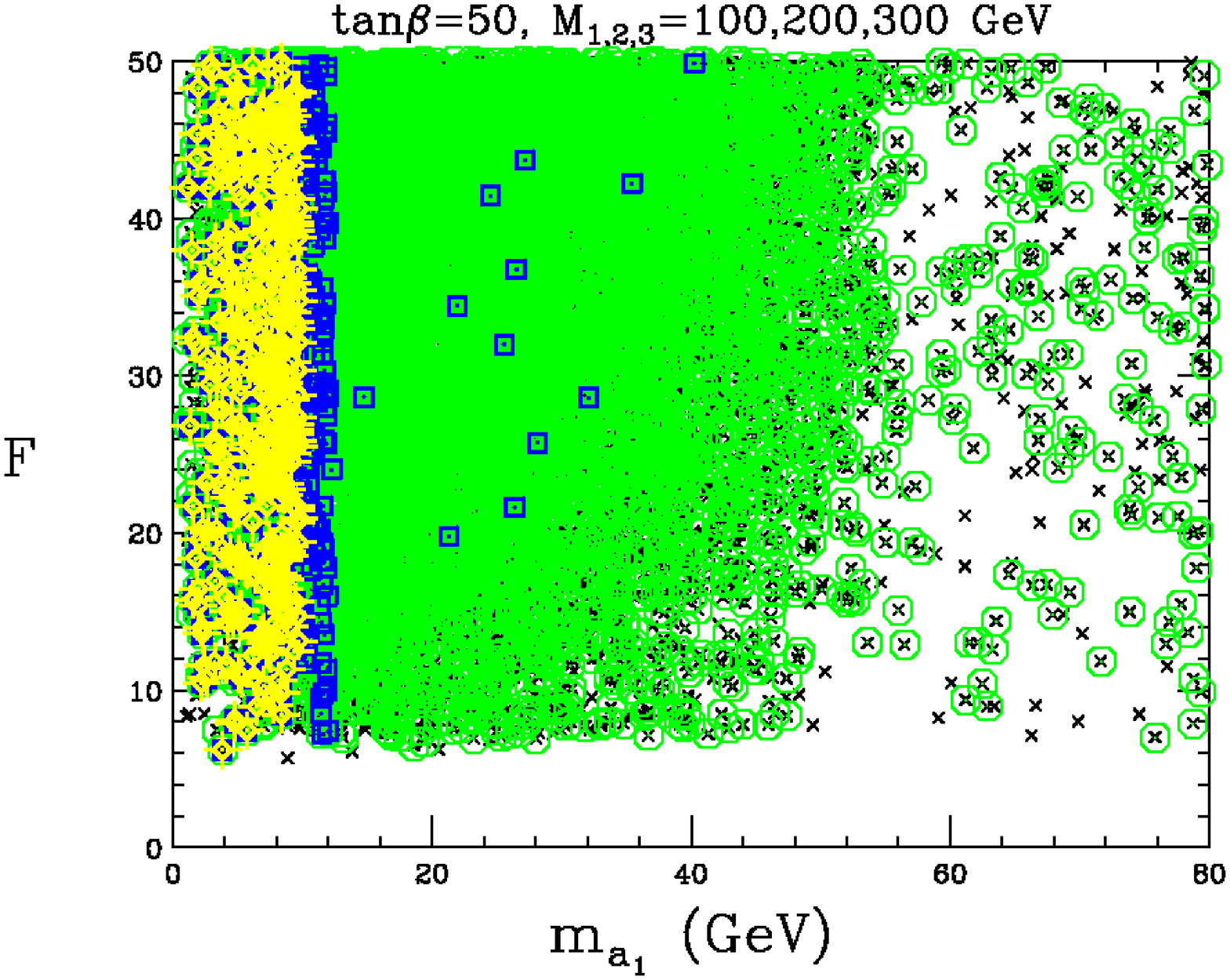}}
  \centerline{\includegraphics[width=2.4in,angle=90]{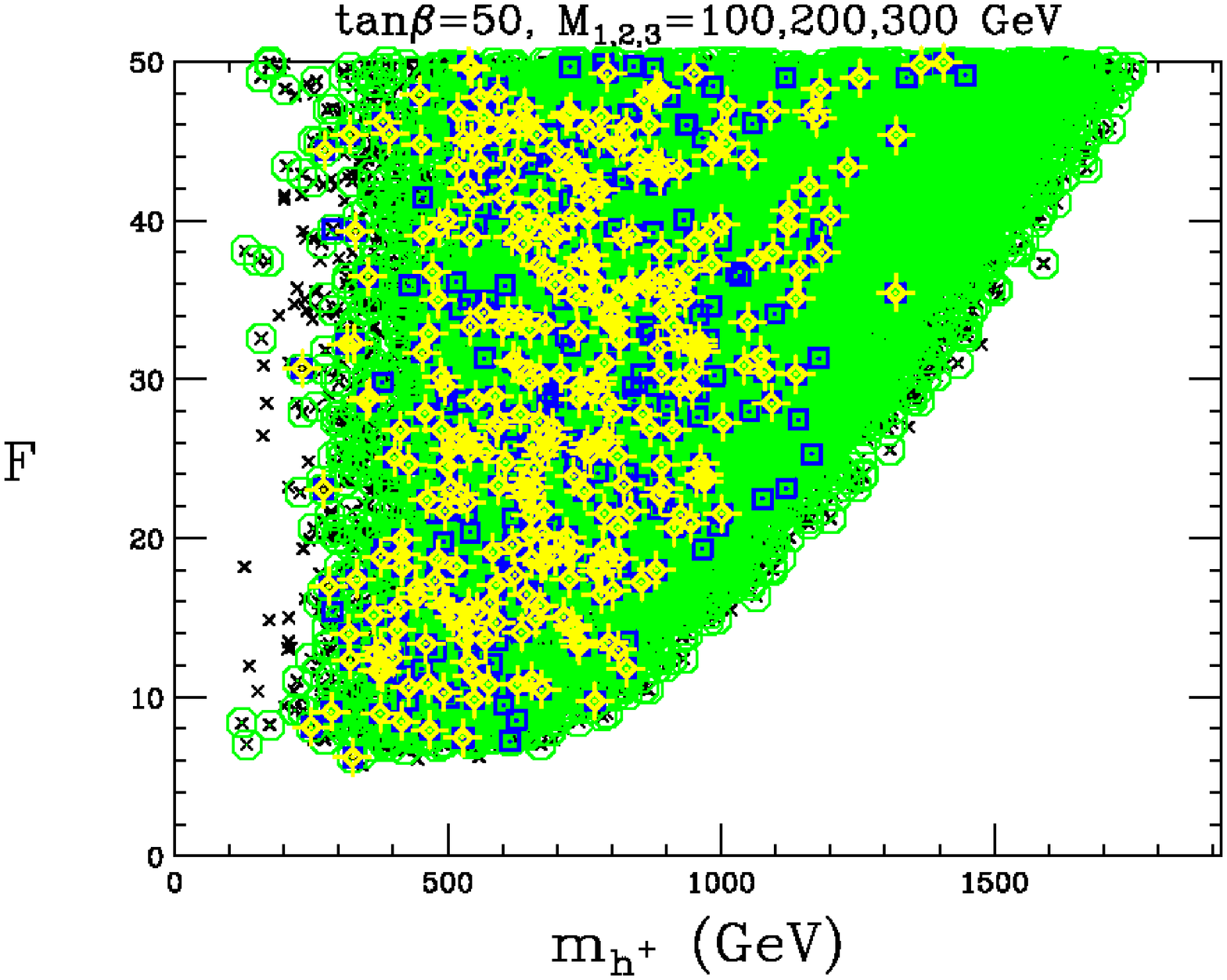}}
  \vspace*{-.1in}
\caption{ Fine tuning vs. $\mai$ (top) and vs. $m_{h^+}$ 
  (bottom) for
  points with $F<50$  taking $M_{1,2,3}(\mz)=100,200,300\gev$
and $\tanb=50$. Point notation as in Fig.~\ref{fvsmh1}.}
\label{fvsma1andmhplustb50}
\vspace*{-.1in}
\end{figure}

\begin{figure}[h!]
  \centerline{\includegraphics[width=2.4in,angle=90]{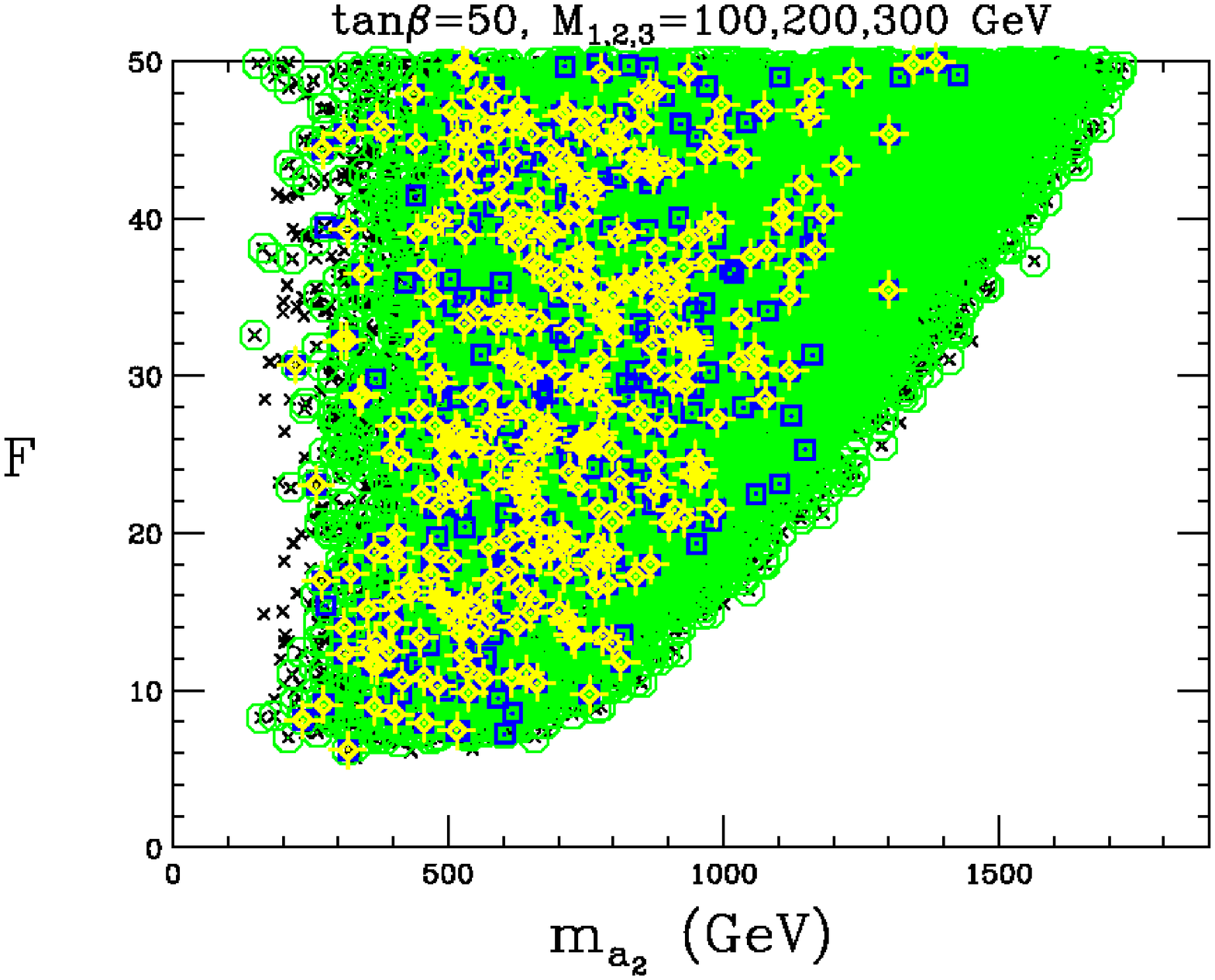}}
  \centerline{\includegraphics[width=2.4in,angle=90]{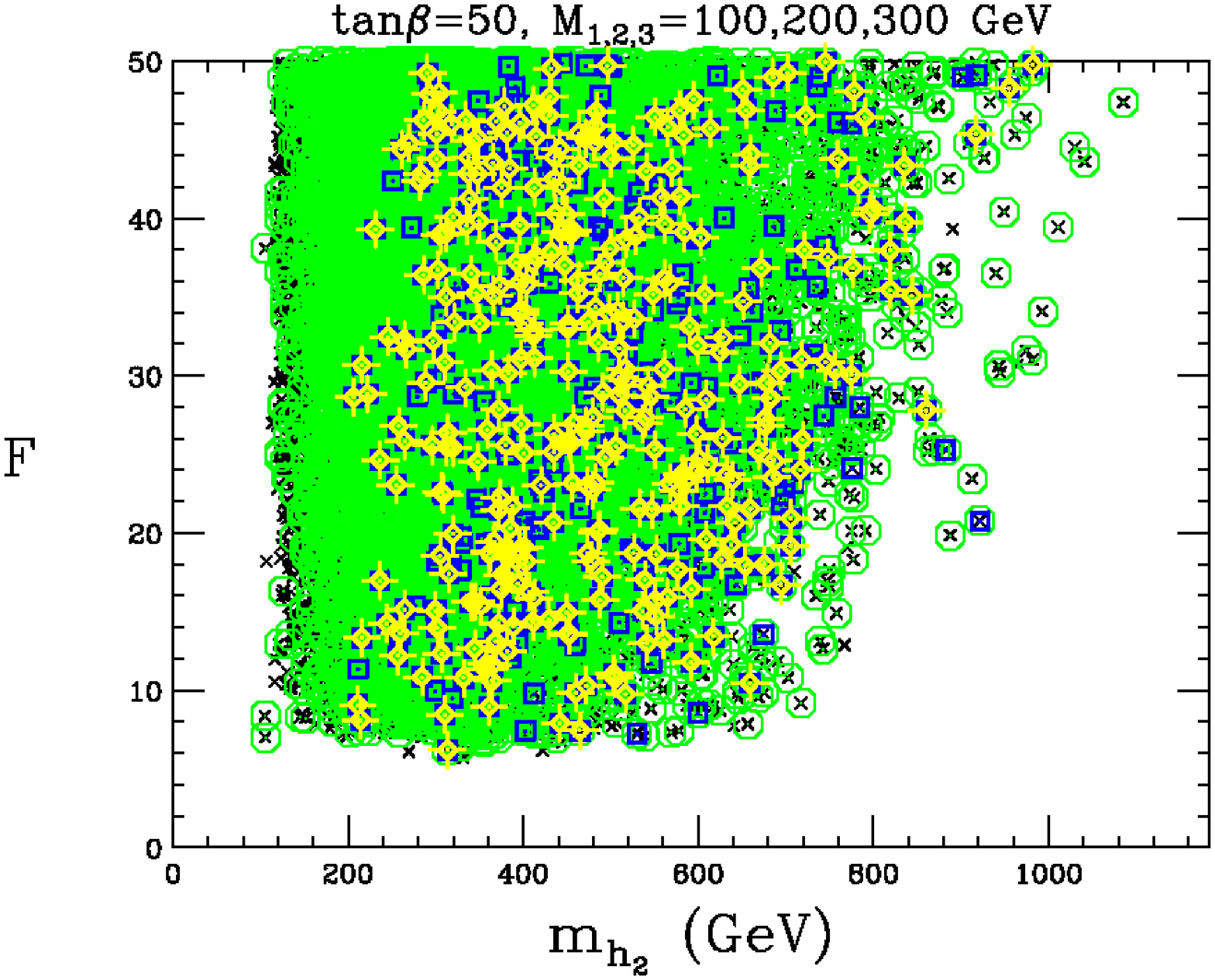}}
  \centerline{\includegraphics[width=2.4in,angle=90]{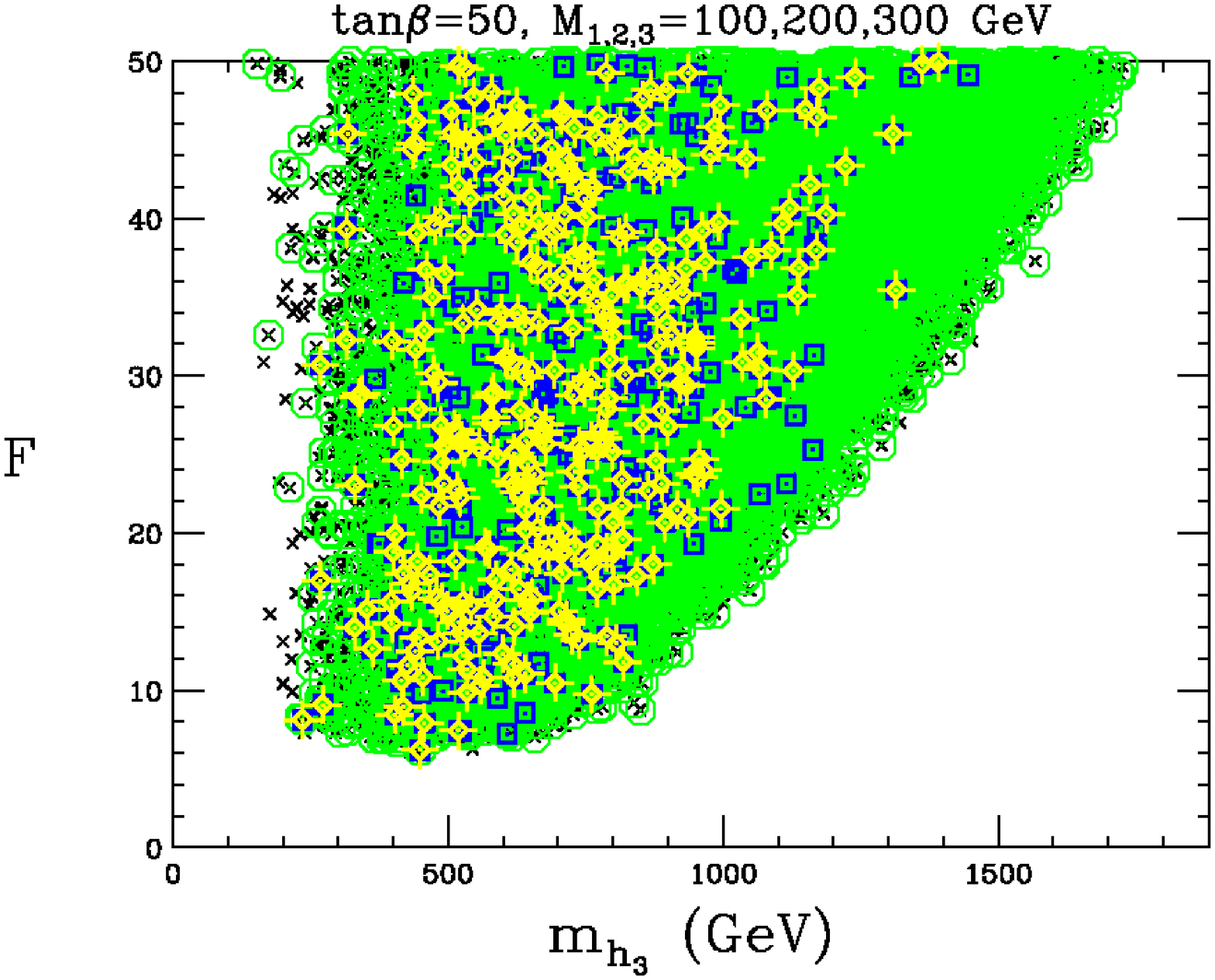}}
  \vspace*{-.1in}
\caption{ Fine tuning vs. $\maii$ (top), $\mhii$ (middle) and 
 $\mhiii$ 
  (bottom) for
  points with $F<50$  taking $M_{1,2,3}(\mz)=100,200,300\gev$
and $\tanb=50$. Point notation as in Fig.~\ref{fvsmh1}.}
\label{fvsma2mh2mh3tb50}
\vspace*{-.1in}
\end{figure}

Another interesting question is whether there is any correlation
between $F$ and $\mai$ or between $F$ and the masses of the heavier
Higgs bosons, $m_{h^+}$, $\maii$, $\mhii$ and $\mhiii$. The plots for
$\tanb=10$ appear in Figs.~\ref{fvsma1andmhplus} and
\ref{fvsma2mh2mh3}.  There, we observe that $F$ depends very weakly on
$\mai$ (once $\mai<2\mb$), but that lower values of
$m_{h^+},\maii,\mhii,\mhiii$ are definitely preferred to obtain small
$F$ and that to obtain fully allowed yellow fancy crosses with
$F<10$ requires $m_{h^+}\in[300\gev,800\gev]$,
$\maii\in[250\gev,750\gev]$, $\mhii\in[200\gev,600\gev]$ and
$\mhiii\in[300\gev,800\gev]$.  The corresponding plots for $\tanb=50$ appear in
Figs.~\ref{fvsma1andmhplustb50} and \ref{fvsma2mh2mh3tb50}.  Again, we
observe that $F$ depends very weakly on $\mai$, and that lower values of
$m_{h^+},\maii,\mhii,\mhiii$ are definitely preferred to obtain small
$F$. For $\tanb=50$, to obtain fully allowed yellow fancy crosses
with $F<10$ requires $m_{h^+}\in[250\gev,750\gev]$,
$\maii\in[250\gev,750\gev]$, $\mhii\in[200\gev,650\gev]$ and
$\mhiii\in[250\gev,750\gev]$.  

These results have implications for the LHC and ILC.  At the LHC, the
main processes for producing and detecting these heavier Higgs bosons
are $gg\to b\anti b H$ (where $H=\aii,\hii,\hiii$) and $gg\to b \anti t
h^++gg\to \anti b t h^-$, with, for example, $H\to \tauptaum$.  One
finds \cite{Gennai:2007ys} that detection becomes possible when the
$b\anti b H$ and $b\anti t h^+$ couplings are enhanced by large
$\tanb$. The mass ranges for the heavier Higgs bosons preferred for
obtaining low $F$ are such that if $\tanb=10$ they will be on the
margin of detectability at the LHC,
whereas if $\tanb=50$ they will certainly be detectable. (At $\tanb=3$
the small-$F$ mass ranges for the $\hii,\hiii,\aii$ are similar, but
$\tanb$ is definitely too small for the above LHC modes to be
detectable.)  For the lowest part of the mass ranges, a signal for
$gg\to b\anti b H$ might also emerge at the Tevatron if $\tanb=50$.
It is also important to note that the low-$F$ mass
ranges of the $\aii,\hii,\hiii,h^+$ are such that their pair
production would mostly be outside the kinematical reach of a $\rts=500\gev$
ILC, but that a substantial portion of the mass ranges are such that
pair production would be possible at a $\rts=1\tev$ ILC.

\subsection{Features and parameter correlations for low-$F$ scenarios}

We next turn to a detailed discussion of various correlations among
the NMSSM parameters that are associated with low-$F$ scenarios having
large $\br(\hi\to\ai\ai)$ and $\mai<2\mb$, \ie\ the points indicated
by the large yellow crosses in the previous figures. We call
such points ``fully ok''. We first
present some figures to illustrate how the fully ok points compare to
points that are either experimentally excluded or else have
sufficiently large $\mhi$ (roughly $\mhi>110-114\gev$) as to avoid LEP
constraints on the $Z+b's$ channel. These latter points are the (blue)
squares, (green) circles and black $\times$'s of the earlier plots.

\begin{figure}[ht!]

  \centerline{\includegraphics[width=2.4in,angle=90]{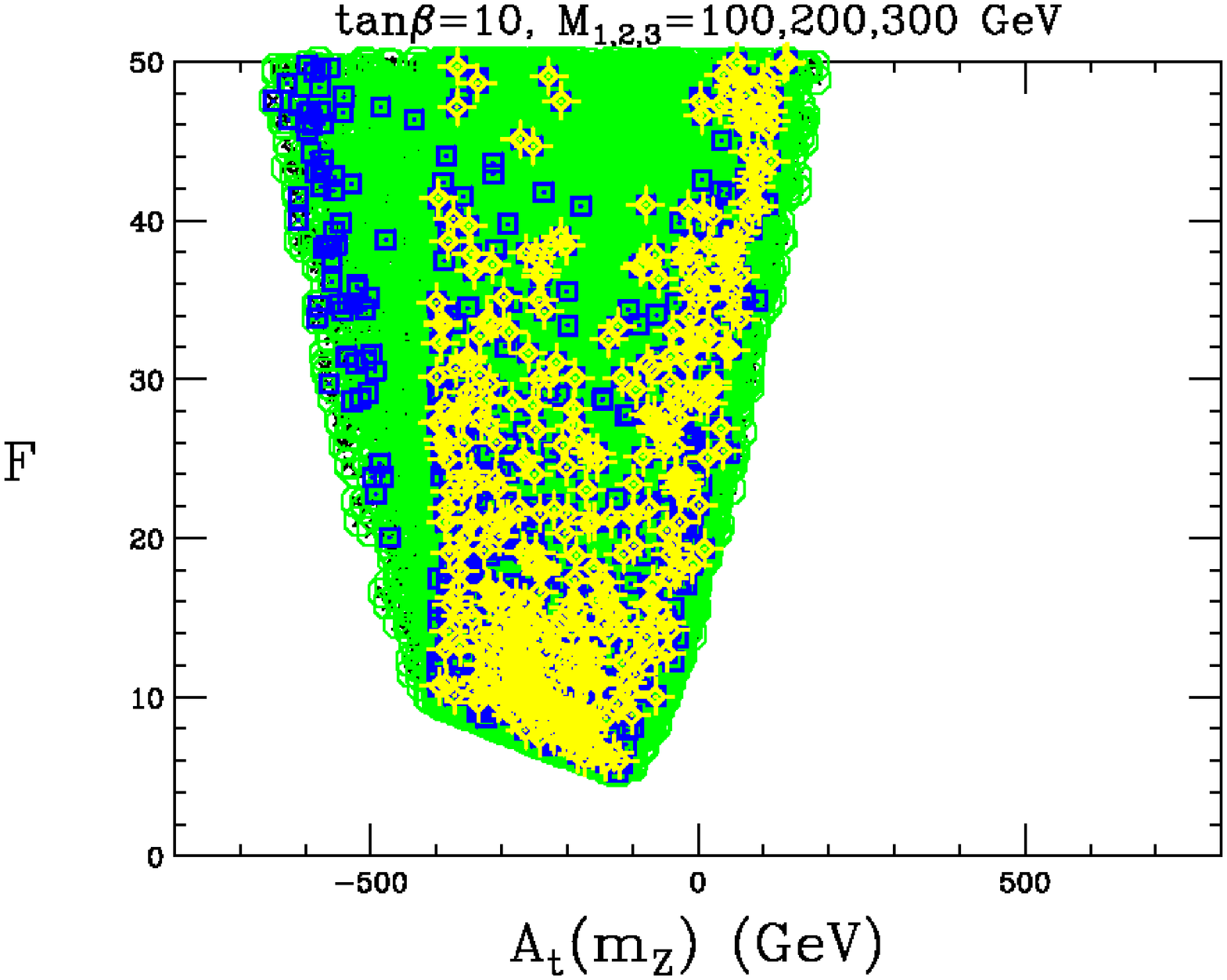}}
  \centerline{\includegraphics[width=2.4in,angle=90]{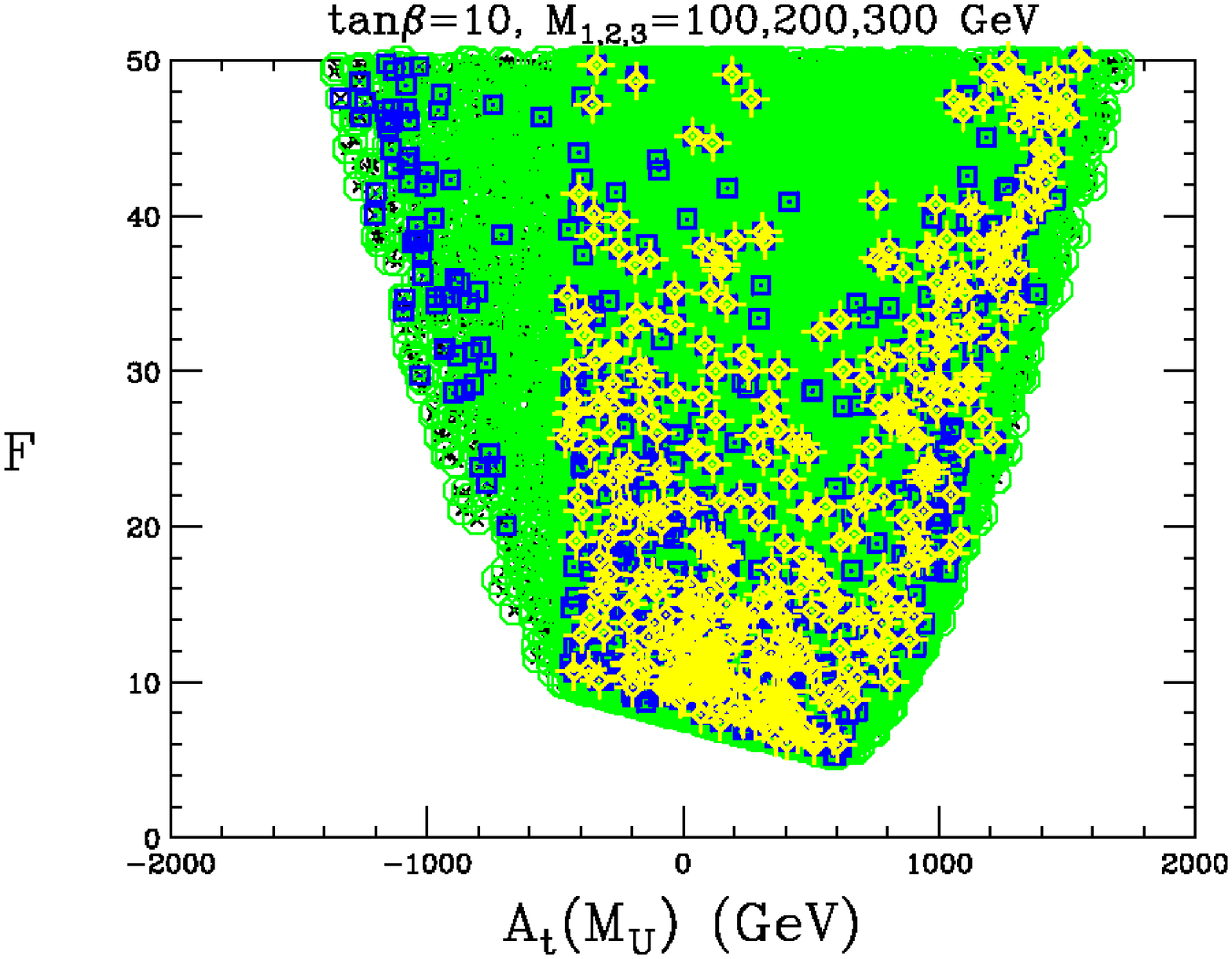}}
  \vspace*{-.1in}
\caption{ Fine tuning vs. $A_t(\mz)$ (top) and vs. $A_t(\mgut)$
  (bottom) for
  points with $F<50$  taking $M_{1,2,3}(\mz)=100,200,300\gev$
and $\tanb=10$. Point notation as in Fig.~\ref{fvsmh1}.}
\label{fvsat}
\vspace*{-.1in}
\end{figure}

\begin{figure}[ht!]

  \centerline{\includegraphics[width=2.4in,angle=90]{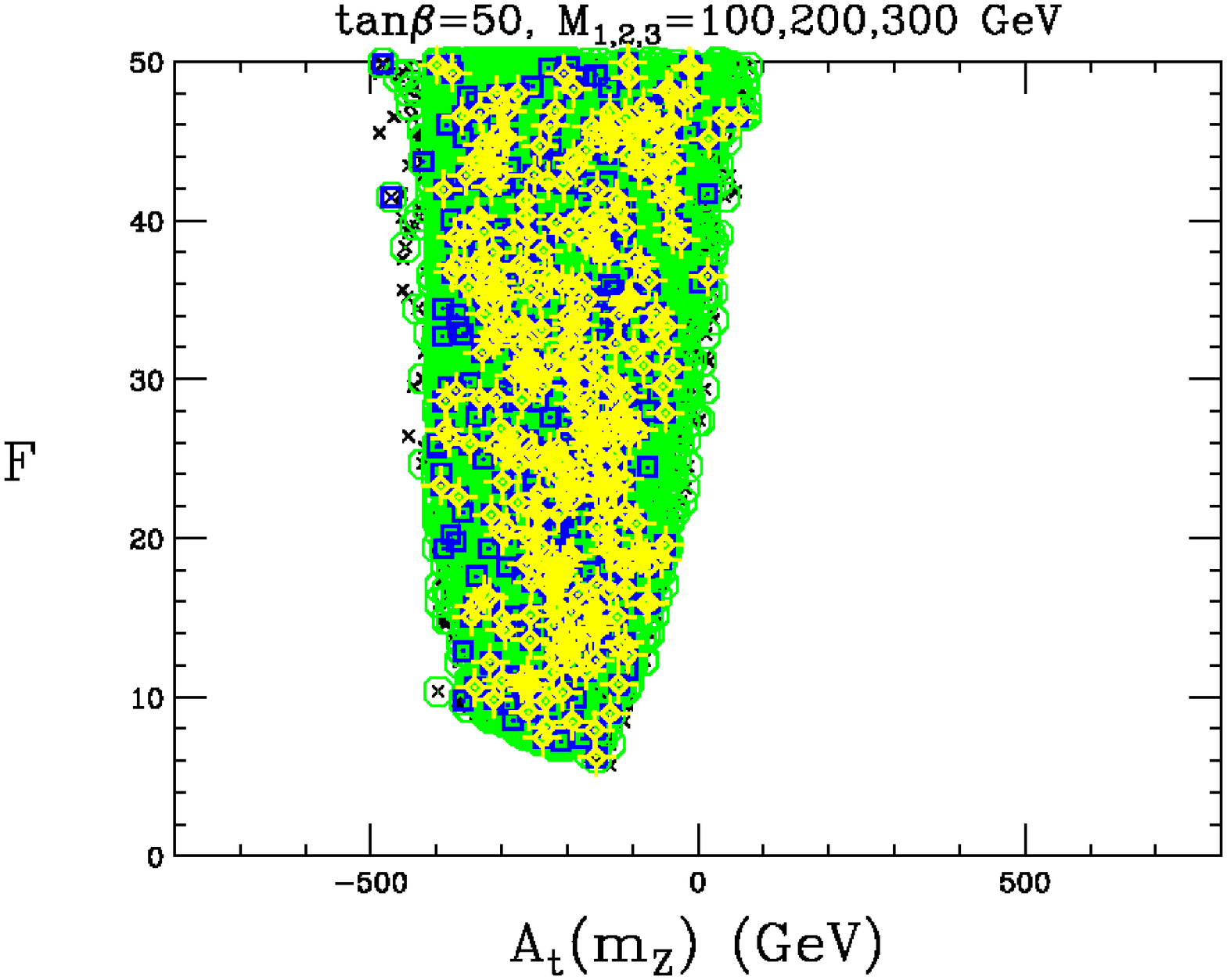}}
  \centerline{\includegraphics[width=2.4in,angle=90]{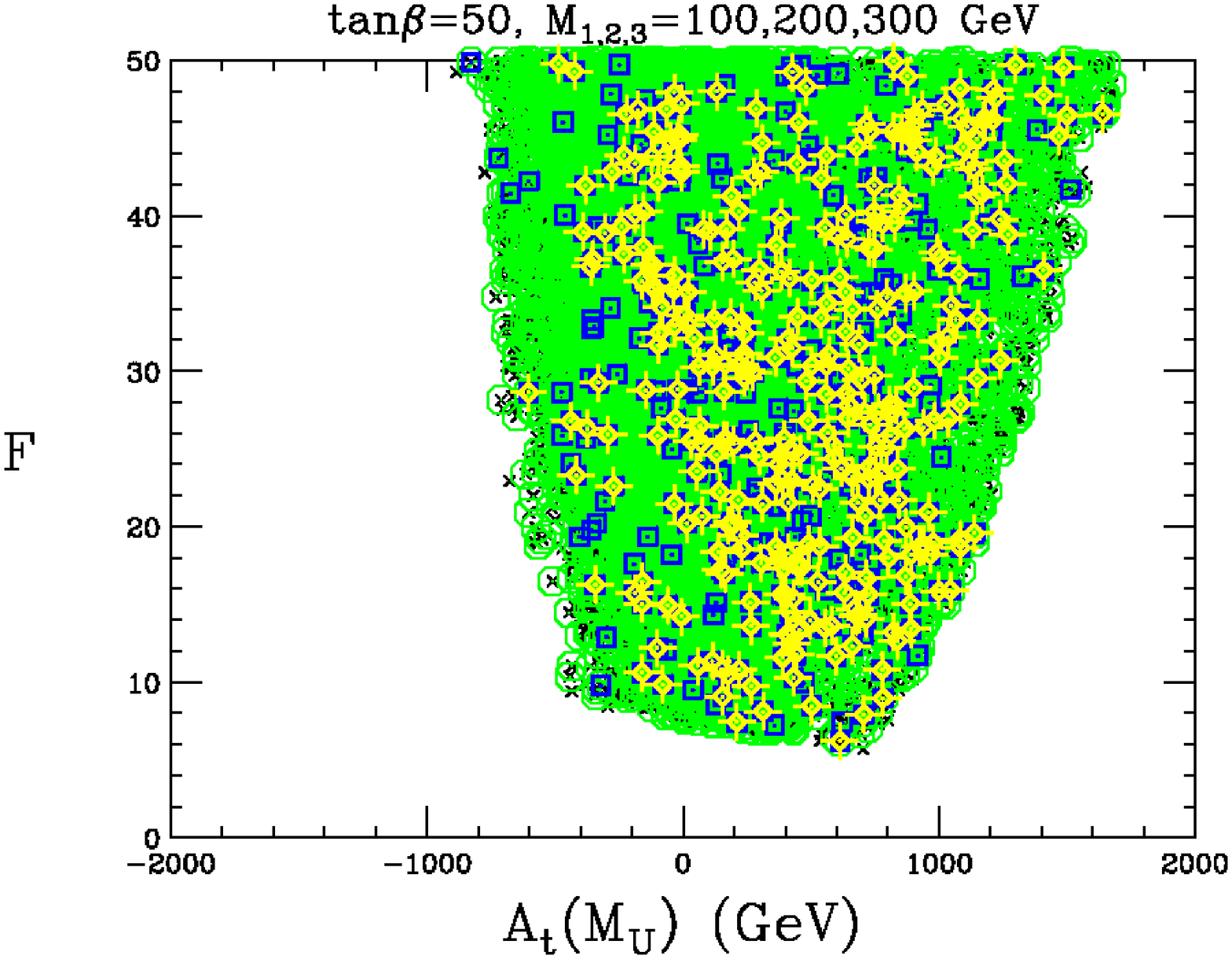}}
  \vspace*{-.1in}
\caption{ Fine tuning vs. $A_t(\mz)$ (top) and vs. $A_t(\mgut)$
  (bottom) for
  points with $F<50$  taking $M_{1,2,3}(\mz)=100,200,300\gev$
and $\tanb=50$. Point notation as in Fig.~\ref{fvsmh1}.}
\label{fvsattb50}
\vspace*{-.1in}
\end{figure}

First consider $A_t$.  Plots of F as a function of $A_t(\mz)$ and
$A_t(\mgut)$ are given in Figs.~\ref{fvsat} and \ref{fvsattb50} for
$\tanb=10$ and $\tanb=50$, respectively.  These show that rather
well-defined (and rather $\tanb$-independent) 
values are needed to achieve the very lowest $F$ values,
especially after imposing Higgs boson experimental limits. 
At scale $\mz$ the preferred $A_t(\mz)$ is of order $-100\gev$.
The corresponding $A_t(\mgut)$ is of order $+600\gev$. The lowest
$F$ values are of course those associated with $\mhi\sim 100\gev$.
This is consistent with our earlier discussion.  The $\tanb=10$ points with large
negative $A_t(\mz)$  values that escape LEP limits by virtue of
$\mhi>114\gev$ are the dark (blue) squares that begin at $F\sim 20$
and $A_t(\mz)\sim -500\gev$.

\begin{figure}[ht!]
  \centerline{\includegraphics[width=2.4in,angle=90]{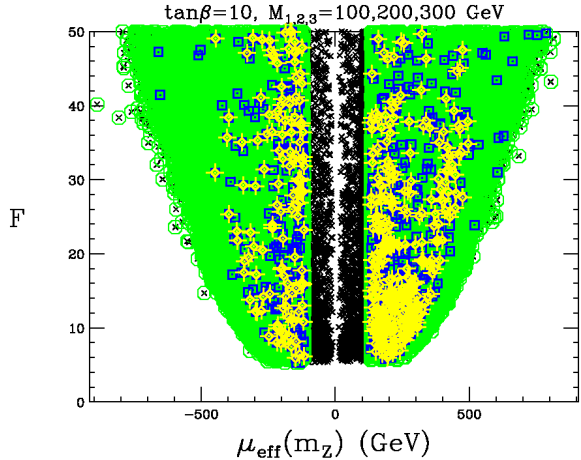}}
  \vspace*{-.1in}
\caption{ Fine tuning vs. $\mueff$ for
  points with $F<50$  taking $M_{1,2,3}(\mz)=100,200,300\gev$
and $\tanb=10$. Point notation as in Fig.~\ref{fvsmh1}.}
\label{fvsmueff}
\vspace*{-.1in}
\end{figure}

\begin{figure}[ht!]
  \centerline{\includegraphics[width=2.4in,angle=90]{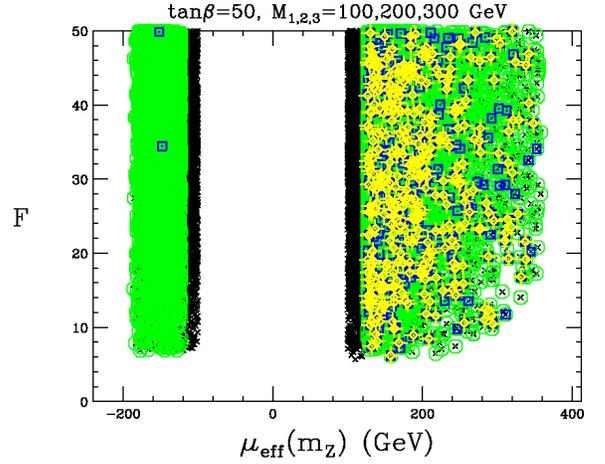}}
  \vspace*{-.1in}
\caption{ Fine tuning vs. $\mueff$ for
  points with $F<50$  taking $M_{1,2,3}(\mz)=100,200,300\gev$
and $\tanb=50$. Point notation as in Fig.~\ref{fvsmh1}.}
\label{fvsmuefftb50}
\vspace*{-.1in}
\end{figure}

In Figs.~\ref{fvsmueff} and \ref{fvsmuefftb50}, we plot $F$ as a function of $\mueff$ (which
in the case of the NMSSM 
is only defined at scale $\mz$ where EWSB has occurred) for the cases
of $\tanb=10$ and $\tanb=50$, respectively. As one could
easily anticipate from Eq.~(\ref{mzsquared}) (with $\mu$ replaced by
$\mueff$), fine-tuning is smallest for the the smallest values of
$\mueff$. This figure also shows that $|\mueff|$ values below
about $100\gev$ are eliminated by the LEP limit on the mass of the lightest chargino.

\begin{figure}[ht!]

  \centerline{\includegraphics[width=2.4in,angle=90]{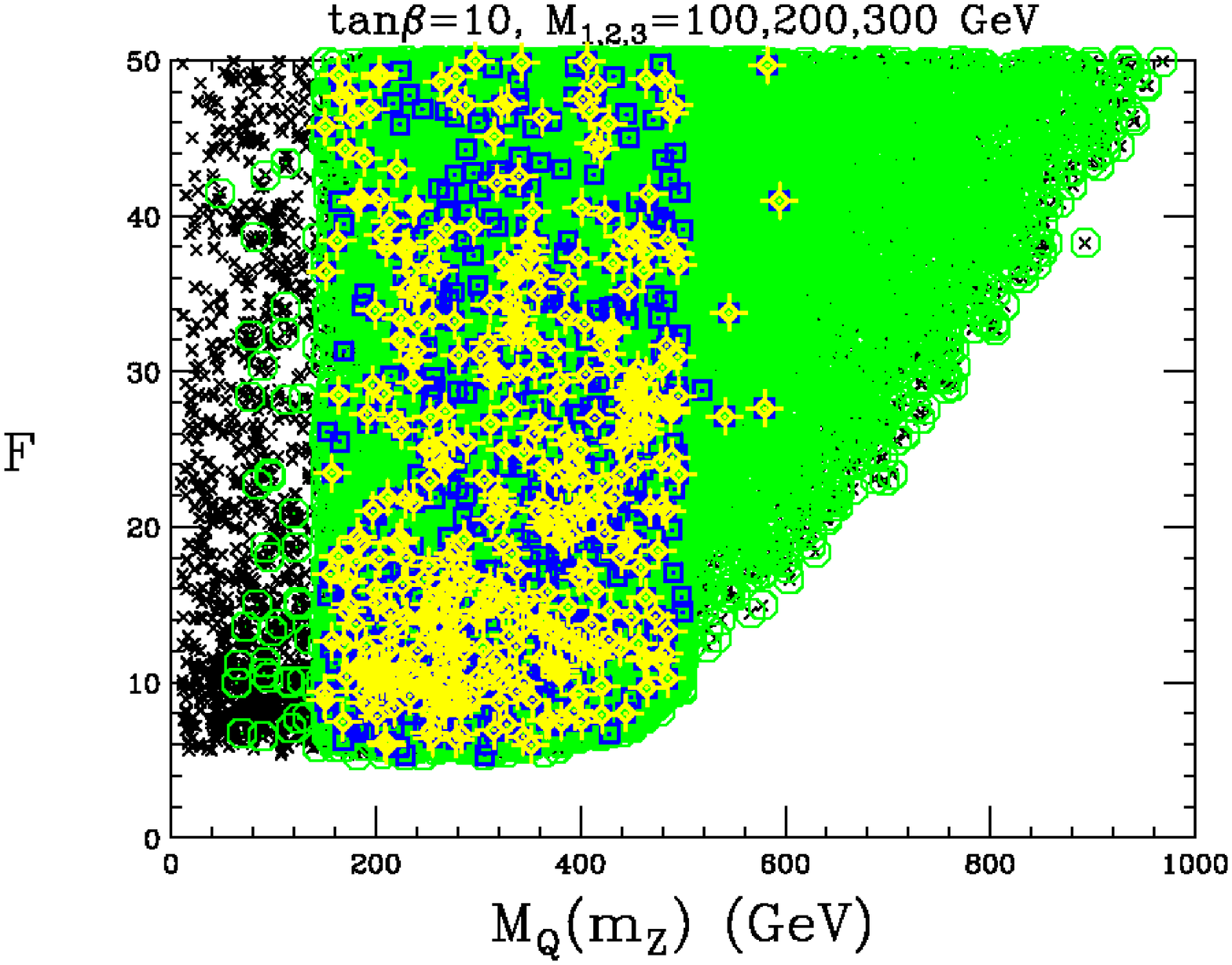}}
  \centerline{\includegraphics[width=2.4in,angle=90]{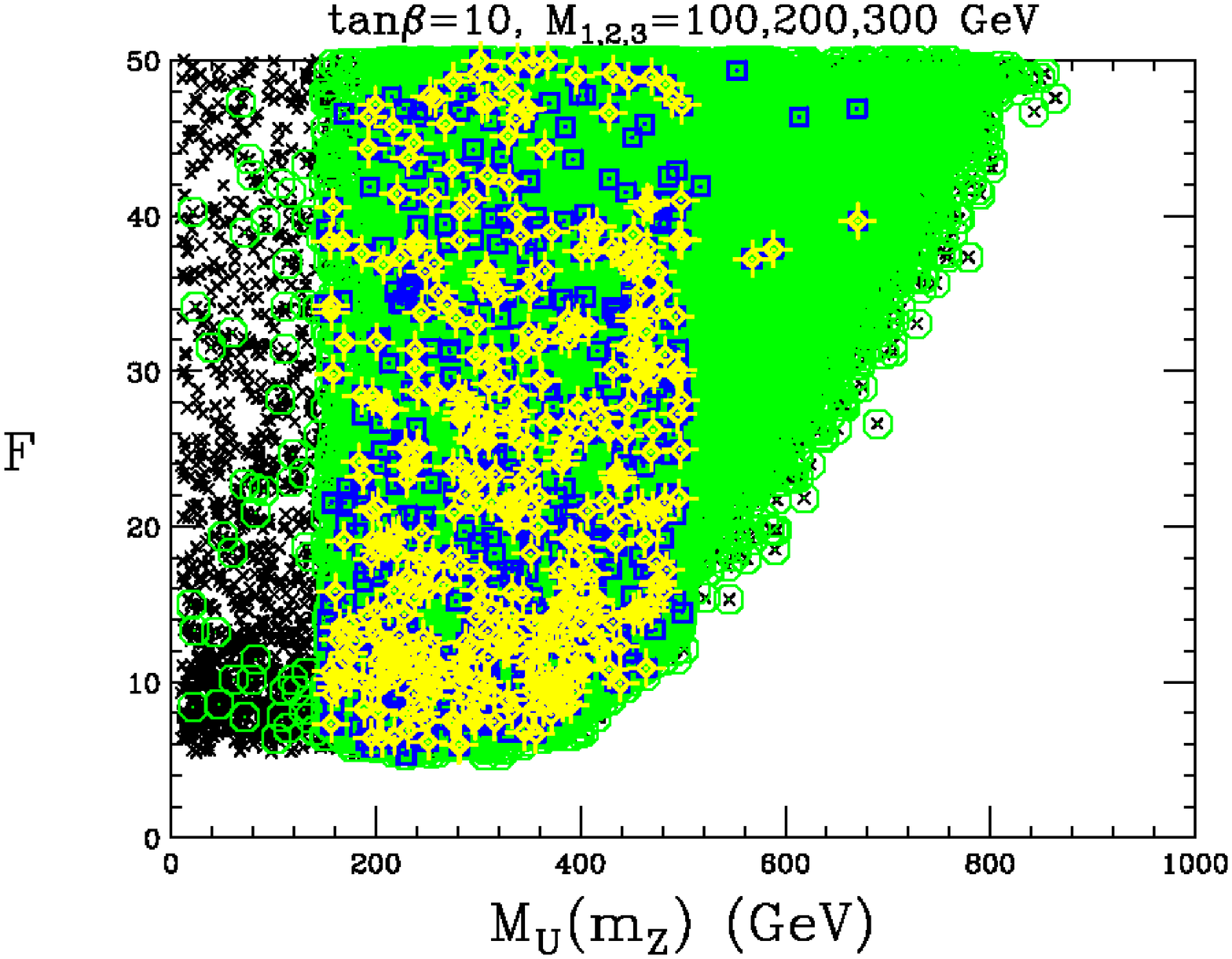}}
  \centerline{\includegraphics[width=2.4in,angle=90]{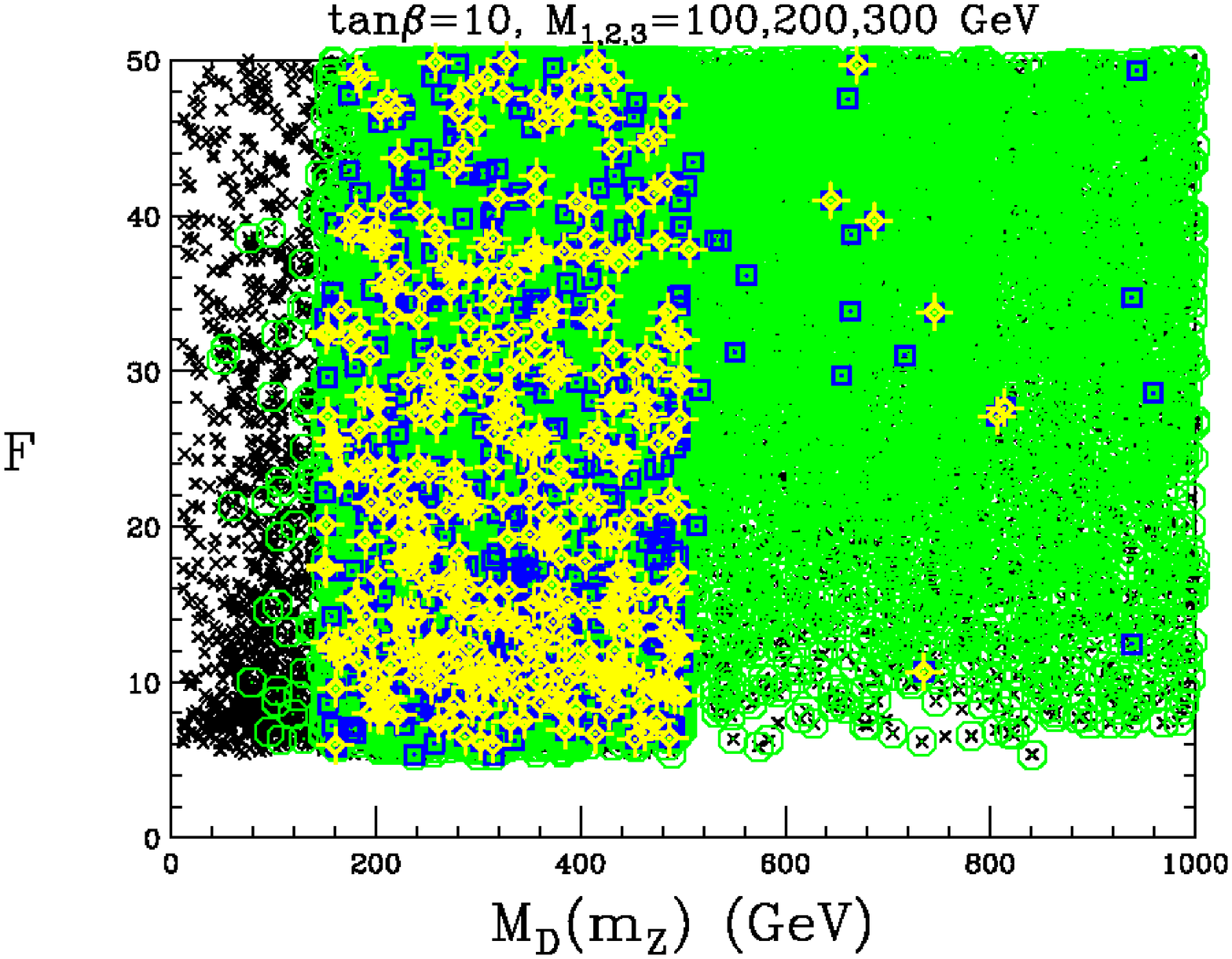}}
%
%
  \vspace*{-.1in}
\caption{ Fine tuning vs. $M_Q(\mz)$, $M_U(\mz)$ and $M_D(\mz)$ 
  for points with $F<50$  taking $M_{1,2,3}(\mz)=100,200,300\gev$
and $\tanb=10$. Point notation as in Fig.~\ref{fvsmh1}.}
\label{fvsmqmz}
\vspace*{-.1in}
\end{figure}

\begin{figure}[ht!]

  \centerline{\includegraphics[width=2.4in,angle=90]{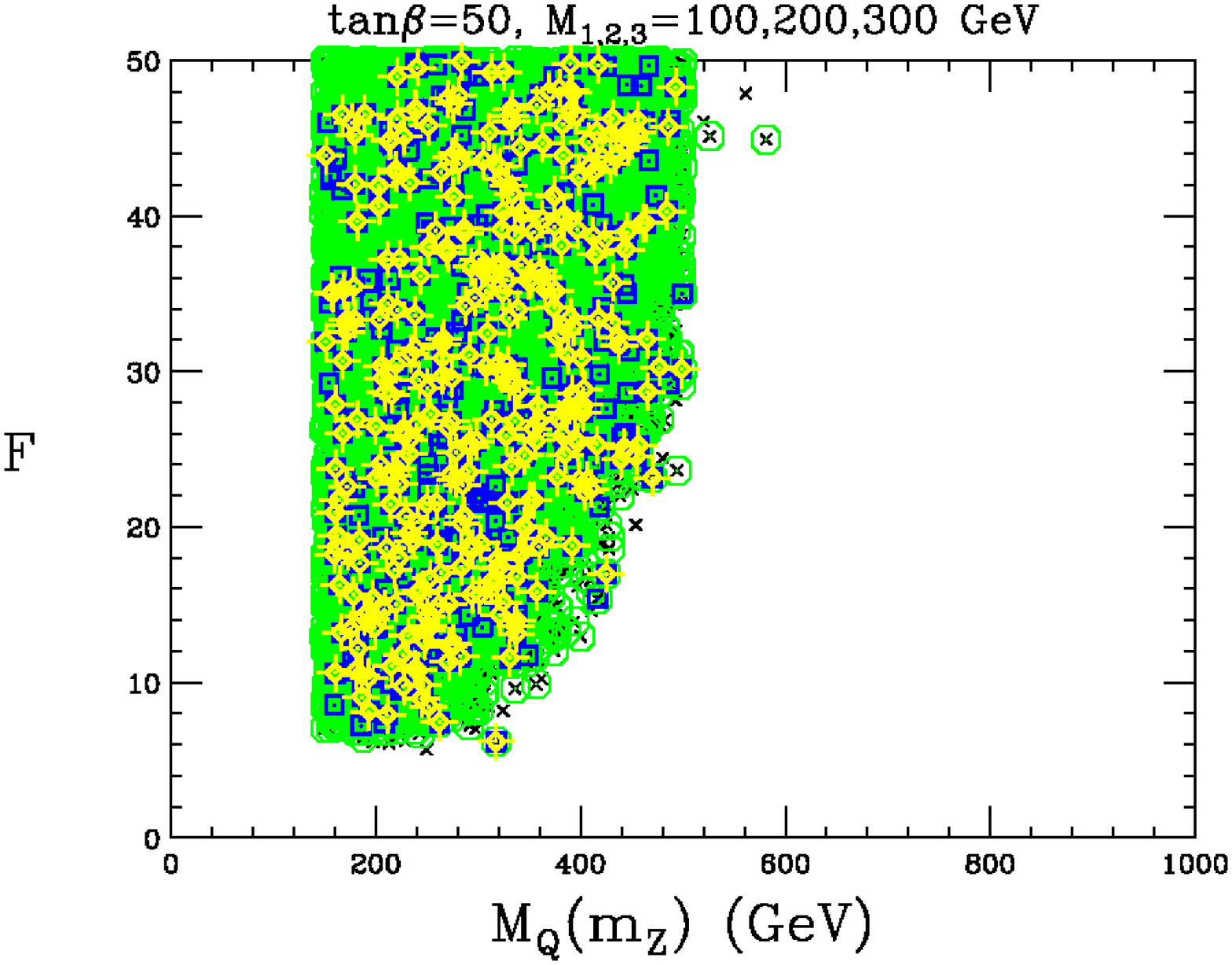}}
  \centerline{\includegraphics[width=2.4in,angle=90]{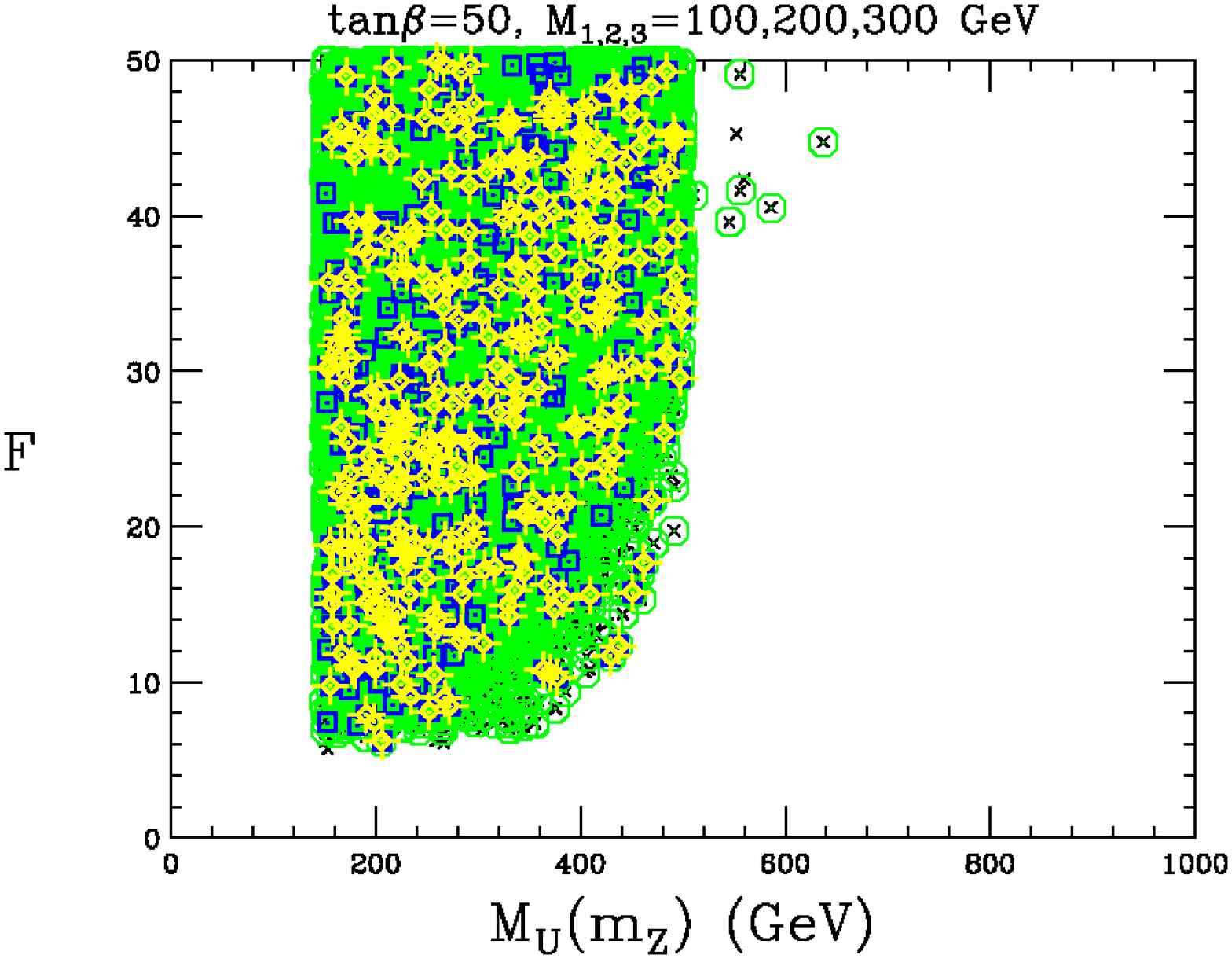}}
  \centerline{\includegraphics[width=2.4in,angle=90]{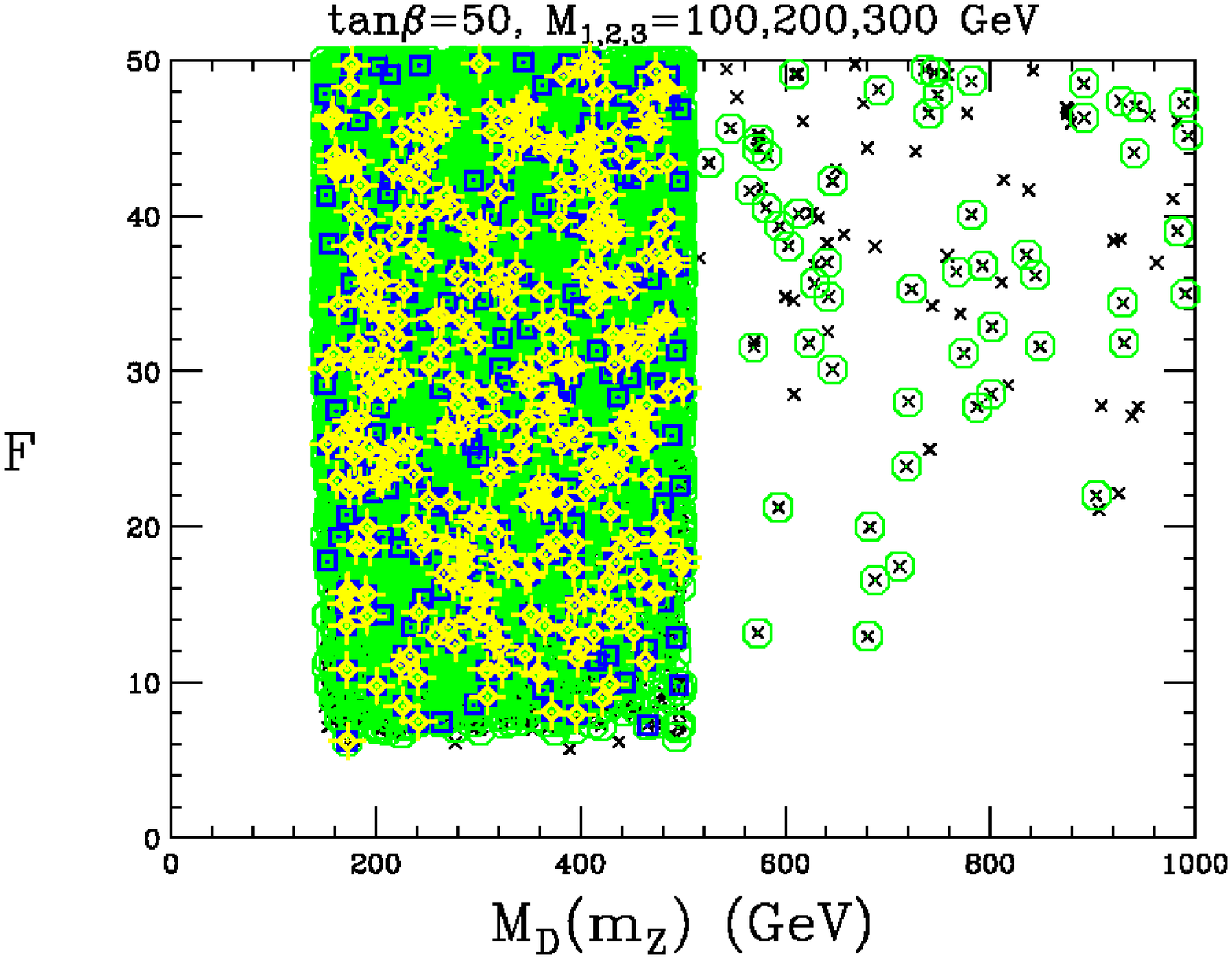}}
%
%
  \vspace*{-.1in}
\caption{ Fine tuning vs. $M_Q(\mz)$, $M_U(\mz)$ and $M_D(\mz)$ 
  for points with $F<50$  taking $M_{1,2,3}(\mz)=100,200,300\gev$
and $\tanb=50$. Point notation as in Fig.~\ref{fvsmh1}.}
\label{fvsmqmztb50}
\vspace*{-.1in}
\end{figure}

\begin{figure}[ht!]

  \centerline{\includegraphics[width=2.4in,angle=90]{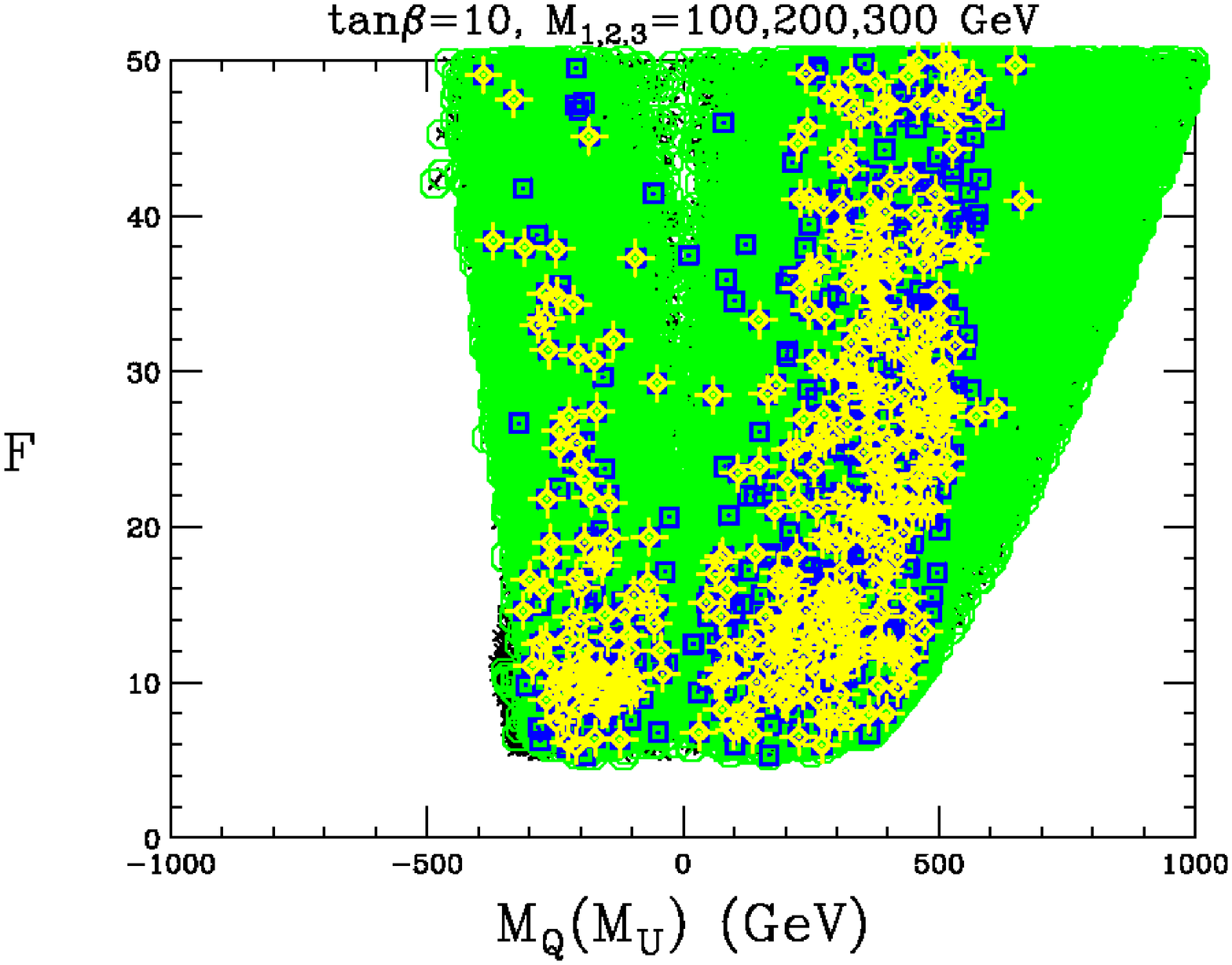}}
  \centerline{\includegraphics[width=2.4in,angle=90]{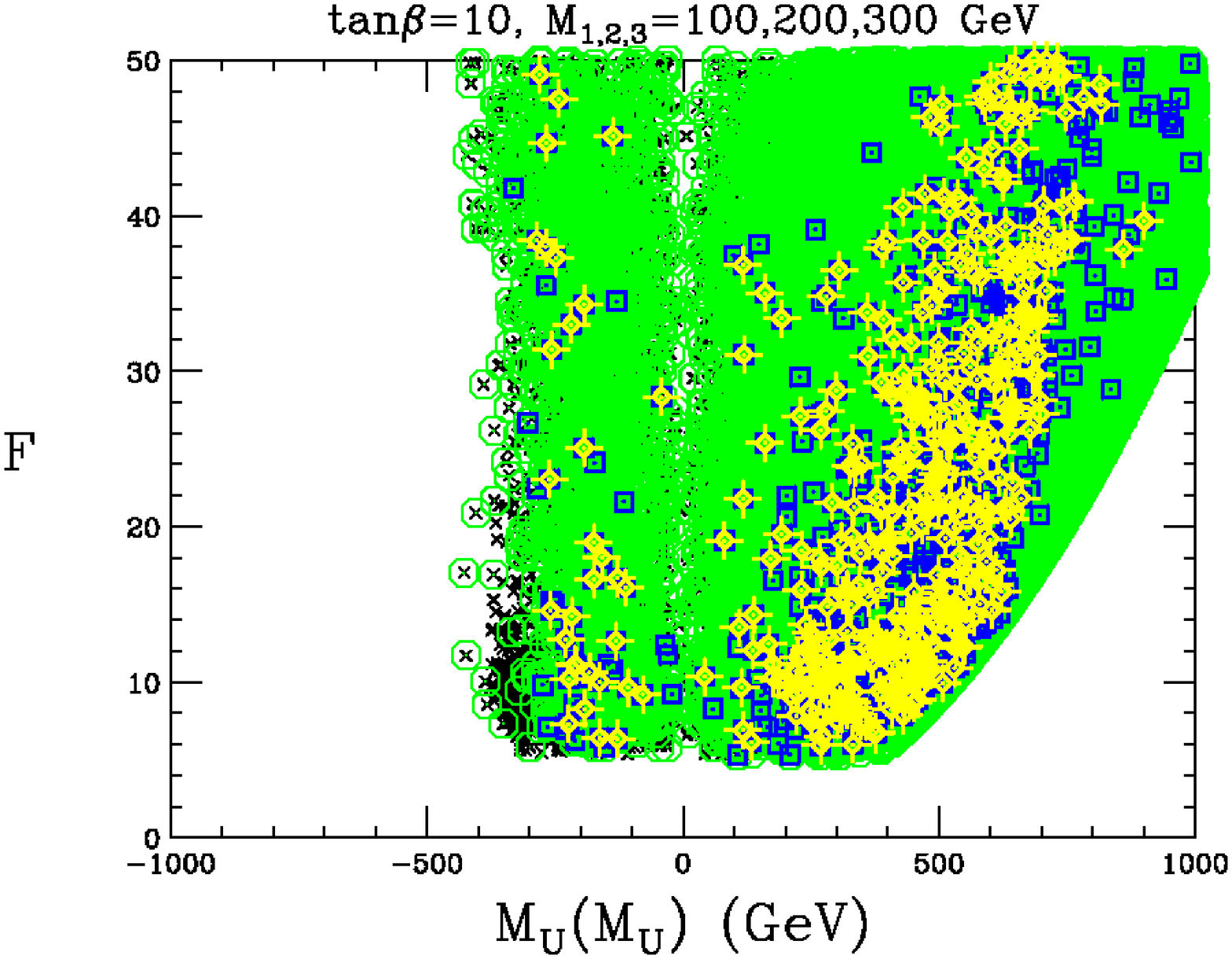}}
  \centerline{\includegraphics[width=2.4in,angle=90]{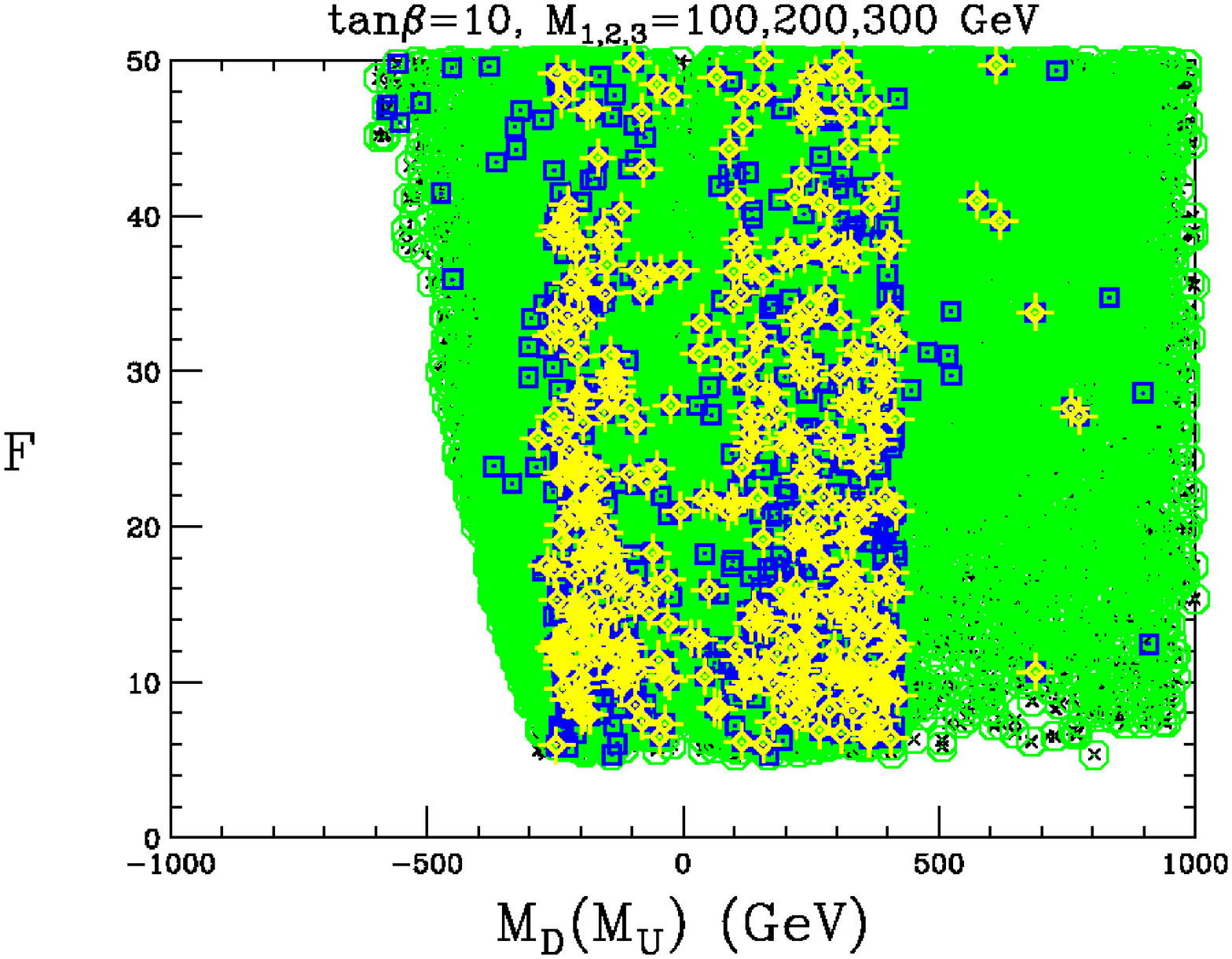}}
  \vspace*{-.1in}
\caption{ Fine tuning vs. $M_Q(\mgut)$, $M_U(\mgut)$ and $M_D(\mgut)$ 
  for points with $F<50$  taking $M_{1,2,3}(\mz)=100,200,300\gev$
and $\tanb=10$. Point notation as in Fig.~\ref{fvsmh1}.
Negative values indicate cases for which $M_{Q,U,D}^2<0$ in which case
the plot gives $-\sqrt{-M_{Q,U,D}^2}$.}
\label{fvsmqgut}
\vspace*{-.1in}
\end{figure}

\begin{figure}[ht!]

  \centerline{\includegraphics[width=2.4in,angle=90]{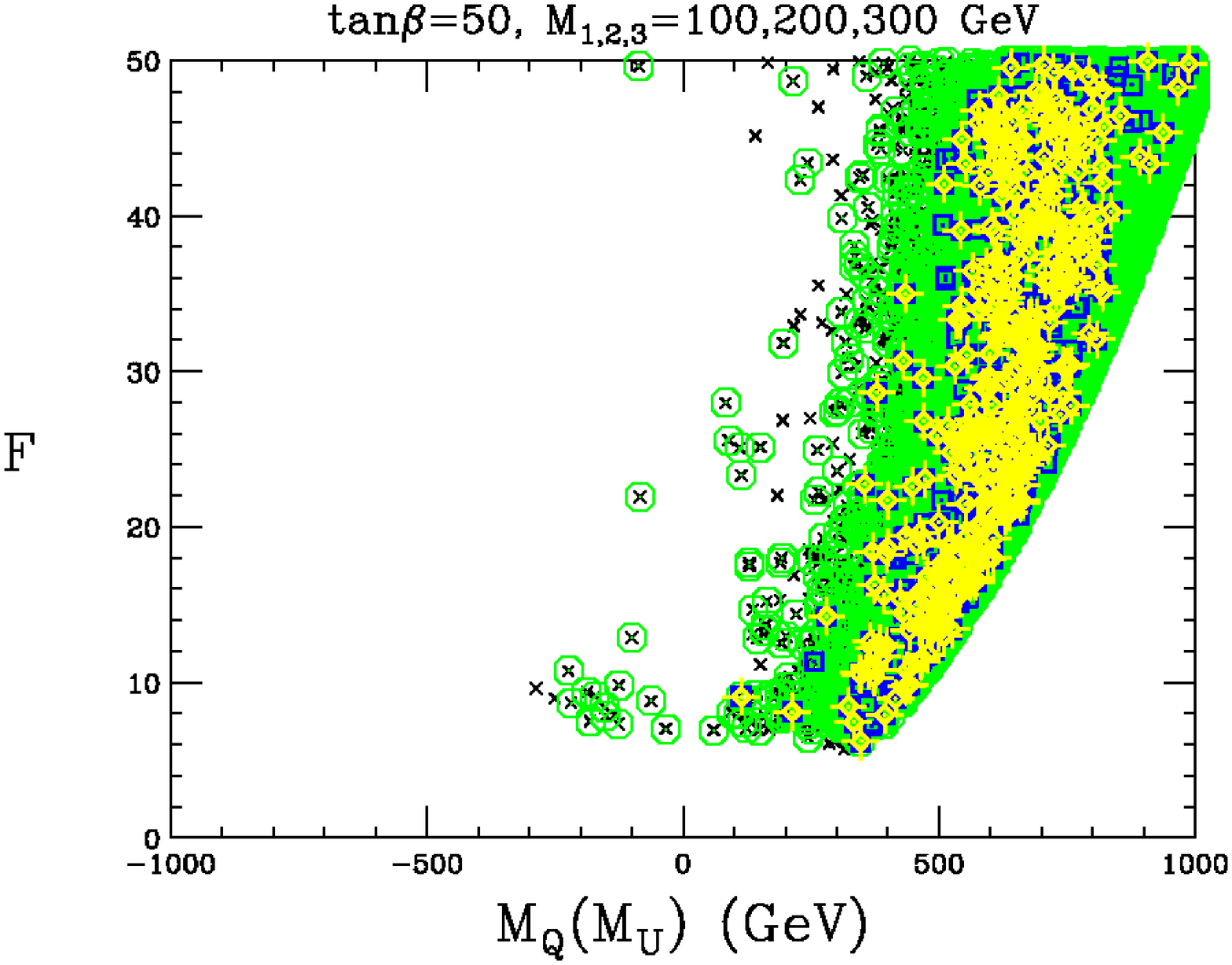}}
  \centerline{\includegraphics[width=2.4in,angle=90]{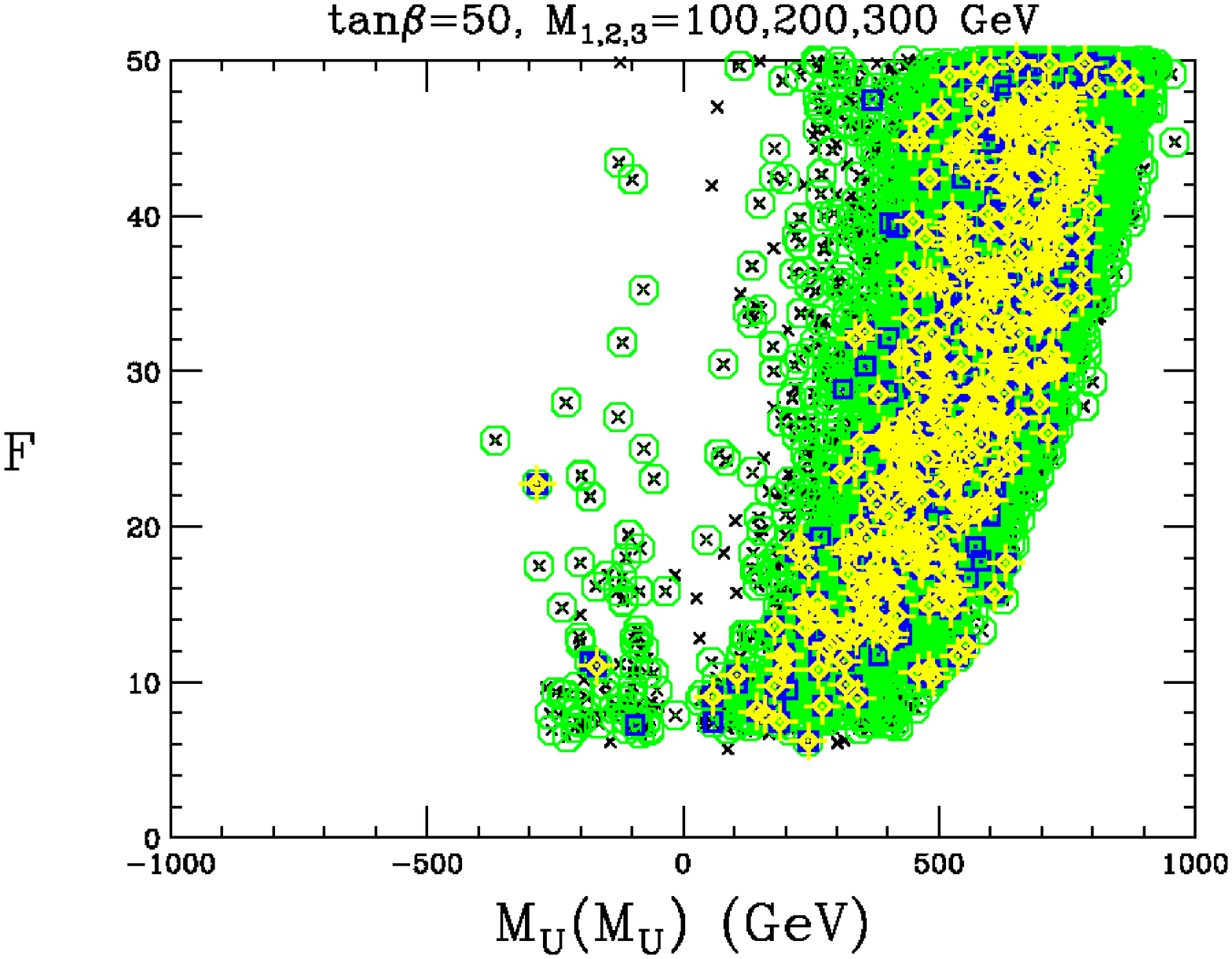}}
  \centerline{\includegraphics[width=2.4in,angle=90]{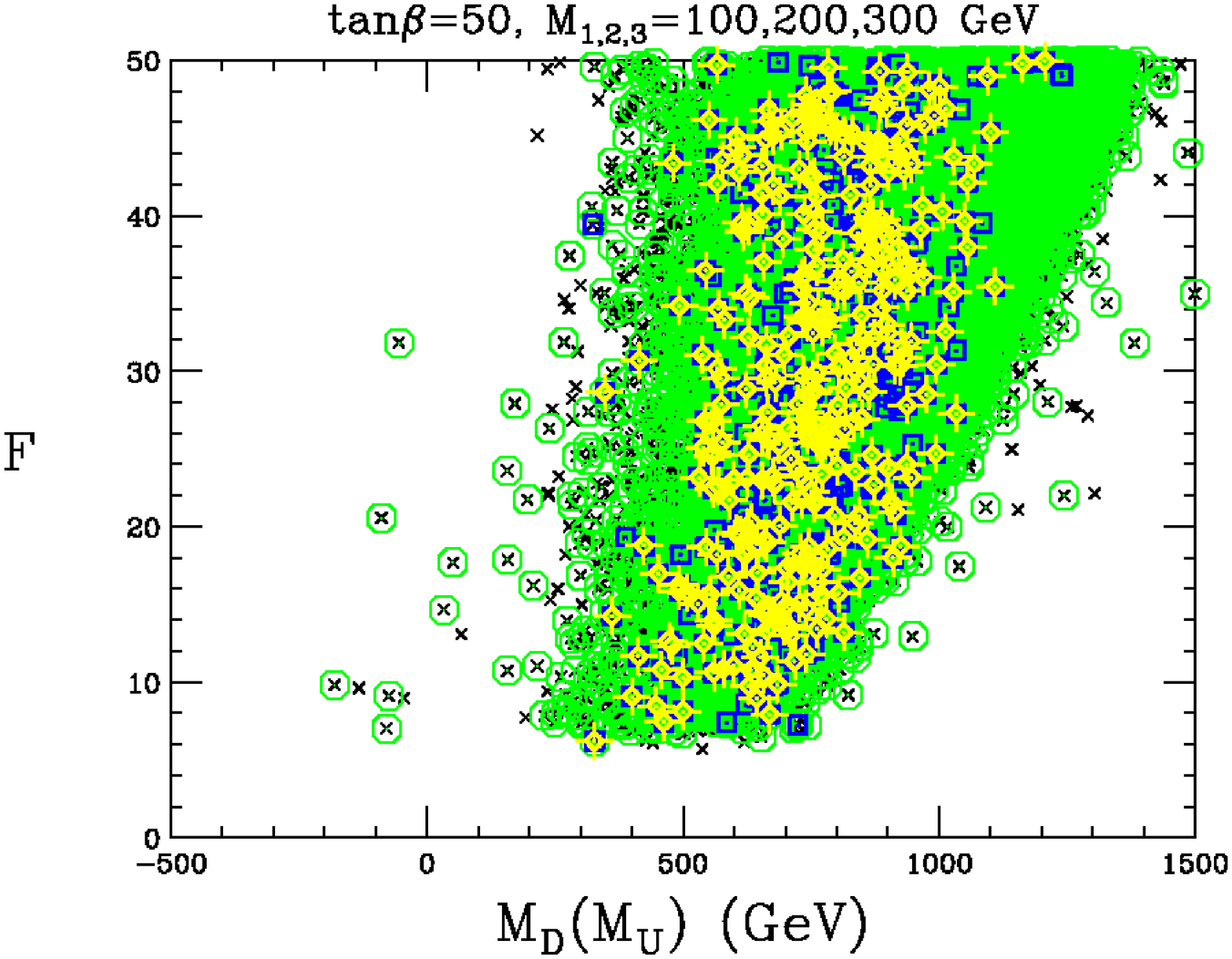}}
  \vspace*{-.1in}
\caption{ Fine tuning vs. $M_Q(\mgut)$, $M_U(\mgut)$ and $M_D(\mgut)$ 
  for points with $F<50$  taking $M_{1,2,3}(\mz)=100,200,300\gev$
and $\tanb=50$. Point notation as in Fig.~\ref{fvsmh1}.
Negative values indicate cases for which $M_{Q,U,D}^2<0$ in which case
the plot gives $-\sqrt{-M_{Q,U,D}^2}$.}
\label{fvsmqguttb50}
\vspace*{-.1in}
\end{figure}

Next, let us examine the soft squark masses --- $M_Q$, $M_U$, and
$M_D$ --- of the third generation.
Values for these at scale $\mz$ are plotted in
Figs.~\ref{fvsmqmz} and \ref{fvsmqmztb50} for $\tanb=10$ and
$\tanb=50$, respectively. We see that to obey limits on stop
masses, there is a fairly definite lower bound on $M_Q(\mz)$ and
$M_D(\mz)$,
although low values for $M_U(\mz)$ are possible. And, to achieve
$F<15$ and satisfy all experimental limits requires all these soft
masses to lie in a very well defined band.  The corresponding
GUT-scale values are given in Figs.~\ref{fvsmqgut} and \ref{fvsmqguttb50}. Points with $F<15$
satisfying all limits again have soft masses squared at the GUT scale that
fall within narrow bands (and are sometimes negative and sometimes positive).

\begin{figure}[ht!]
  \centerline{\includegraphics[width=2.4in,angle=90]{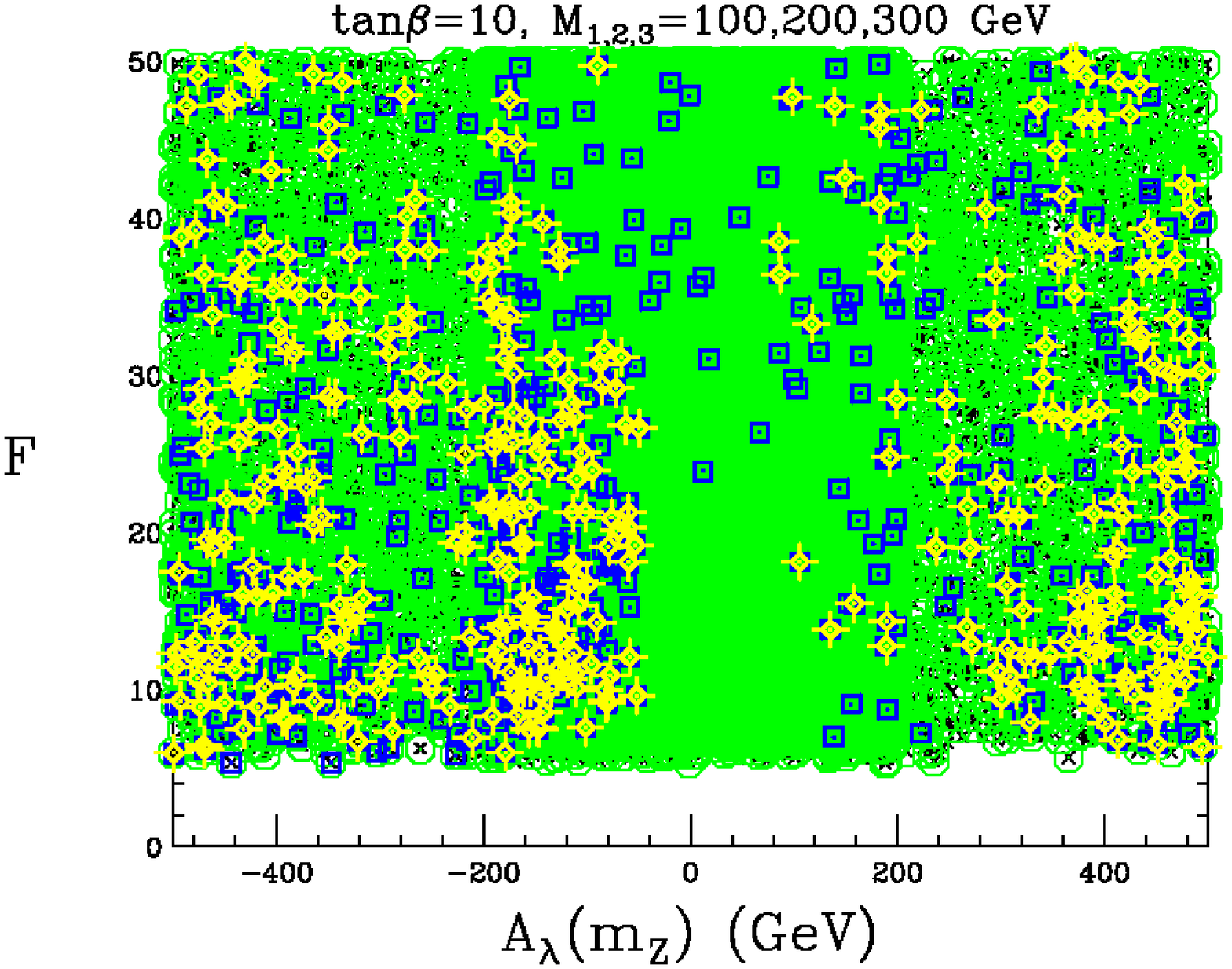}}
  \centerline{\includegraphics[width=2.4in,angle=90]{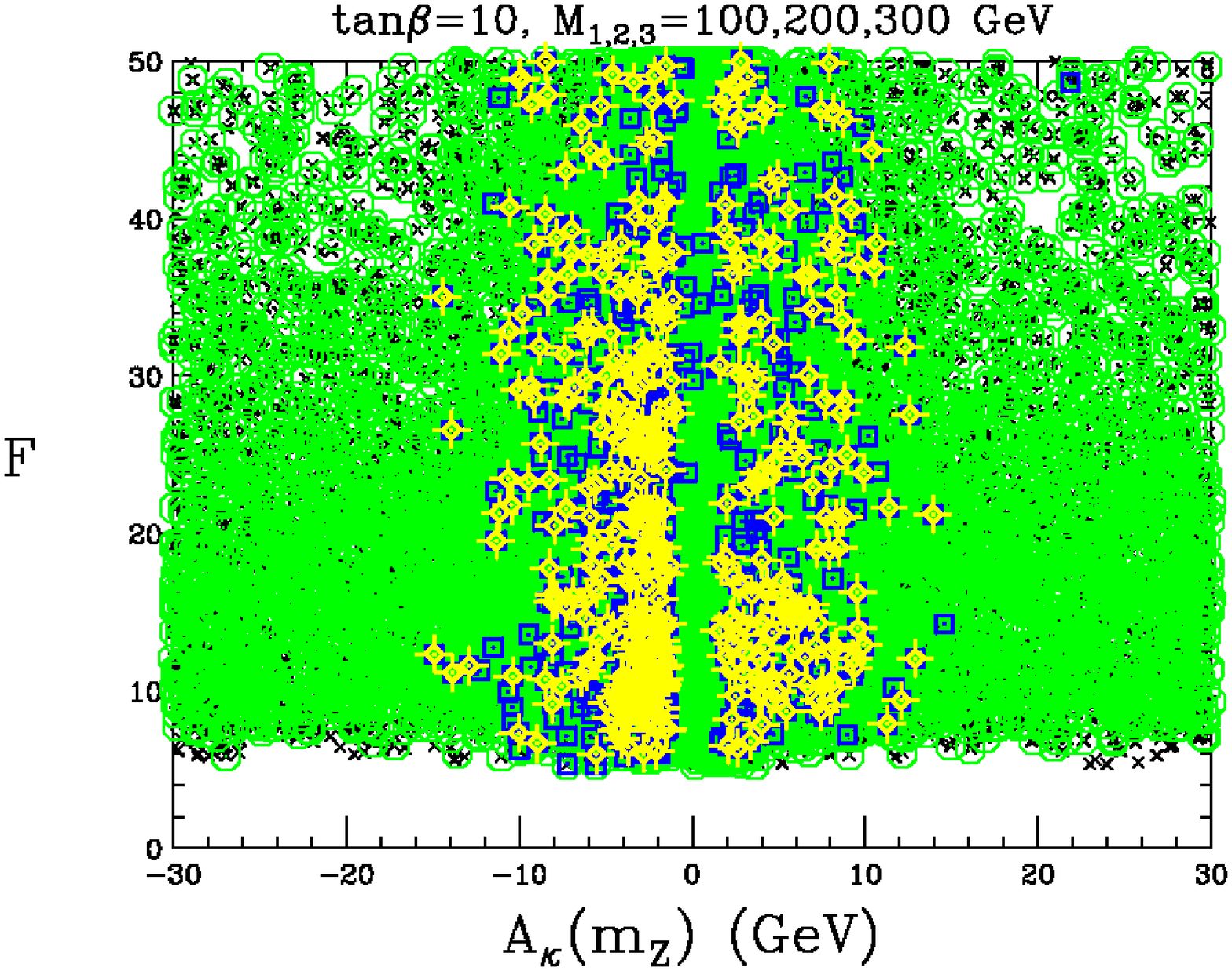}}
  \centerline{\includegraphics[width=2.4in,angle=90]{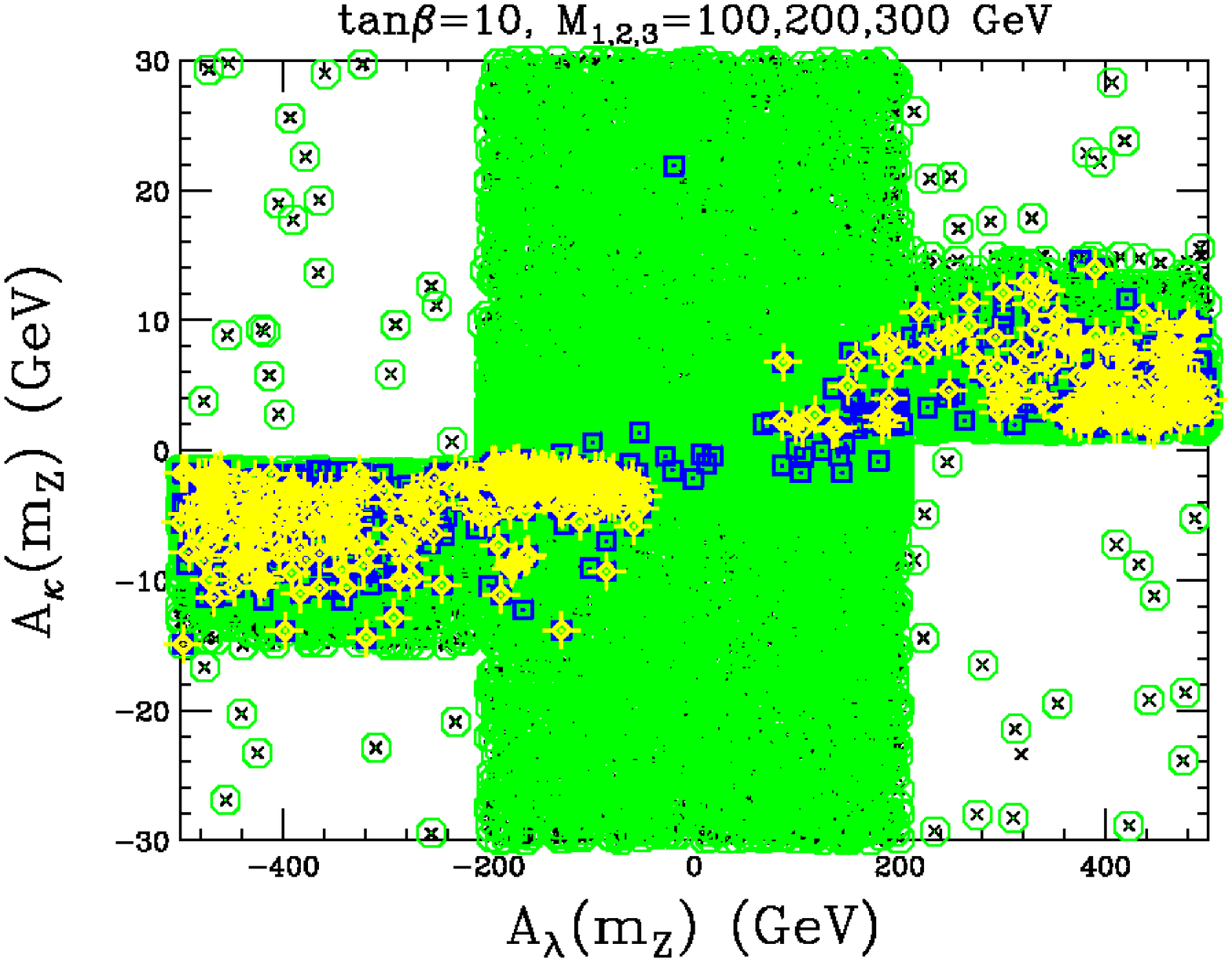}}
  \vspace*{-.1in}
\caption{ $F$ vs. $\alam(\mz)$ and $\akap(\mz)$ (upper two plots)
and $\akap(\mz)$ vs. $\alam(\mz)$ (lower plot) 
  for points with $F<50$  taking $M_{1,2,3}(\mz)=100,200,300\gev$
and $\tanb=10$. Point notation as in Fig.~\ref{fvsmh1}.
}
\label{akmzalmzplotsflt50}
\end{figure}

Next, we examine, for the case of $\tanb=10$, $\alam$ and $\akap$ at scale $\mz$.
Fig.~\ref{akmzalmzplotsflt50} gives some results.  The upper plot
shows that $F<15$ can be achieved for a wide range of $\alam(\mz)$,
with points that obey all limits requiring a minimum value of
$|\alam(\mz)|\gsim 40\gev$.  The middle plot shows that fully ok
points require $\akap(\mz)$ in a rather narrow band with
$0.8<|\akap(\mz)|<15$. The bottom plot shows the correlation between
$\akap(\mz)$ and $\alam(\mz)$ that is required to get small $\mai<2\mb$, as
discussed earlier.  Note that either both must be negative or both
positive for any point that is fully consistent with experimental
limits. The lower bounds on their absolute values for the fully ok
points --- the
large yellow crosses --- are those required to have large
enough $\br(\hi\to\ai\ai)$ to escape the $Z+b's$ LEP limits for
$\mhi\sim 100\gev$. Similar results are obtained for $\tanb=50$.  
All these results can be understood analytically as discussed in \cite{Dermisek:2006wr}.

\begin{figure}[ht!]
  \centerline{\includegraphics[width=2.4in,angle=90]{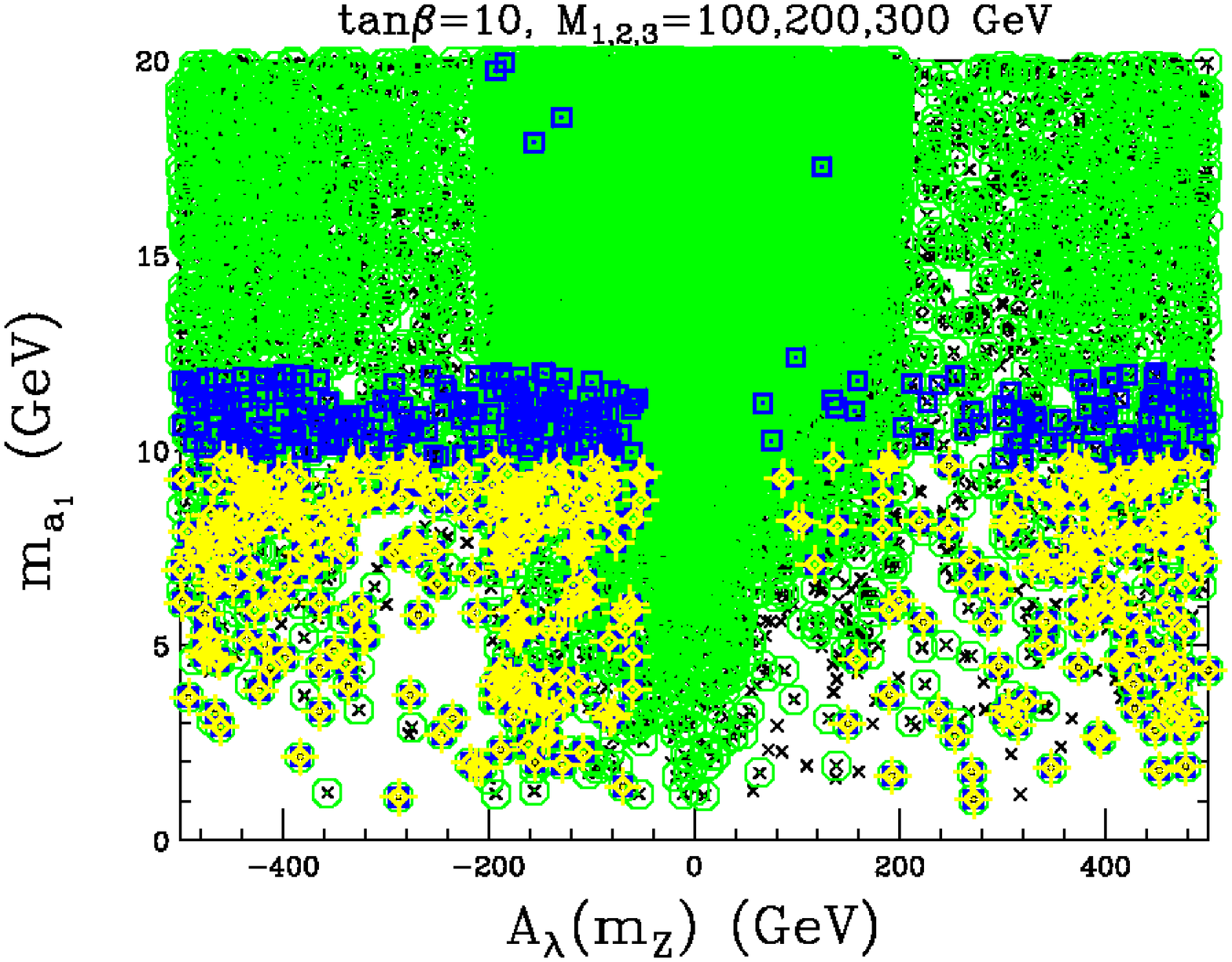}}
  \centerline{\includegraphics[width=2.4in,angle=90]{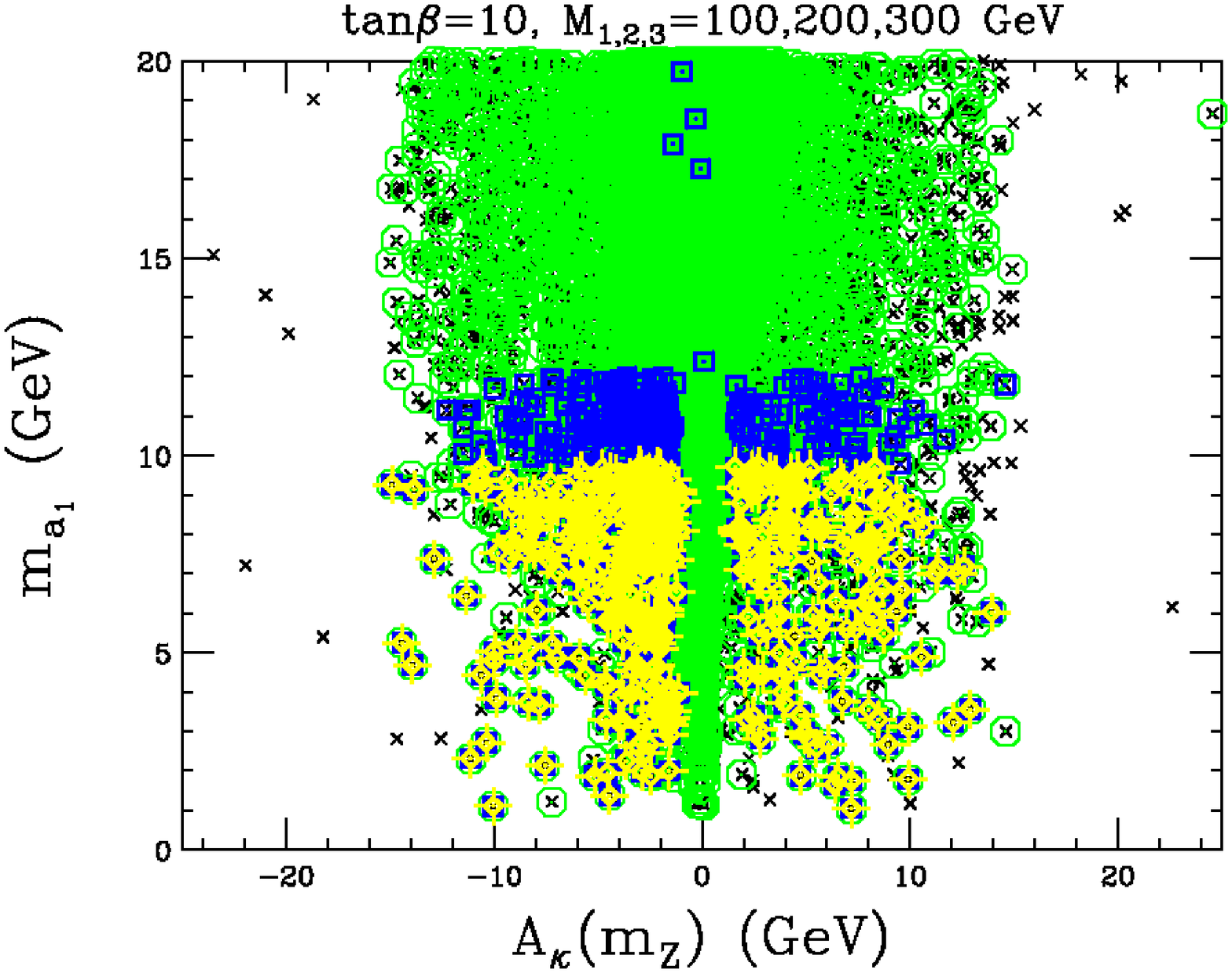}}
  \vspace*{-.1in}
\caption{ $\mai$ vs. $\alam(\mz)$ and $\akap(\mz)$ 
  for points with $F<50$  taking $M_{1,2,3}(\mz)=100,200,300\gev$
and $\tanb=10$. Point notation as in Fig.~\ref{fvsmh1}.
}
\label{ma1vsalakmzplotsflt50}
\end{figure}

Fig.~\ref{ma1vsalakmzplotsflt50} shows the dependence on $\mai$ on
$\alam(\mz)$ and $\akap(\mz)$ in the case of $\tanb=10$.  One observes
that large $\mai$ can be achieved for these same ranges of
$\alam(\mz)$ and $\akap(\mz)$ just as easily as small $\mai$.  It is
just that cases with large $\mai>2\mb$ and small $F$, which requires
$\mhi\sim 100\gev$, are not consistent with LEP limits on the net
$Z+b's$ channel, as we have discussed. Similar results are found for
$\tanb=50$.

\begin{figure}[ht!]
  \centerline{\includegraphics[width=2.4in,angle=90]{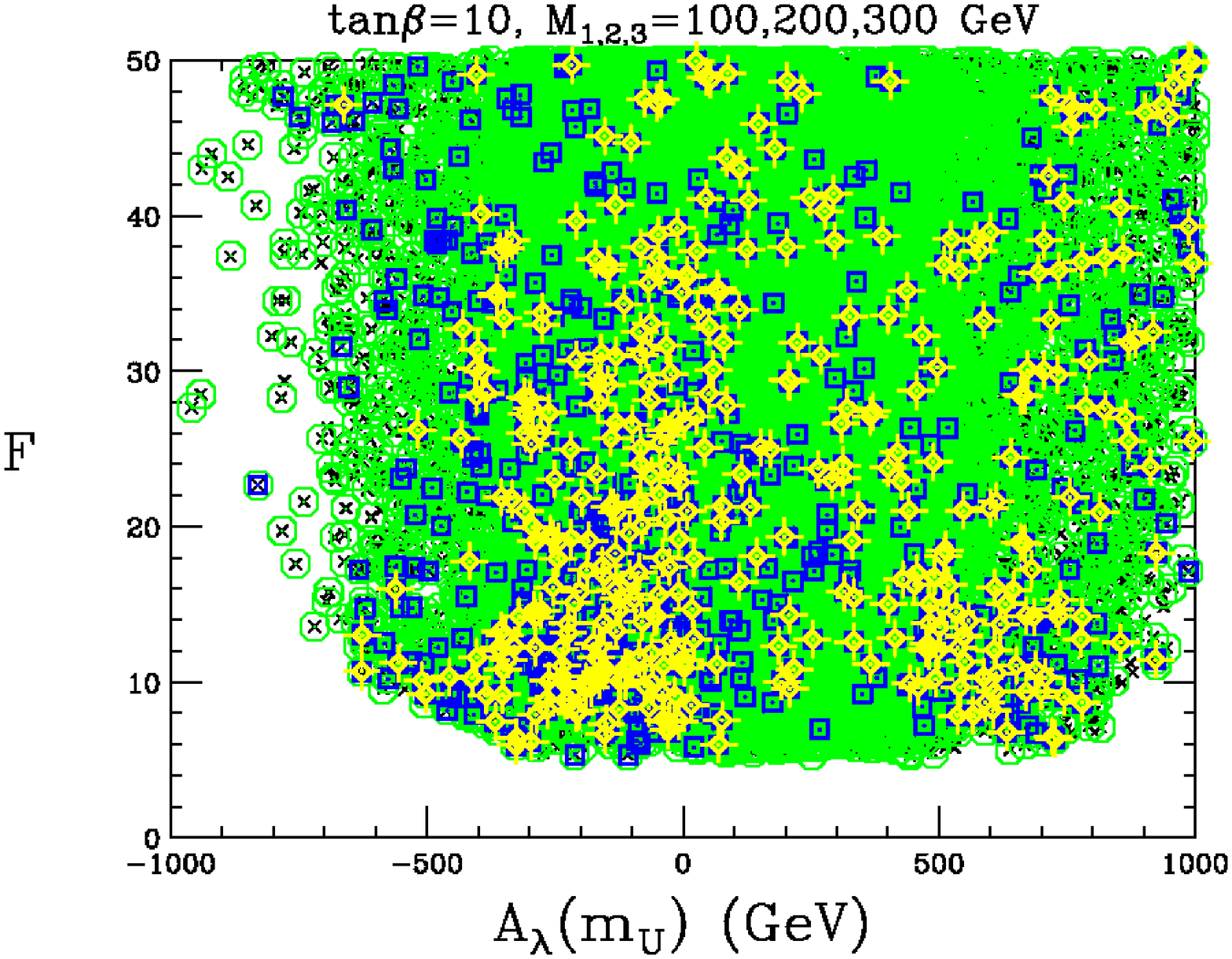}}
  \centerline{\includegraphics[width=2.4in,angle=90]{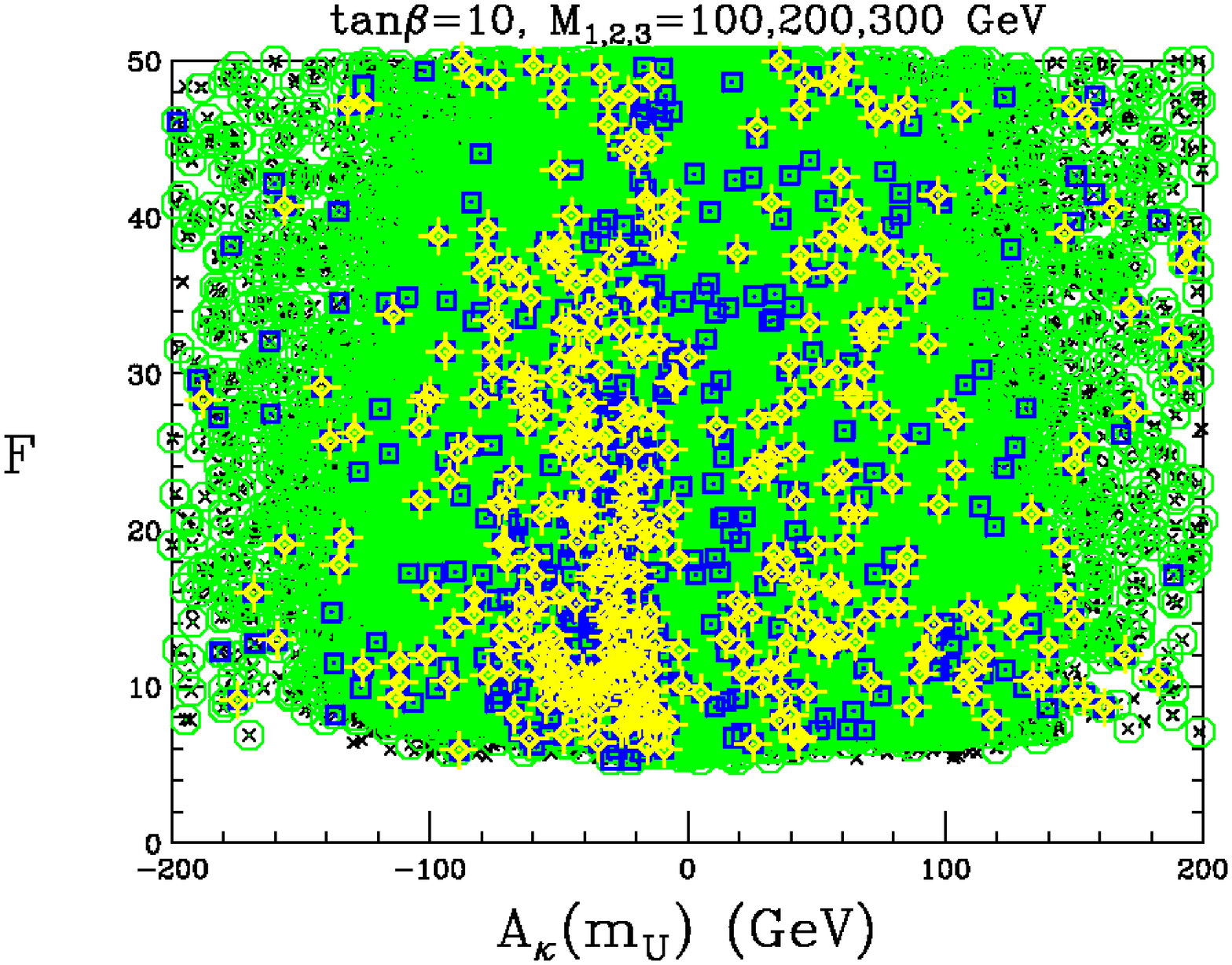}}
  \centerline{\includegraphics[width=2.4in,angle=90]{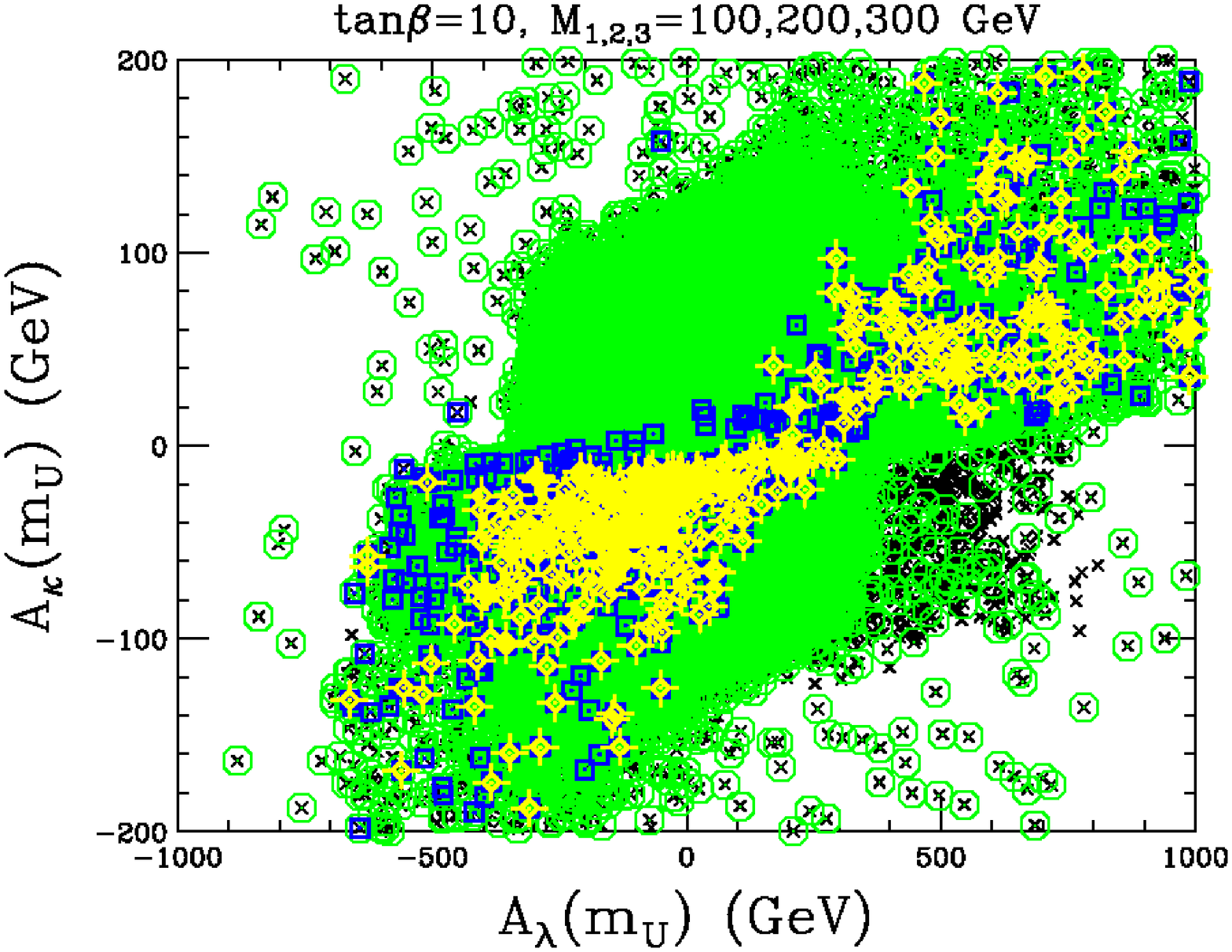}}
  \vspace*{-.1in}
\caption{ $F$ vs. $\alam(\mgut)$ and $\akap(\mgut)$ (upper two plots)
and $\akap(\mgut)$ vs. $\alam(\mgut)$ (lower plot) 
  for points with $F<50$  taking $M_{1,2,3}(\mz)=100,200,300\gev$
and $\tanb=10$. Point notation as in Fig.~\ref{fvsmh1}.
}
\label{akgutalgutplotsflt50}
\end{figure}

\begin{figure}[ht!]

  \centerline{\includegraphics[width=2.4in,angle=90]{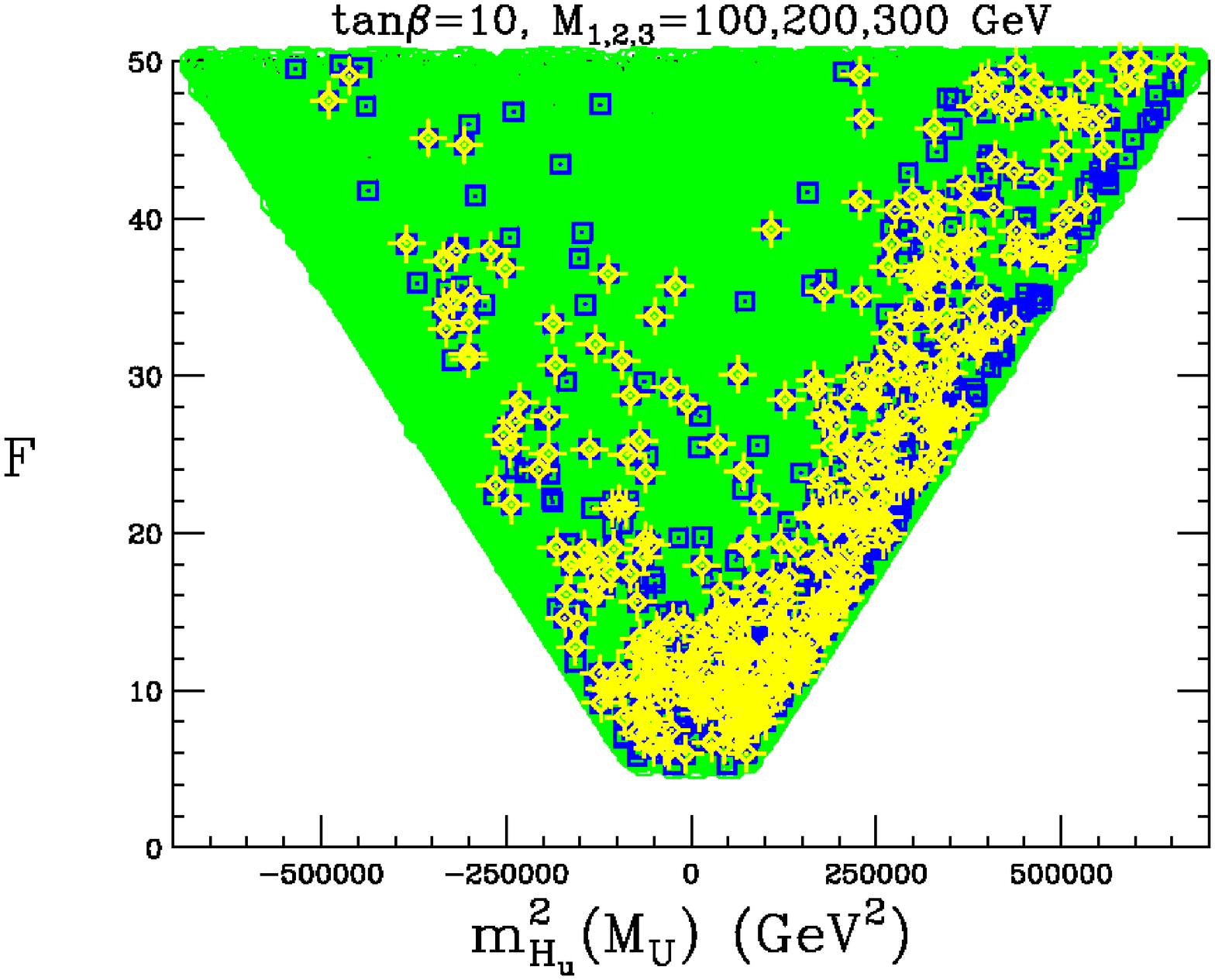}}
  \centerline{\includegraphics[width=2.4in,angle=90]{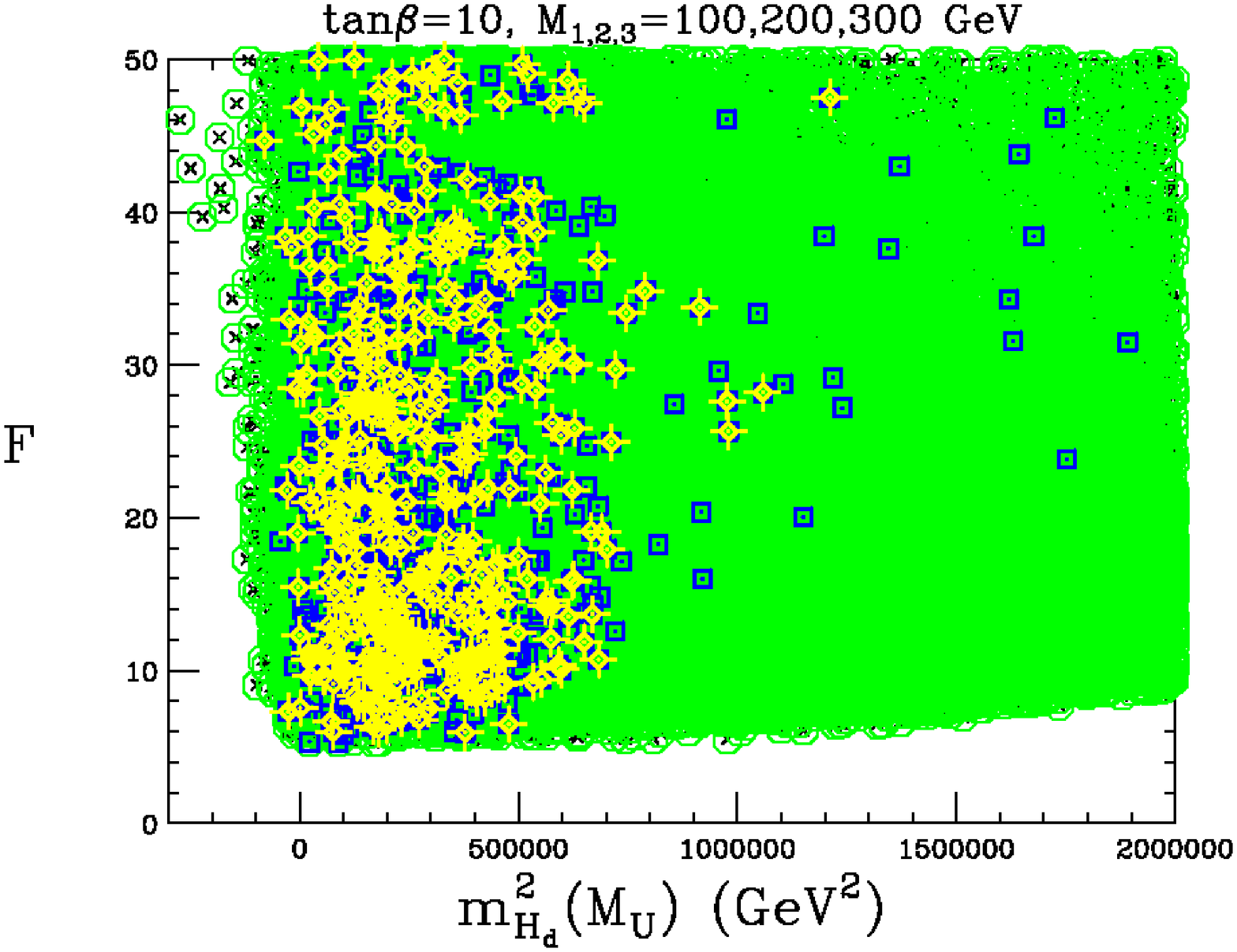}}
  \centerline{\includegraphics[width=2.4in,angle=90]{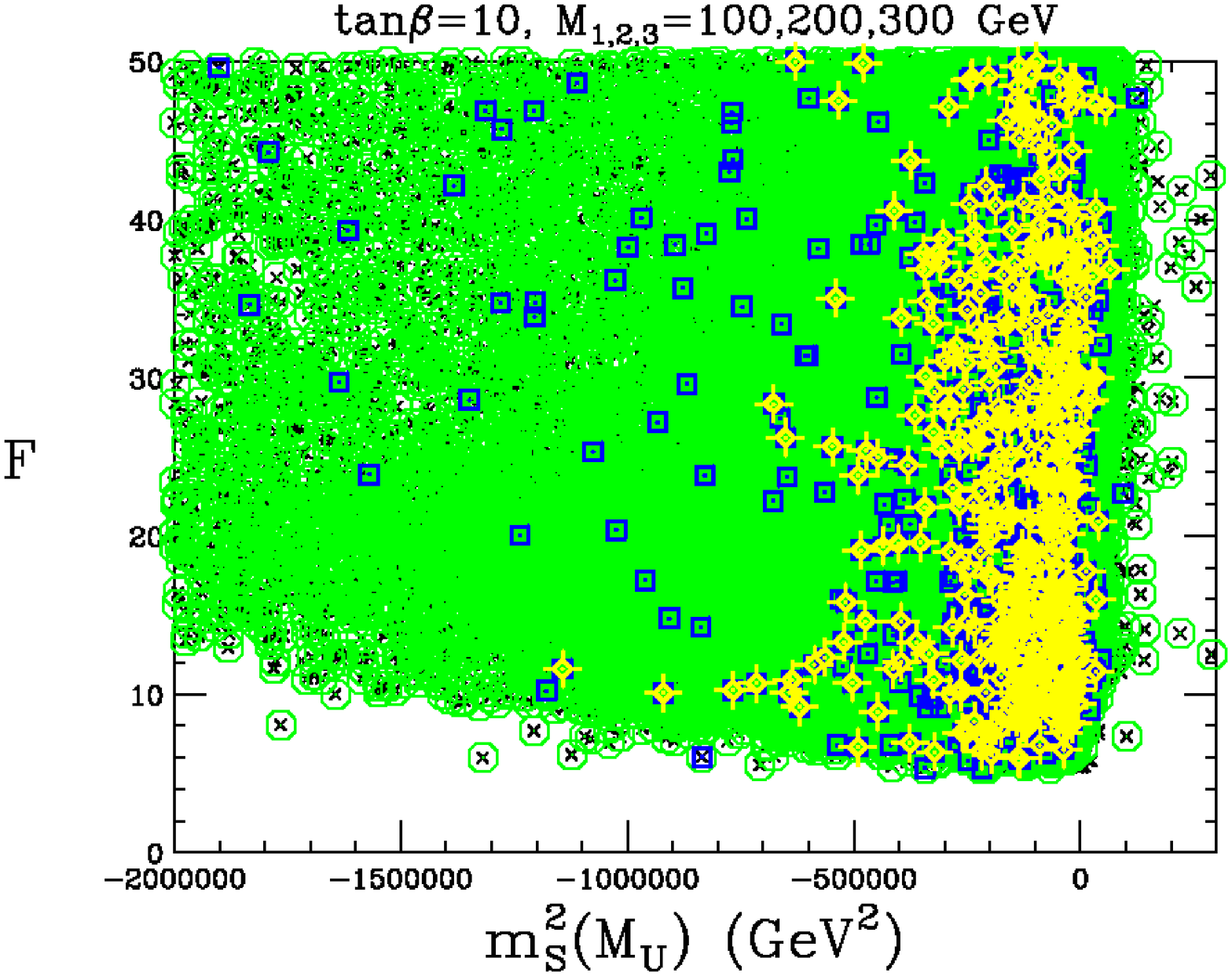}}
  \vspace*{-.1in}
\caption{ Fine tuning vs. $\mhusq(\mgut)$, $\mhdsq(\mgut)$ and $\mssq(\mgut)$ 
  for points with $F<50$  taking $M_{1,2,3}(\mz)=100,200,300\gev$
and $\tanb=10$. Point notation as in Fig.~\ref{fvsmh1}.
}
\label{fvsmhsqgutflt50}
\end{figure}

Plots of the GUT-scale parameters, $\alam(\mgut)$ and $\akap(\mgut)$,
appear in Fig.~\ref{akgutalgutplotsflt50}.  These show that the
lowest-$F$ scenarios that are fully consistent with experiment are
often achieved for small values of these parameters. In terms of model
building, these soft-SUSY-breaking parameters are thus close to values
associated with 'no-scale' soft-SUSY-breaking.

\begin{figure}[ht!]

  \centerline{\includegraphics[width=2.4in,angle=90]{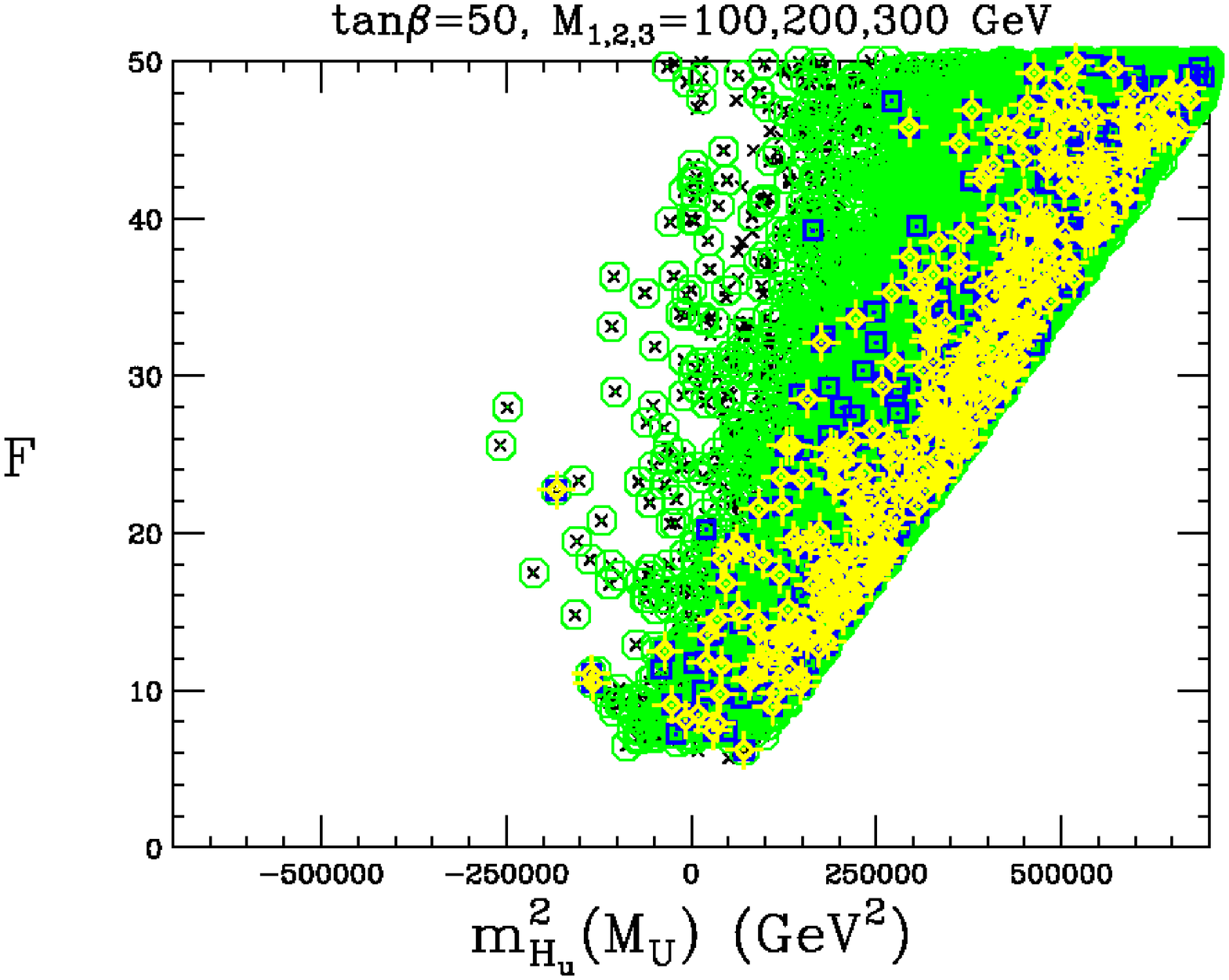}}
  \centerline{\includegraphics[width=2.4in,angle=90]{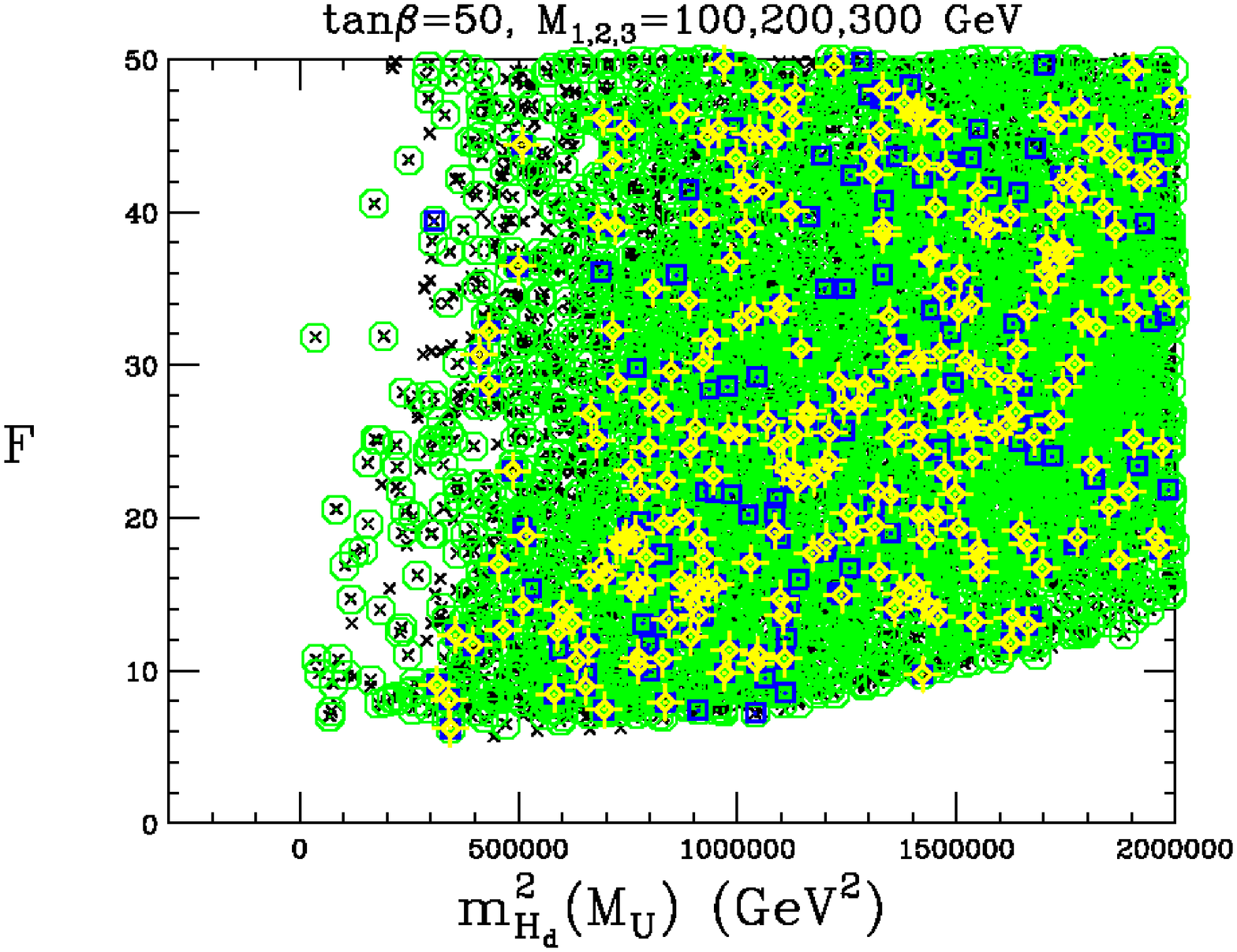}}
  \centerline{\includegraphics[width=2.4in,angle=90]{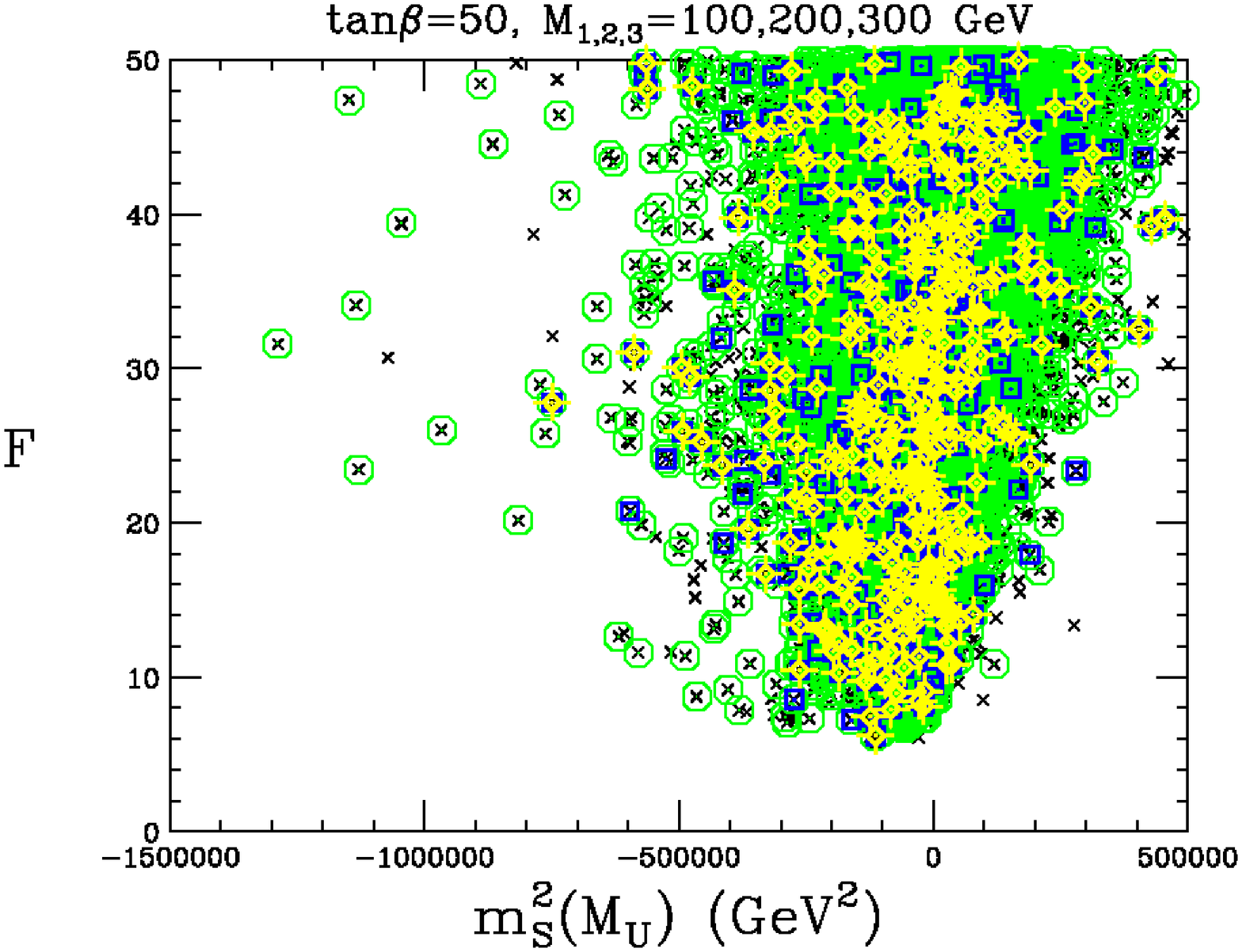}}
  \vspace*{-.1in}
\caption{ Fine tuning vs. $\mhusq(\mgut)$, $\mhdsq(\mgut)$ and $\mssq(\mgut)$ 
  for points with $F<50$  taking $M_{1,2,3}(\mz)=100,200,300\gev$
and $\tanb=50$. Point notation as in Fig.~\ref{fvsmh1}.
}
\label{fvsmhsqgutflt50tb50}
\end{figure}

Probably the most interesting parameter correlation is that regarding
the soft-SUSY-breaking Higgs masses squared at the GUT scale.  These
are plotted in Figs.~\ref{fvsmhsqgutflt50} and
\ref{fvsmhsqgutflt50tb50}
for the cases of $\tanb=10$ and $\tanb=50$, respectively. These plots show 
that the fully ok scenarios with smallest $F$ have very modest 
soft masses squared at the GUT scale, especially in the case of $\mhusq$.
Thus, something close to a 'no-scale' model for soft Higgs masses
squared at the GUT scale is preferred for low $F$.

\begin{figure}[ht!]
  \centerline{\includegraphics[width=2.4in,angle=90]{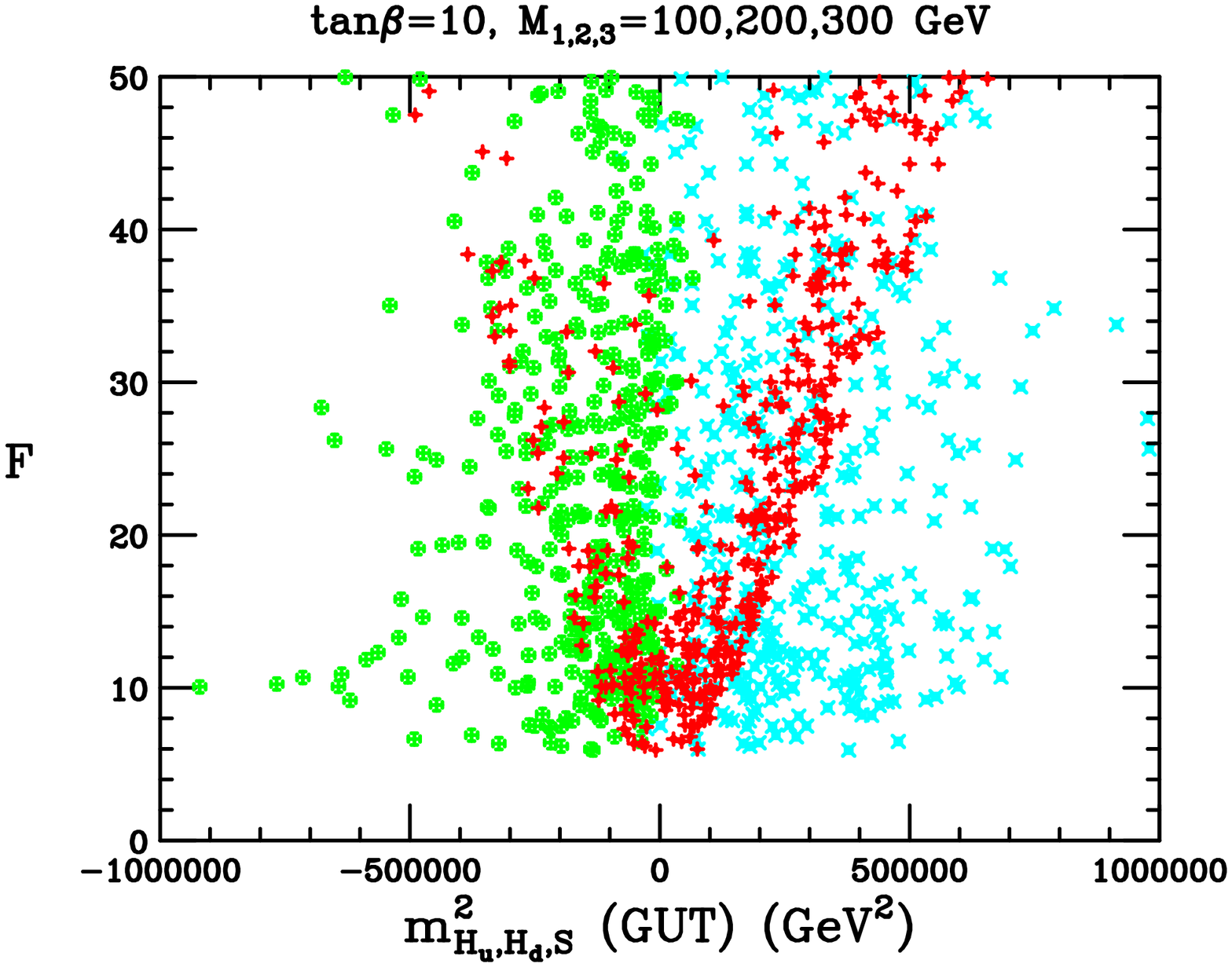}}
\caption{ Fine tuning vs. the GUT-scale soft Higgs masses squared for
  points with $F<50$  taking $M_{1,2,3}(\mz)=100,200,300\gev$
and $\tanb=10$. Point notation: dark (red)
$\mhusq(\mgut)$; light grey (cyan) $\mhdsq(\mgut)$; 
darker grey (green) $\mssq(\mgut)$. Points plotted are the 
yellow fancy cross points from Fig.~\ref{fvsmhsqgutflt50}.}
\label{fvsmhsqgut}
\end{figure}

This preference for a 'no-scale' type of boundary condition for the
Higgs soft masses squared at the GUT-scale is further emphasized by
the $\tanb=10$ plot of Fig.~\ref{fvsmhsqgut}, where we overlap the
values of $\mhusq(\mgut)$, $\mhdsq(\mgut)$ and $\mssq(\mgut)$ for $F<50$ 
yellow fancy cross scenarios of Fig.~\ref{fvsmhsqgutflt50}.

\begin{figure}[ht!]
  \centerline{\includegraphics[width=2.4in,angle=90]{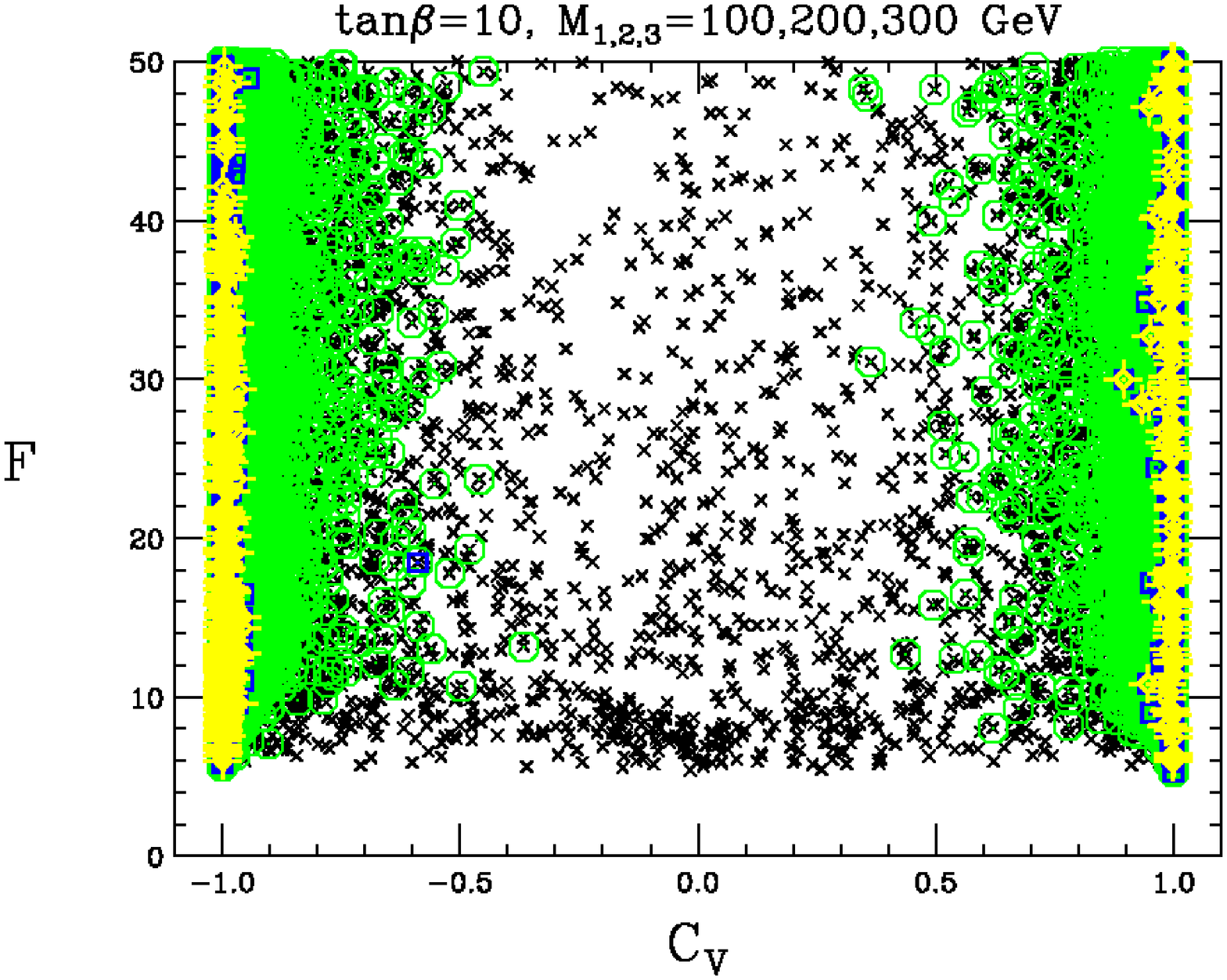}}
\caption{ Fine tuning vs. the relative coupling strength,
$C_V\equiv g_{\hi WW}/g_{\hsm WW}$ for $\tanb=10$. Point notation as in Fig.~\ref{fvsmh1}.}
\label{fvscv}
\end{figure}



We have said in many places that for the fully ok scenarios the $\hi$ 
has quite SM-like $\hi ZZ$ coupling.  This is illustrated in
Fig.~\ref{fvscv} where we plot $F$ as a function of 
\beq
C_V\equiv {g_{\hi ZZ}\over g_{\hsm ZZ}}\,,
\eeq
for the case of $\tanb=10$ (results for $\tanb=50$ are similar).
We see that the fully ok yellow fancy crosses all have
$|C_V|\sim 1$. In fact, for $F<50$ scenarios, 
$|C_V|\sim 1$ also for the (blue) square points that are not also
yellow fancy crosses, \ie\ those points obtained
if one only requires that the scenario is consistent with
experimental limits that include the $Z+2b$ channel (that is,
before requiring $\mai<2\mb$ as needed to avoid the limits on the
combined $Z+b's$ channel). Suppressed $|C_V|$ values only appear in
these plots if the
Higgs experimental limits are removed. As discussed later, there are
some very special points for which this is not true that will be
considered in a follow-up paper.

\subsection{Parameter correlations for the very lowest $F$ points}

In this subsection, we consider at a still more detailed level 
the fully ok yellow points having large $\br(\hi\to\ai\ai)$ and
$\mai<2\mb$ that also have $F<10$. We will present results only for
  the case of $\tanb=10$ and
$M_1,M_2,M_3=100,200,300\gev$.  In general, the other $\tanb$ values
give similar correlations aside from the shift in the value of $\mhi$
that gives the lowest $F$ value. In the plots presented in 
this section we will use blue $+$'s in place of
the yellow crosses, since the latter do not display well on their own.
Hopefully, there are few enough points on the following plots that the
reader can match points from one plot to another.

\begin{figure}[h!]
\centerline{\includegraphics[width=2.4in,angle=90]{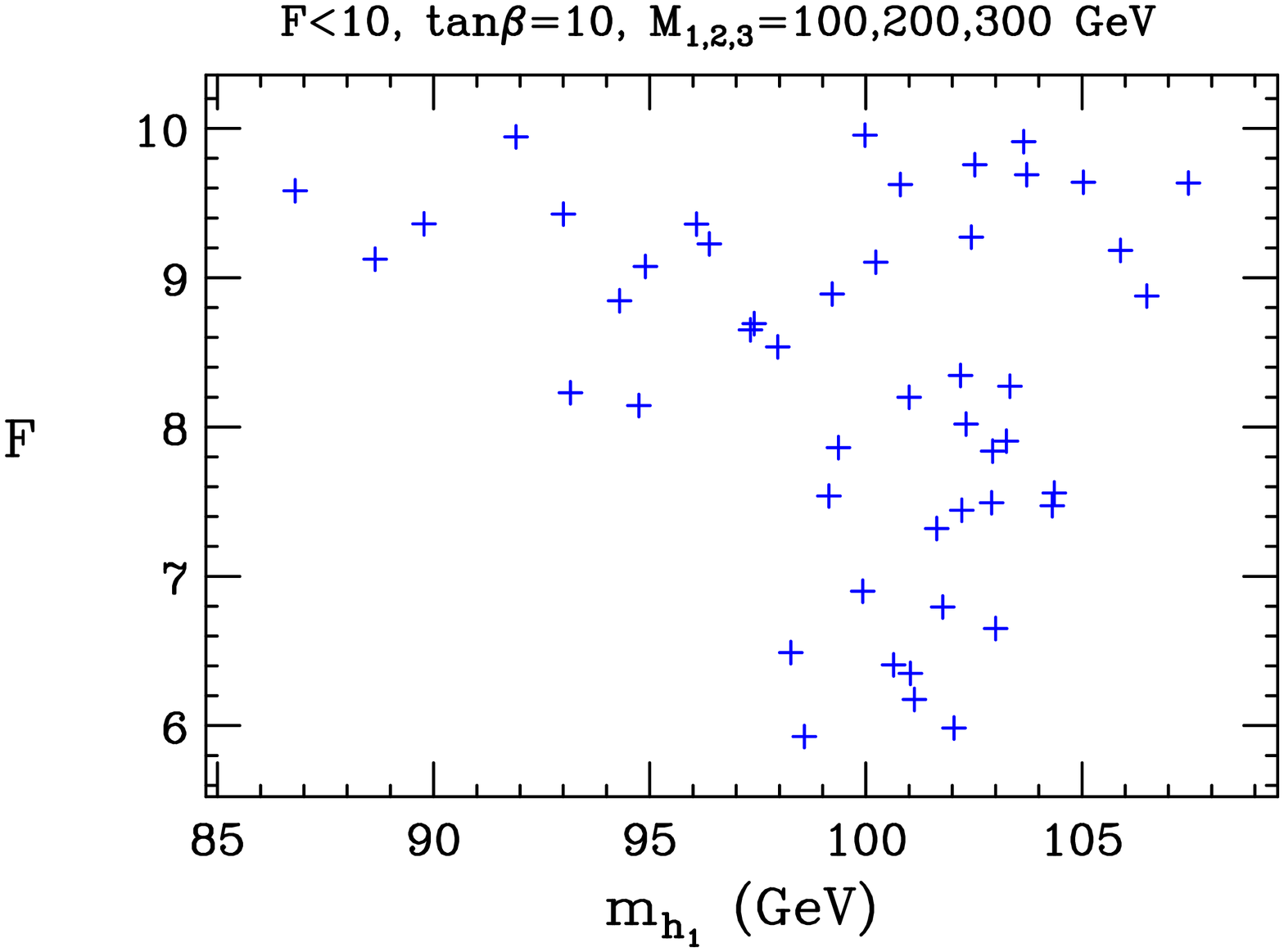}}
\vspace*{.15in}
\centerline{\includegraphics[width=2.4in,angle=90]{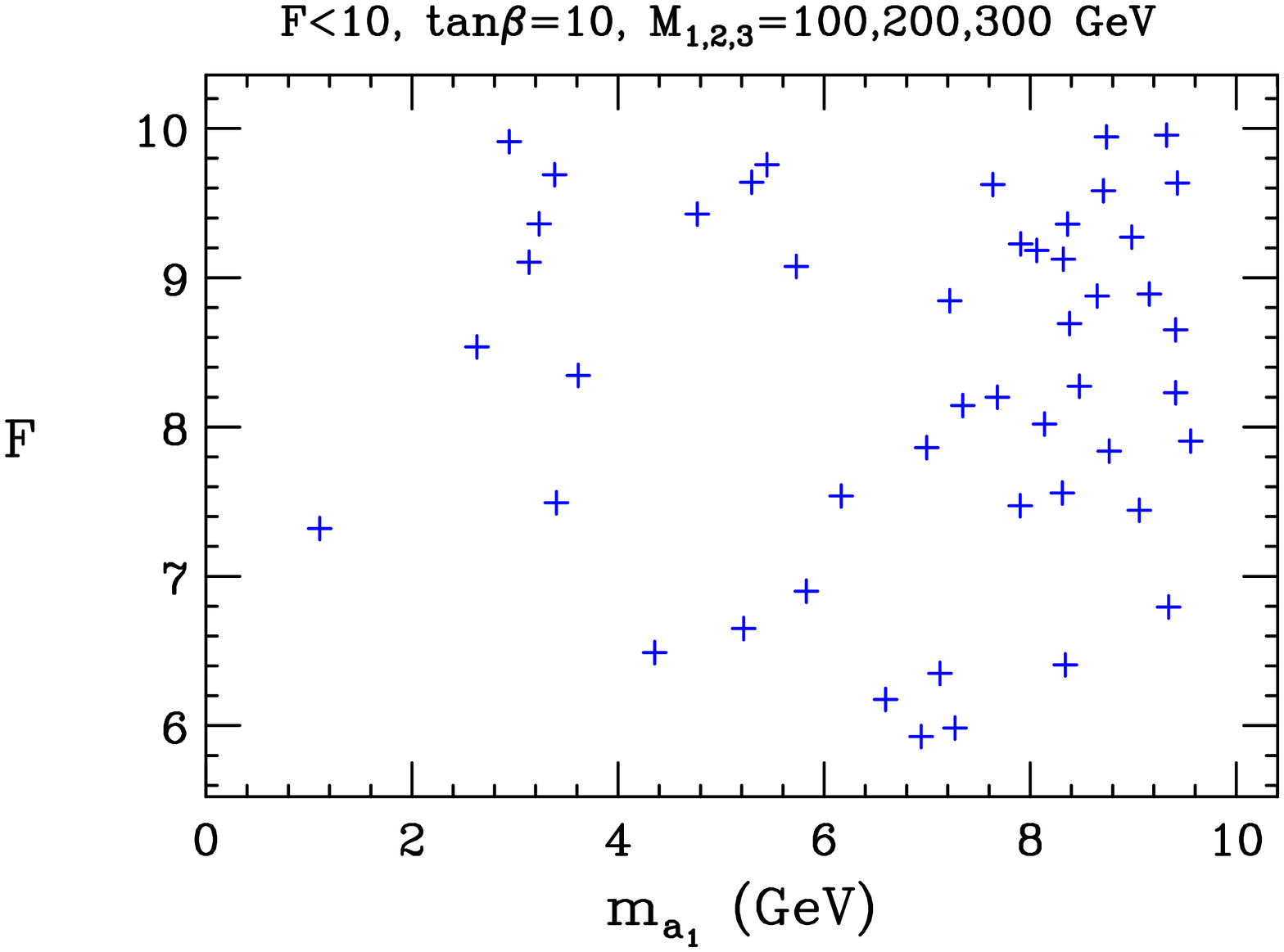}}
\caption{The upper plot shows fine tuning vs. $\mhi$ for the large
  (yellow) cross   points (for clarity, we use blue $+$'s in their place in this
  and succeeding plots)
 with $F<10$  taking $M_{1,2,3}(\mz)=100,200,300\gev$
and $\tanb=10$. The lower plot shows $F$ vs. $\mai$. }
\label{fvsmhimai}
\end{figure}
First, we show $F$ as a function of $\mhi$ and $\mai$ in Fig.~\ref{fvsmhimai}.
In the upper plot, we see again the preference for $\mhi$ near $100\gev$.
The lower plot shows that the very smallest $F$ values occur at $\mai$
values above $2\mtau$, implying that the $\ai\to \tauptaum$ channel is
the dominant $\ai$ decay. We note that the $2\mtau<\mai$ part of the
$\mai<2\mb$ fully ok zone was also found in the companion paper
\cite{Dermisek:2006wr} to be preferred in order to avoid
fine-tuning associated with getting small $\mai$.

\begin{figure}[h!]
\centerline{\includegraphics[width=2.4in,angle=90]{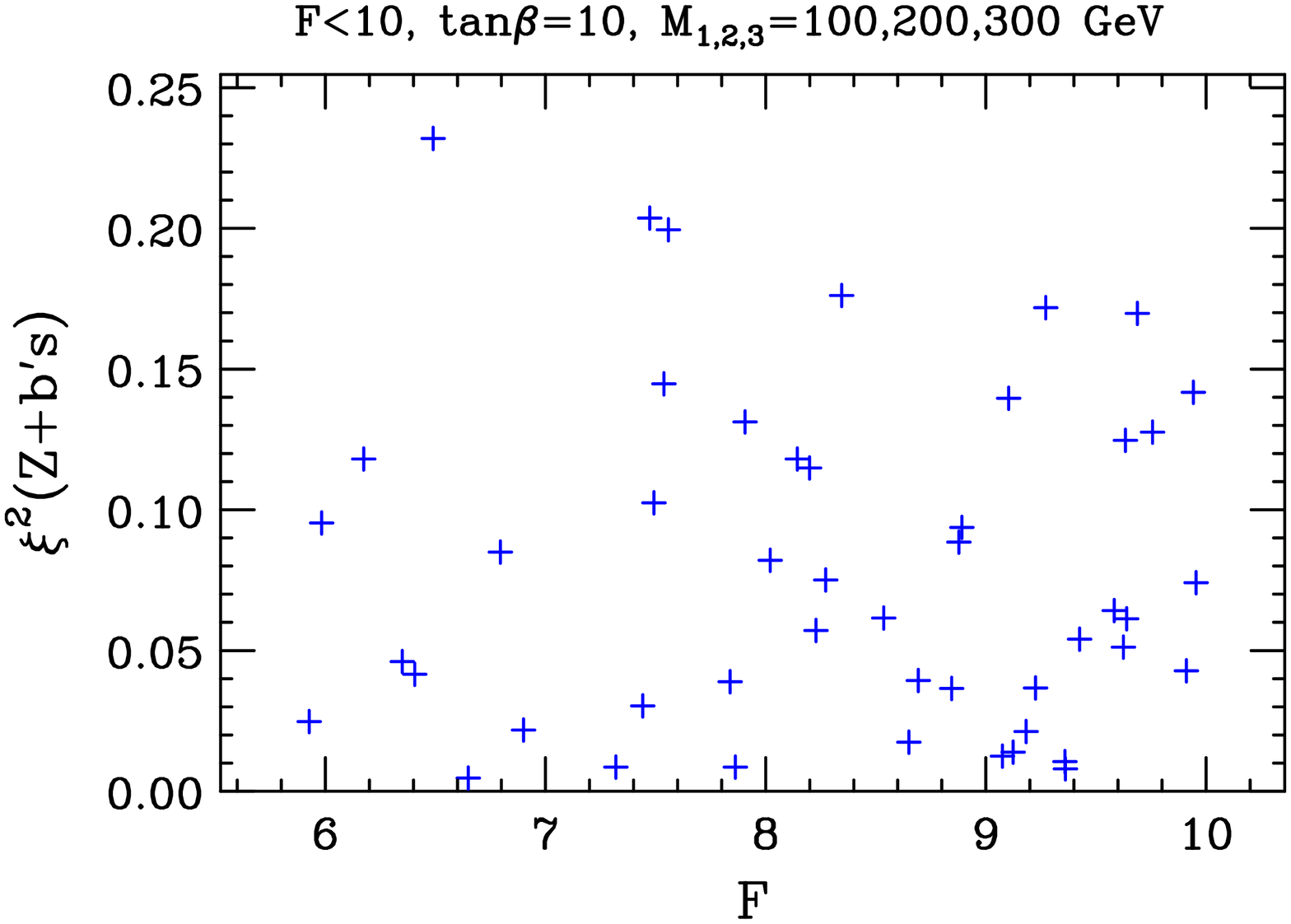}}
\caption{The  plot shows $\xi^2(Z+b's)$ (which equals $\cbb$ 
since $\cbbbb=0$ when $\mai<2\mb$) vs. $F$  for the large
  (yellow) cross   points 
 with $F<10$  taking $M_{1,2,3}(\mz)=100,200,300\gev$
and $\tanb=10$.}
\label{xisqvsf}
\end{figure}
One important prediction of any given parameter set is that for
$\xi^2(Z+b's)$, where
\bea 
\xi^2(Z+b's)&\equiv& {g_{ZZ\hi}^2\over
  g_{ZZ\hsm}^2}\Bigl[\br(\hi\to b \anti b)\cr
&&+\br(\hi\to\ai\ai)\left[\br(\ai\to b\anti b)\right]^2\Bigr]\,. 
\label{xisqdef}
\eea
For the fully ok points one has $\mai<2\mb$ and thus
$\xi^2(Z+b's)=\cbb$.  More generally, $\xi^2(Z+b's)$ is the net rate
for LEP production of $Z+2b$ {\it and} $Z+4b$ final states relative to
the rate that one would obtain for a SM Higgs boson which decayed
entirely to $b\anti b$. Of particular interest is the correlation
between $\xi^2(Z+b's)$ and $F$.  Thus, an important comparison is the
model prediction for $\xi^2(Z+b's)$ relative to the excess found at
LEP in the vicinity of $b\anti b$ mass $\sim 100\gev$.  The value of
$\xi^2(Z+b's)=\cbb$ as a function of $\mhi$ is plotted in
Fig.~\ref{xisqvsf}.  We observe that the points with $F<10$ lie in the
range $\cbb\lsim 0.2$ with many of the very lowest $F$ points having
$\cbb\in[0.07,0.13]$, the range most consistent with the LEP excess at
$bb$ mass $\sim 98\gev$.

\begin{figure}[h!]
  \centerline{\includegraphics[width=2.4in,angle=90]{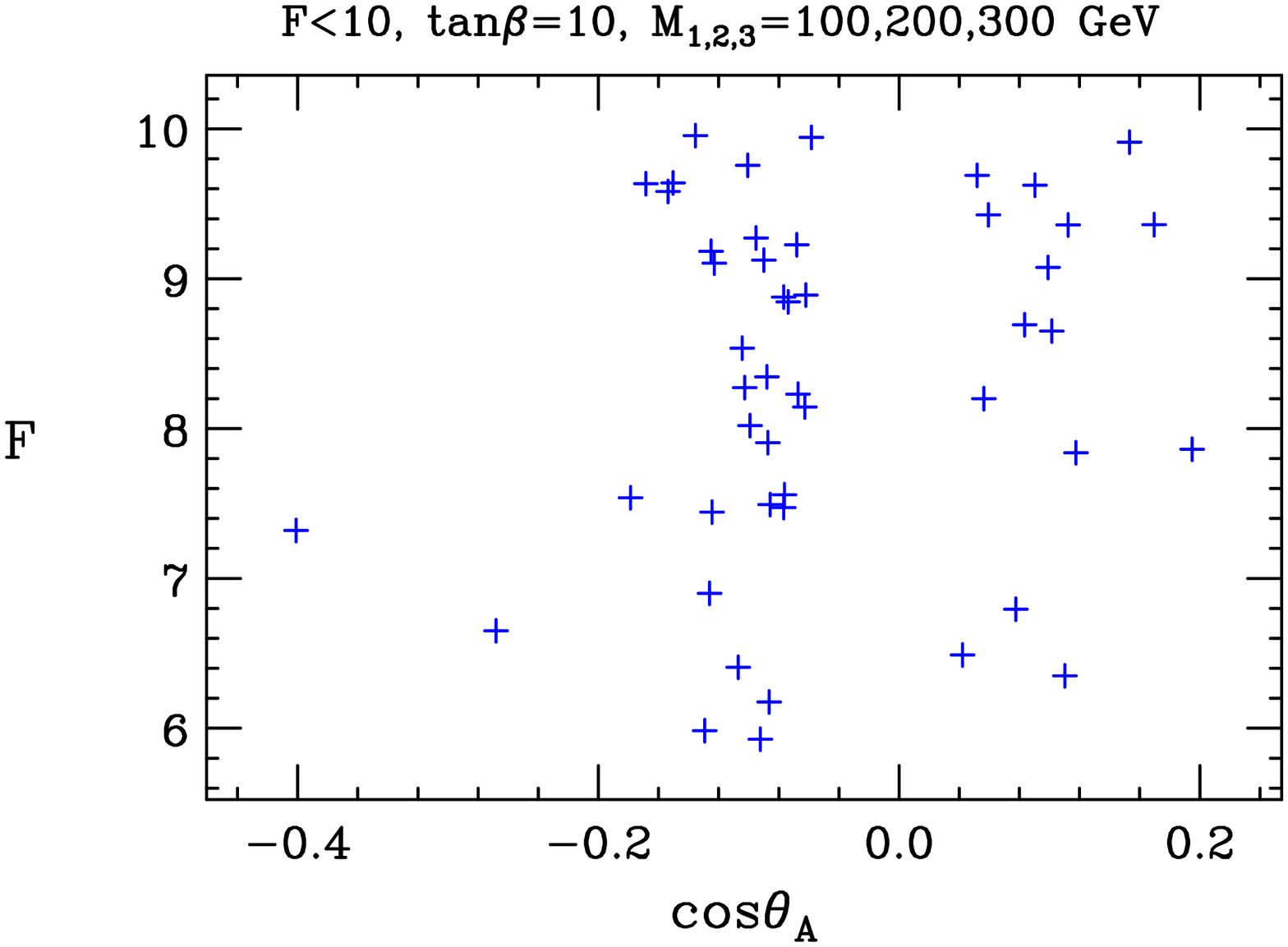}}
\vspace*{.15in}
  \centerline{\includegraphics[width=2.4in,angle=90]{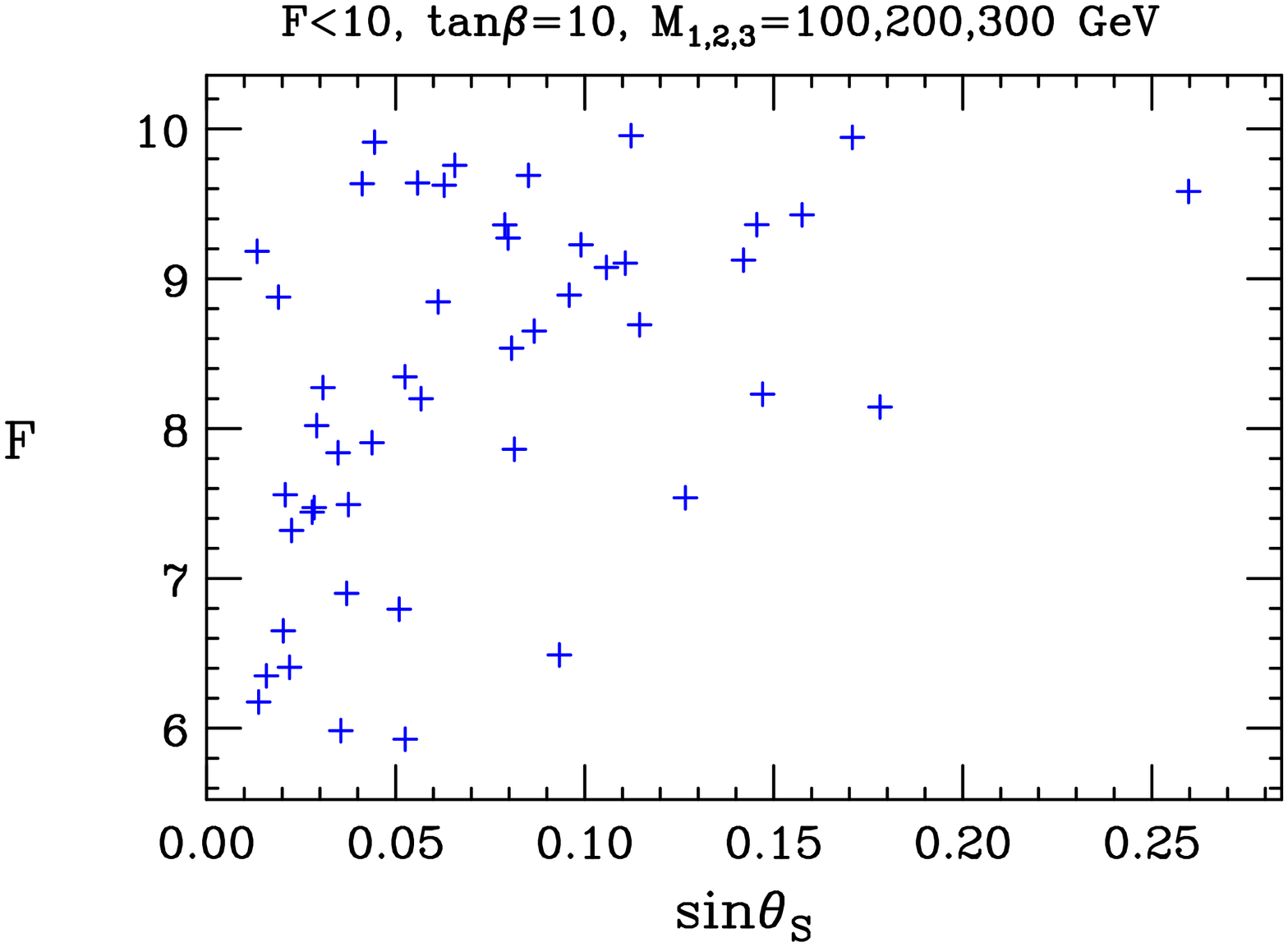}}
\caption{The plots show  $F$ vs. $\cta$ (upper)
  and $F$ vs. $\sin\theta_S$ (lower) for the large (yellow) cross
  points with $F<10$ taking $M_{1,2,3}(\mz)=100,200,300\gev$ and $\tanb=10$.}
\label{fvscta}
\end{figure}

Next, we wish to illustrate the relative MSSM vs. singlet composition
of the $\ai$ and $\hi$ for the $F<10$ points. This composition has
obvious implications for their couplings to SM particles.  The more
pure MSSM the $\hi$ is, the more SM-like will be its couplings.  The
more singlet the $\ai$ is, the more weakly it will be coupled to SM
particles. In particular, its couplings to SM down-type fermions and
leptons are given by $\tanb\cta$ times the SM-like weight in which
$\anti b 1 b$ is replaced by $\anti b i\gam_5 b$, for example.  These
compositions are shown in Fig.~\ref{fvscta}. The upper plot
illustrates that there is a lower bound on $|\cta|$ that arises from
the joint requirements of large $\br(\hi\to\ai\ai)$ and $\mai<2\mb$.
As noted earlier, this guarantees that the $b\anti b$ (and
$\tauptaum$) coupling strengths of the $\ai$ are sufficiently large
that the decays of the $\ai$ are dominated by the heaviest fermionic
states, \ie\ $\ai\to\tauptaum$ for $2\mtau<\mai<2\mb$ and $\ai\to
c\anti c$ for $2m_c<\mai<2\mtau$, with $\ai\to gg$ also being
important. For $\mai<2m_c$, $\ai\to gg$ is the dominant decay.  Note
the preference for $\cta\sim -0.1$ for the very lowest $F$ points.
The lower plot of Fig.~\ref{fvscta} shows the singlet component,
$\sin\theta_S$, of the $\hi$ for the fully ok solutions.  The $\hi$
can be as much as $20\%$ singlet at the amplitude level, but this
means it is still $96\%$ non-singlet in the amplitude-squared sense.
As a result, all plotted points have $|C_V|\sim 1$. The very lowest
$F$ points are clearly associated with very small $\sin\theta_S$.

We now turn to the GUT-scale parameters associated with $F<10$ large
yellow fancy cross points (plotted as blue $+$'s for these figures) that pass
all experimental constraints and the correlations among them.

\begin{figure}[h!]
\centerline{\includegraphics[width=2.4in,angle=90]{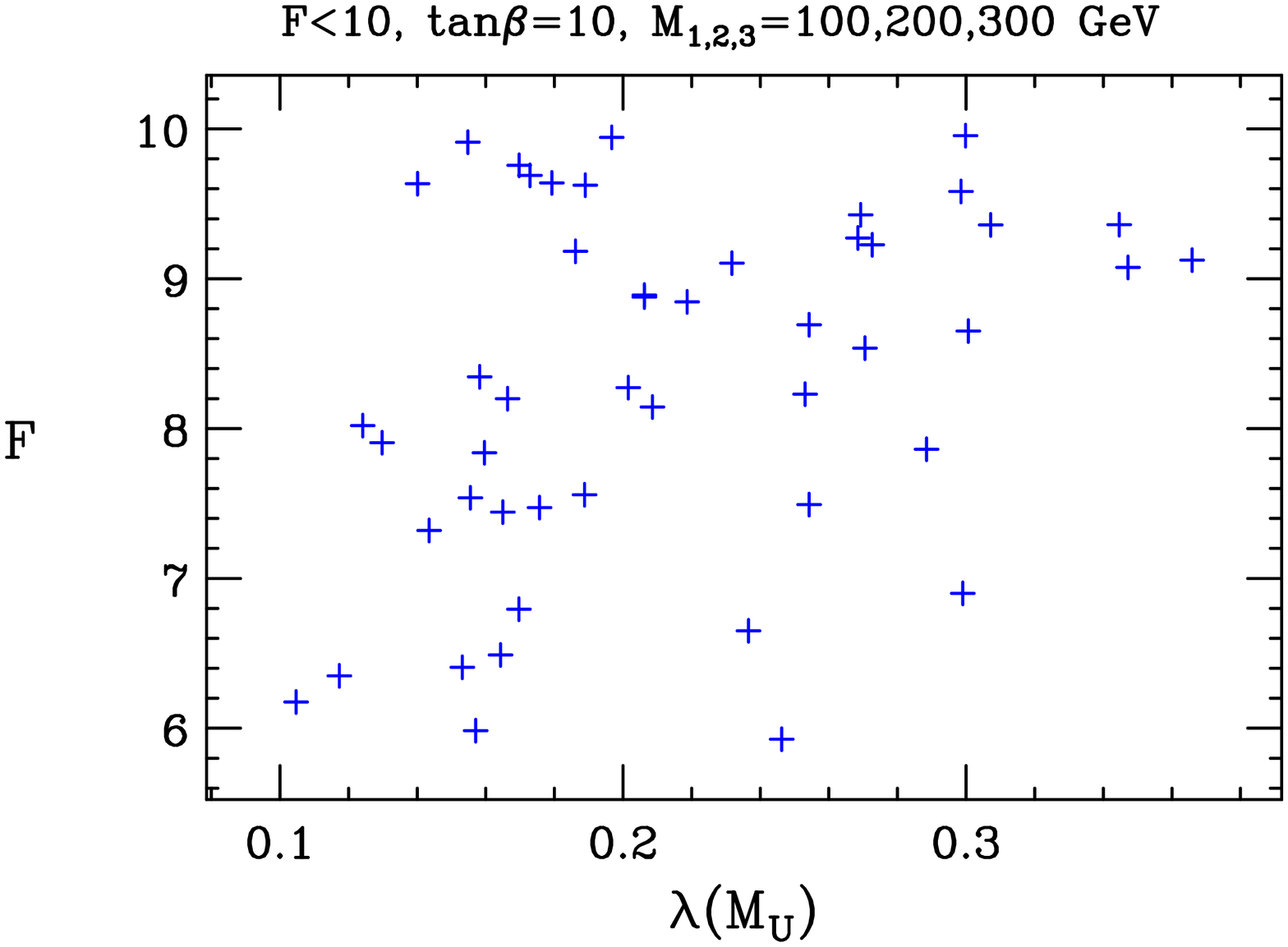}}
\vspace*{.15in}
\centerline{\includegraphics[width=2.4in,angle=90]{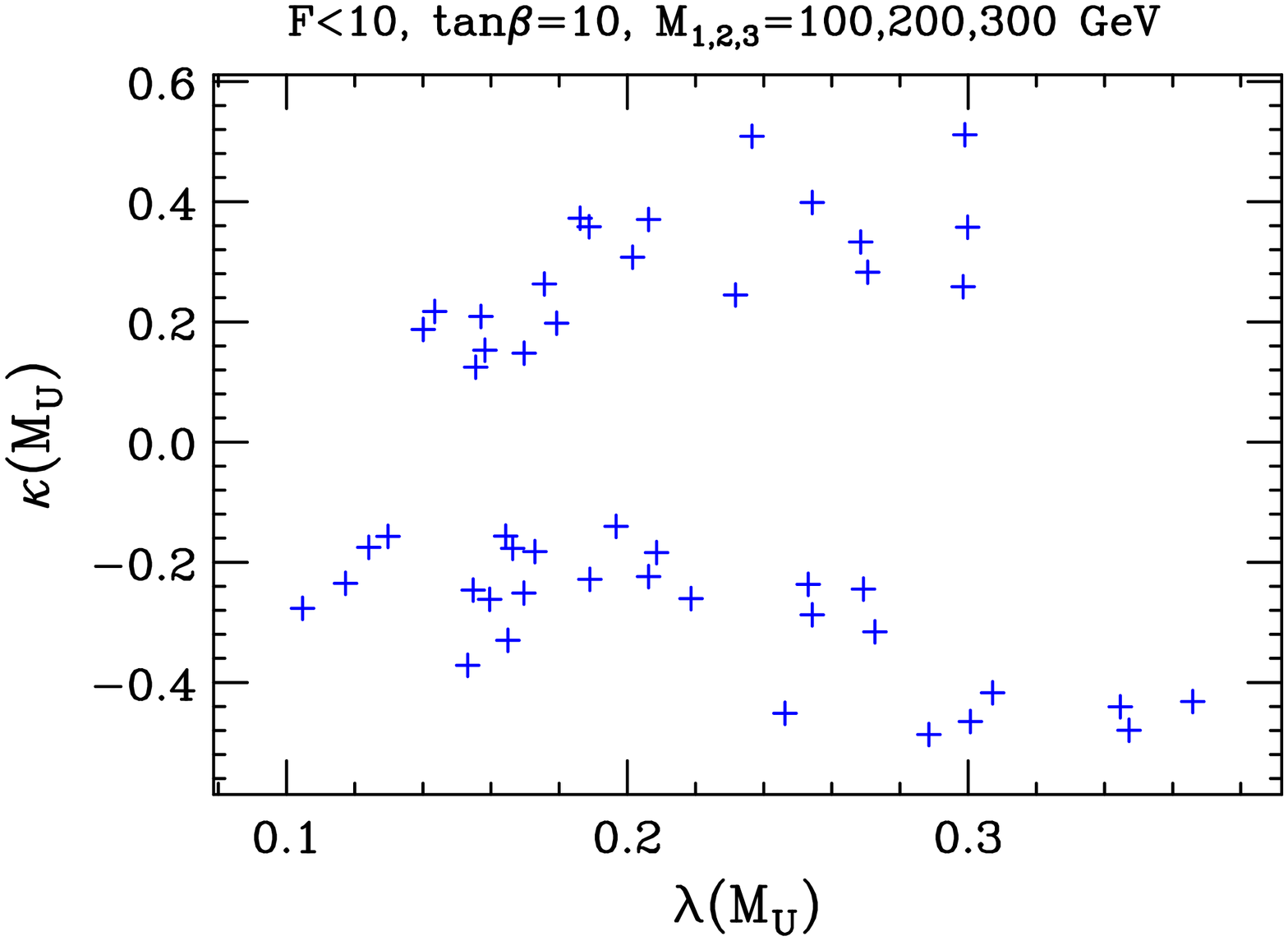}}
\caption{The upper plot shows fine tuning vs. $\lam(\mgut)$ for large
  (yellow) cross   points  
 with $F<10$  taking $M_{1,2,3}(\mz)=100,200,300\gev$
and $\tanb=10$. 
The lower plot shows $\kap(\mgut)$ as a function of $\lam(\mgut)$.}
\label{fvslgut}
\end{figure}
First, we consider $\kap(\mgut)$ and $\lam(\mgut)$ in Fig.~\ref{fvslgut}.
We see that the very lowest $F$ values have fairly small $\lam(\mgut)$
and significantly larger $\kap(\mgut)\sim -0.4$.

\begin{figure}[h!]
\centerline{\includegraphics[width=2.4in,angle=90]{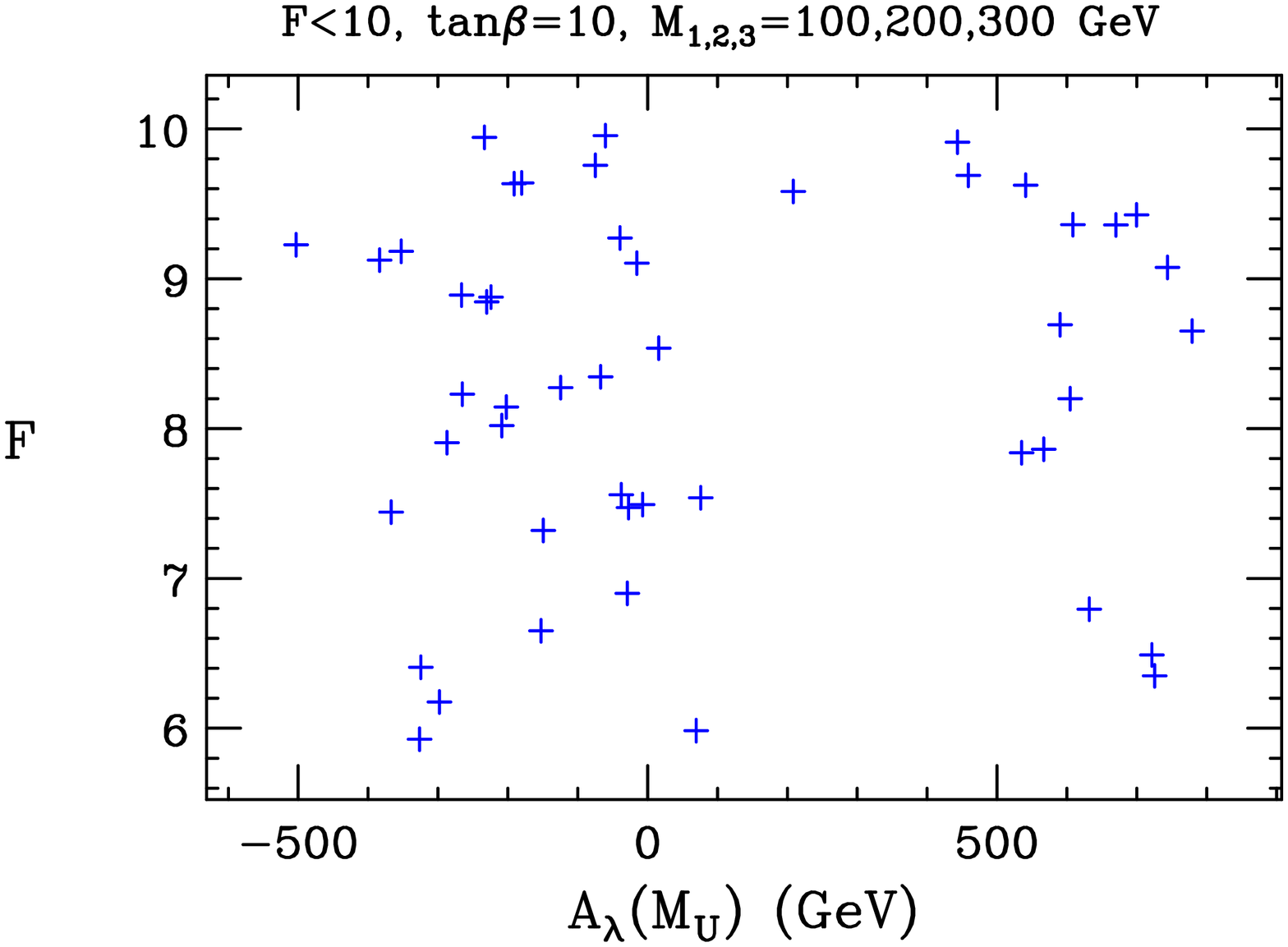}}
\vspace*{.15in}
\centerline{\includegraphics[width=2.4in,angle=90]{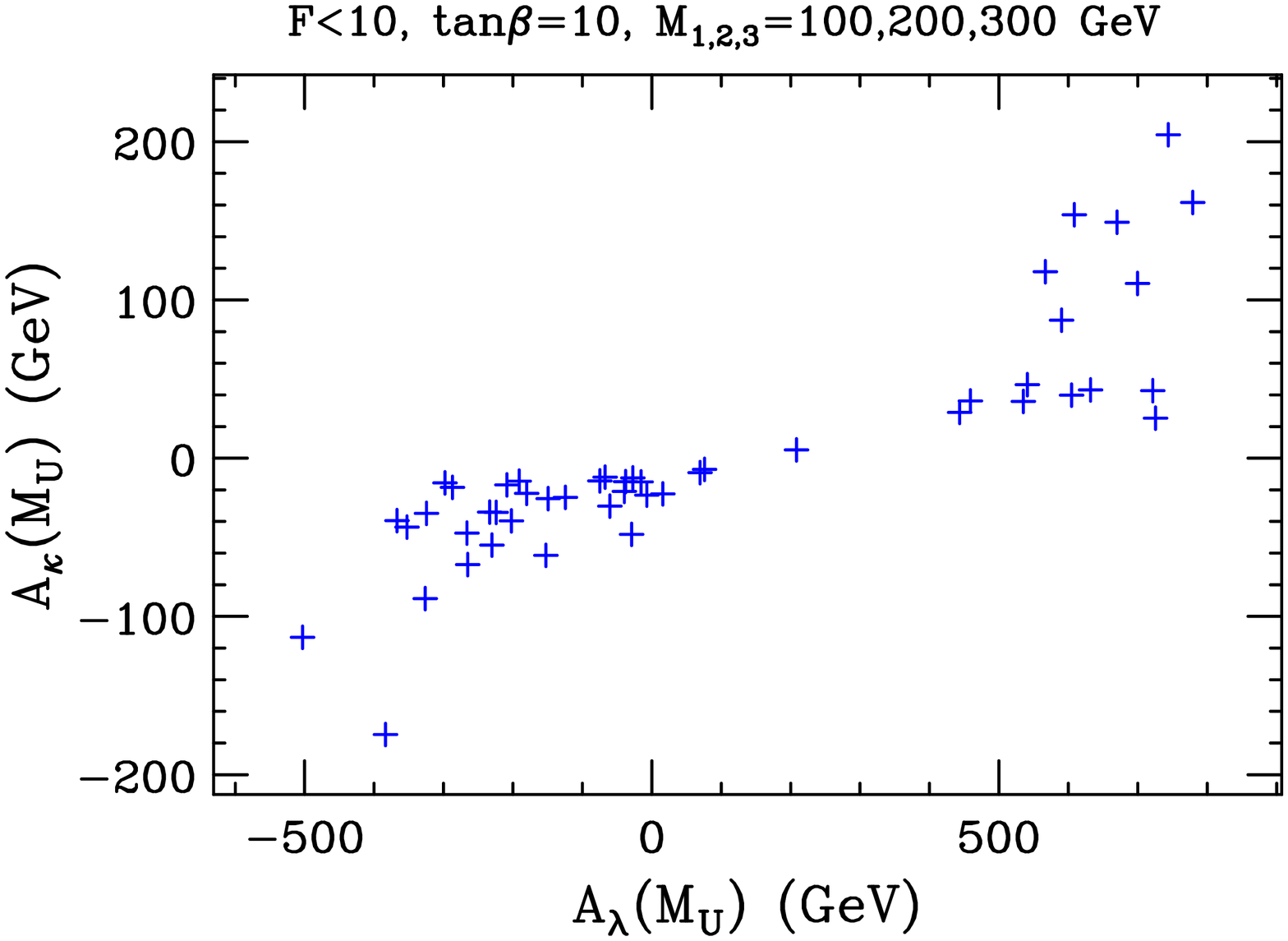}}
\caption{The upper plot shows fine tuning vs. $\alam(\mgut)$ for 
fully ok points with $F<10$  taking $M_{1,2,3}(\mz)=100,200,300\gev$
and $\tanb=10$. 
The lower plot shows $\akap(\mgut)$ as a function of $\alam(\mgut)$.}
\label{fvsalgut}
\end{figure}
We consider $\akap(\mgut)$ and $\alam(\mgut)$ in Fig.~\ref{fvslgut}.
We see that the very lowest $F$ values have fairly small $\alam(\mgut)$
and  $\akap(\mgut)$, \ie, as noted earlier, both are close to being consistent
with no-scale boundary conditions at $\mgut$.

\begin{figure}[h!]
\centerline{\includegraphics[width=2.4in,angle=90]{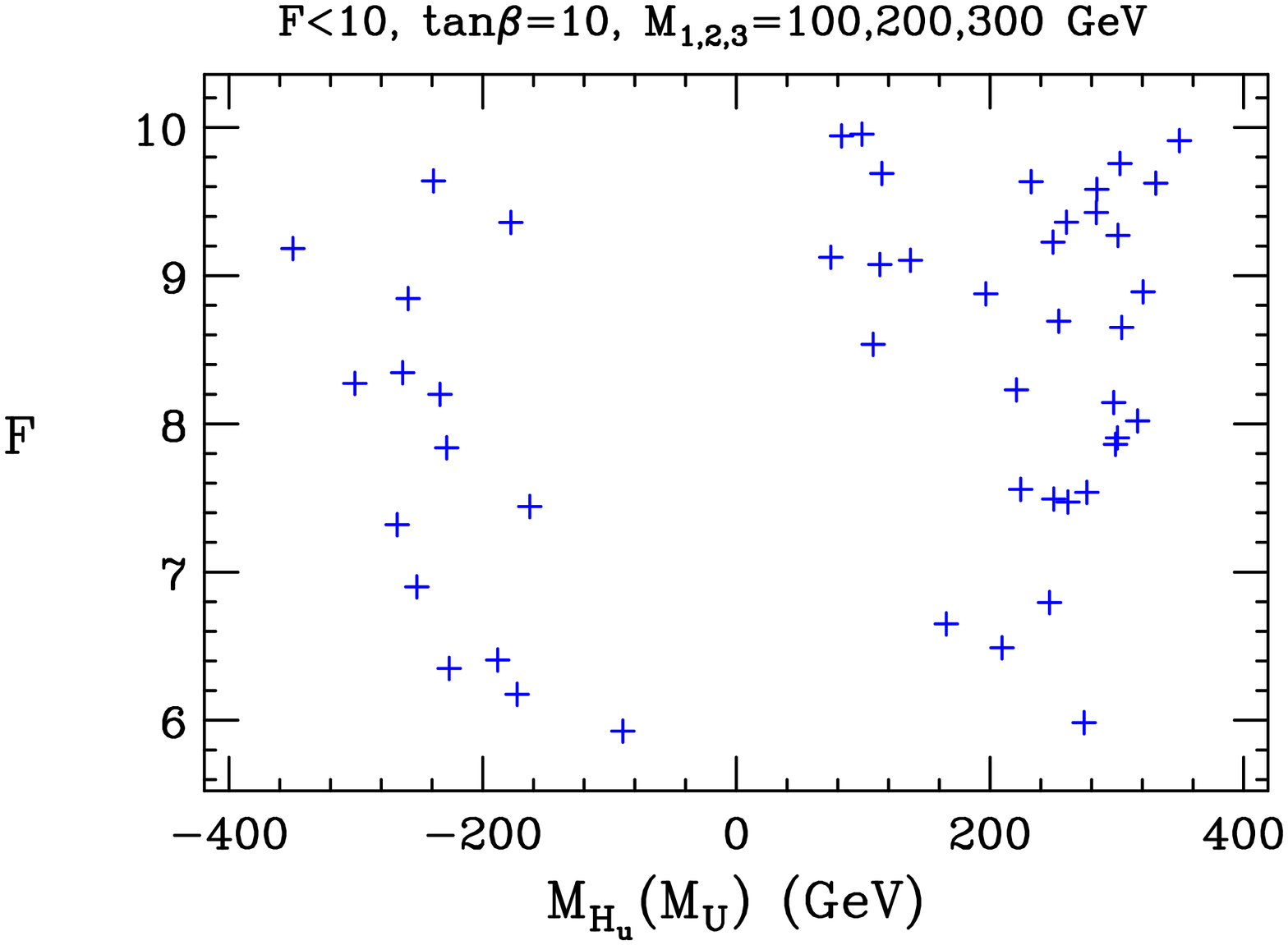}}
\vspace*{.15in}
\centerline{\includegraphics[width=2.4in,angle=90]{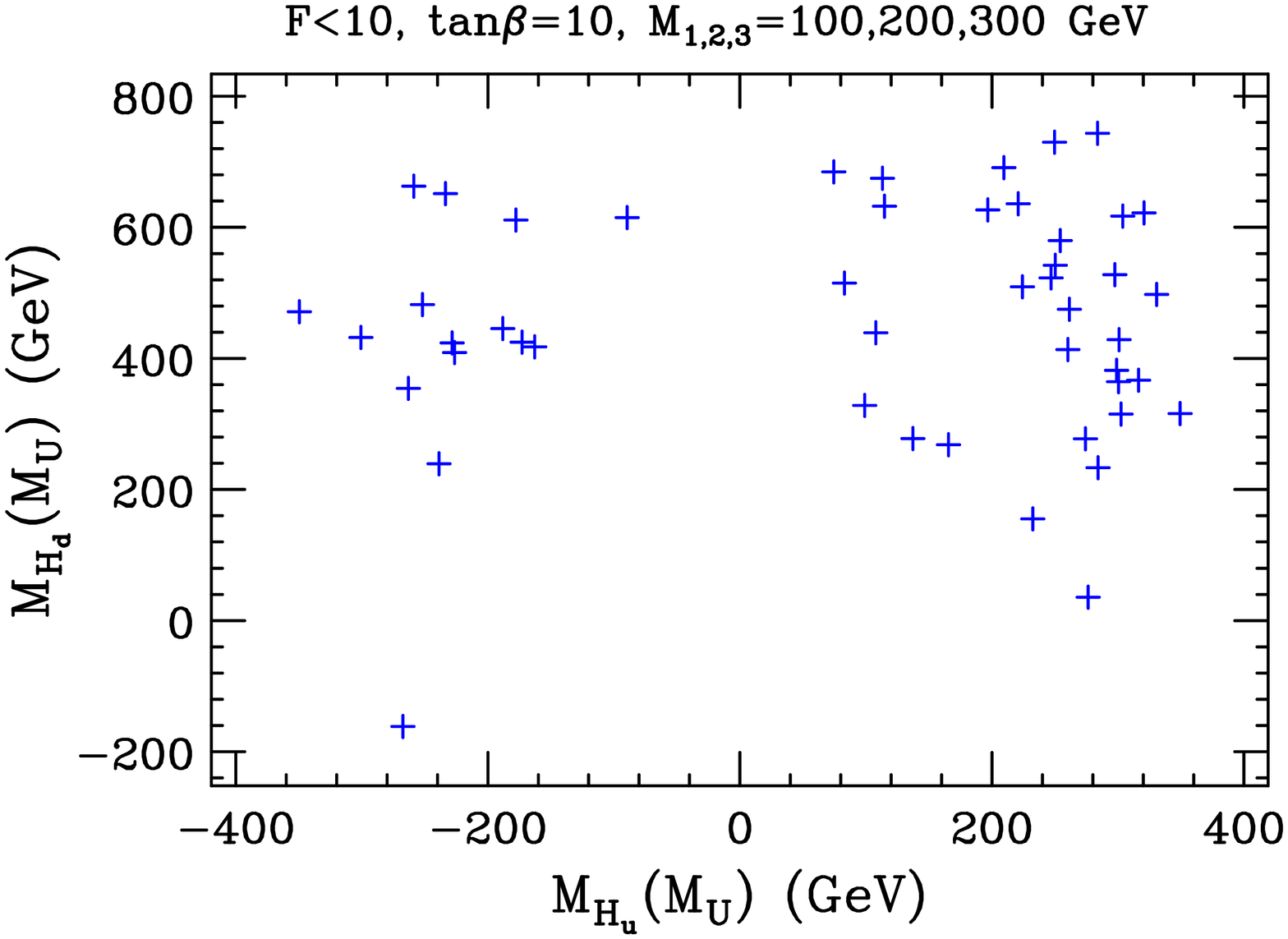}}
\vspace*{.15in}
\centerline{\includegraphics[width=2.4in,angle=90]{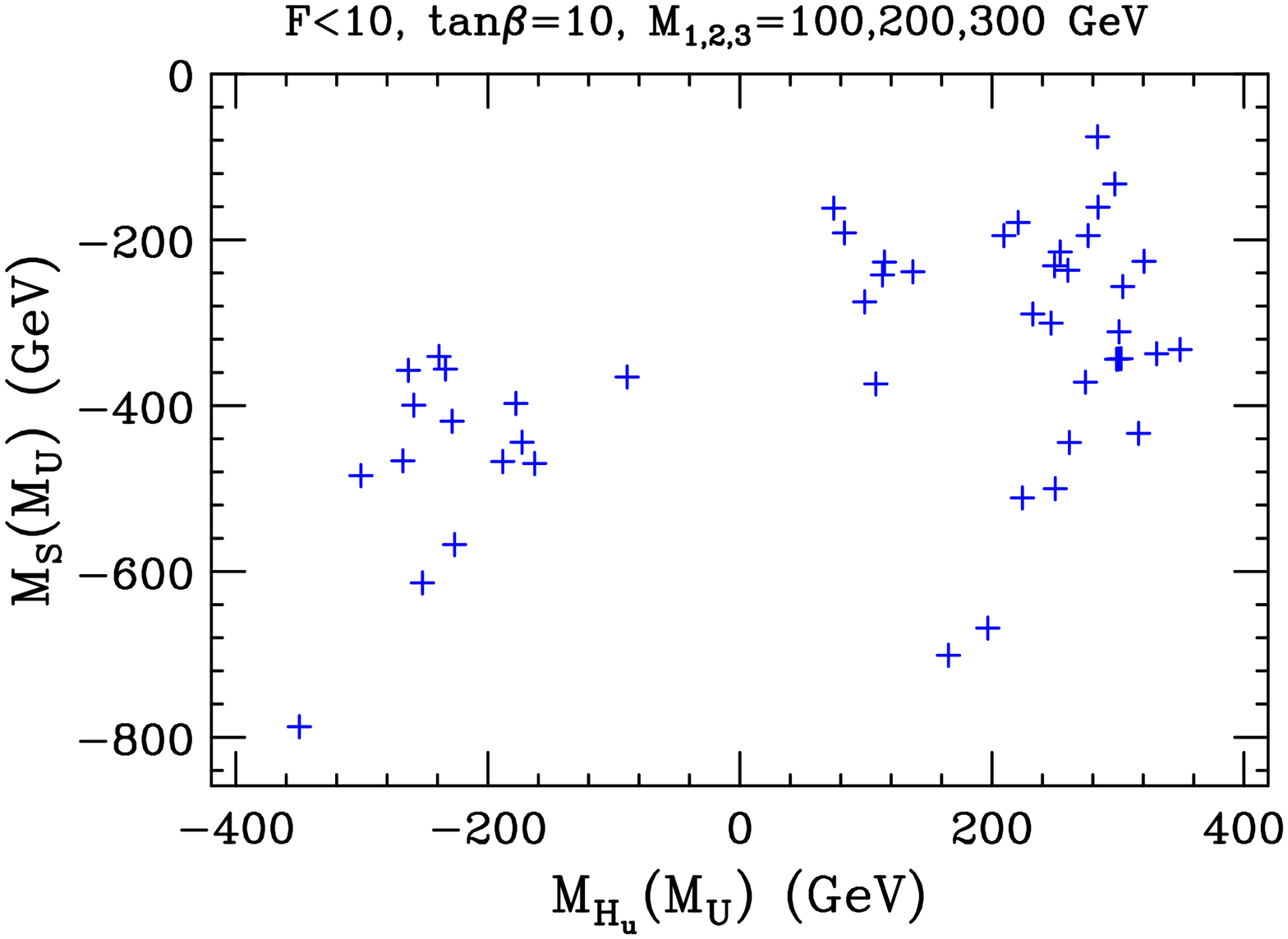}}
\caption{The upper plot shows fine tuning vs. $\mhu(\mgut)$ for 
fully ok points with $F<10$  taking $M_{1,2,3}(\mz)=100,200,300\gev$
and $\tanb=10$. 
The middle plot shows $\mhd(\mgut)$ as a function of $\mhu(\mgut)$.
The bottom plot shows $\ms(\mgut)$ as a function of $\mhu(\mgut)$.
Our convention is that if an $M^2$ is negative then we plot $-\sqrt{-M^2}$.}
\label{fvsmsqgut}
\end{figure}
We consider $\mhu(\mgut)$, $\mhd(\mgut)$ and $\ms(\mgut)$ in Fig.~\ref{fvsmsqgut}.
We see that the very lowest $F$ values have fairly small GUT-scale
values for all the scalar Higgs mass squared values, 
again close to being consistent
with no-scale boundary conditions at $\mgut$.
However, our scans did not locate any fully ok points for which
$\mhu(\mgut)$, $\mhd(\mgut)$ and $\ms(\mgut)$ were all simultaneously
small. We are unsure at this time as to whether this is an artifact of
limited computer time for scanning or something deeper.

\section{Moderately low $\boldmath F$ points with dominant $\boldmath \ai\to \gam\gam$ decays}

\begin{figure}[b!]
\vspace*{.1in}
  \centerline{\includegraphics[width=2.2in,angle=90]{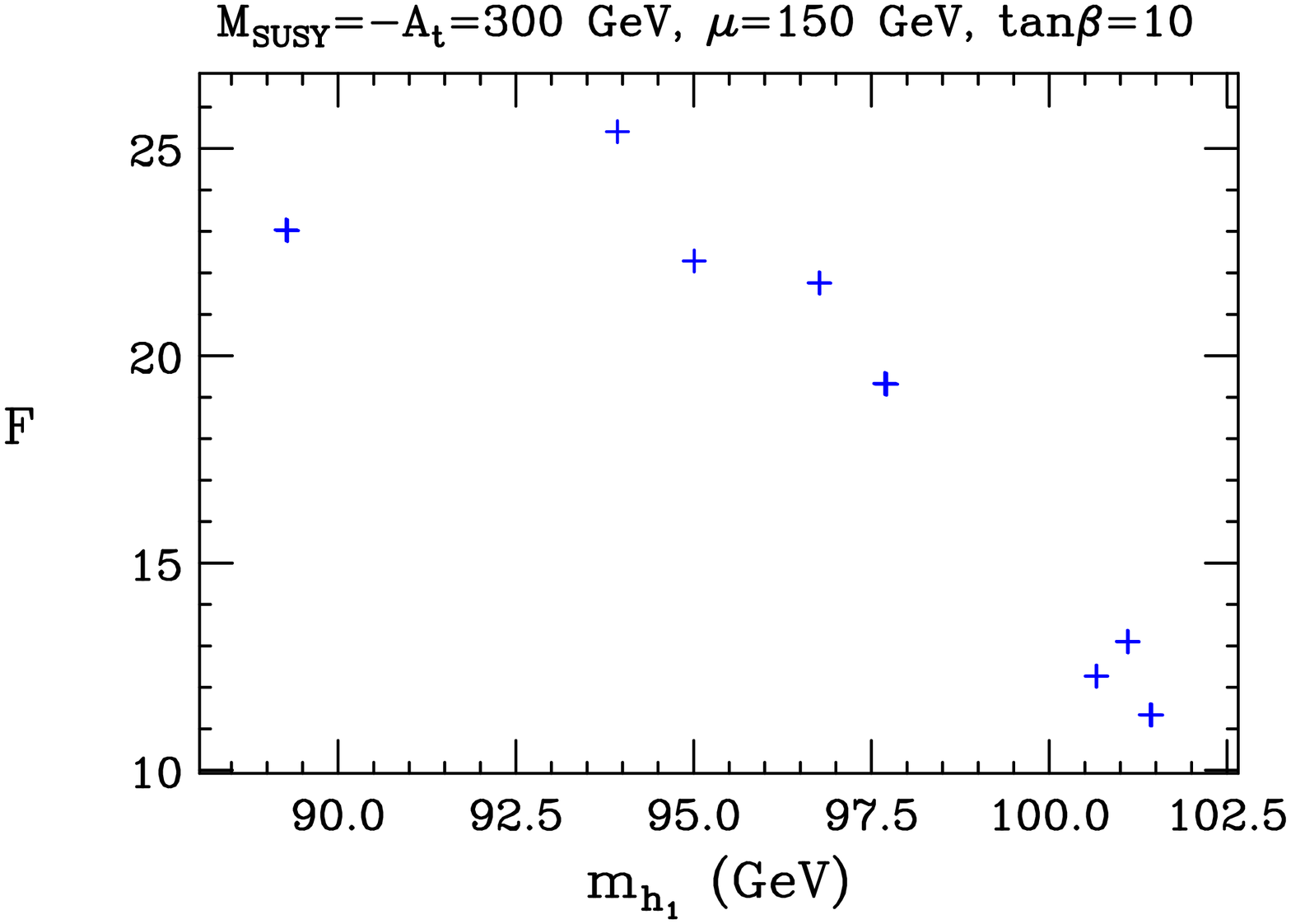}}
  \centerline{\includegraphics[width=2.2in,angle=90]{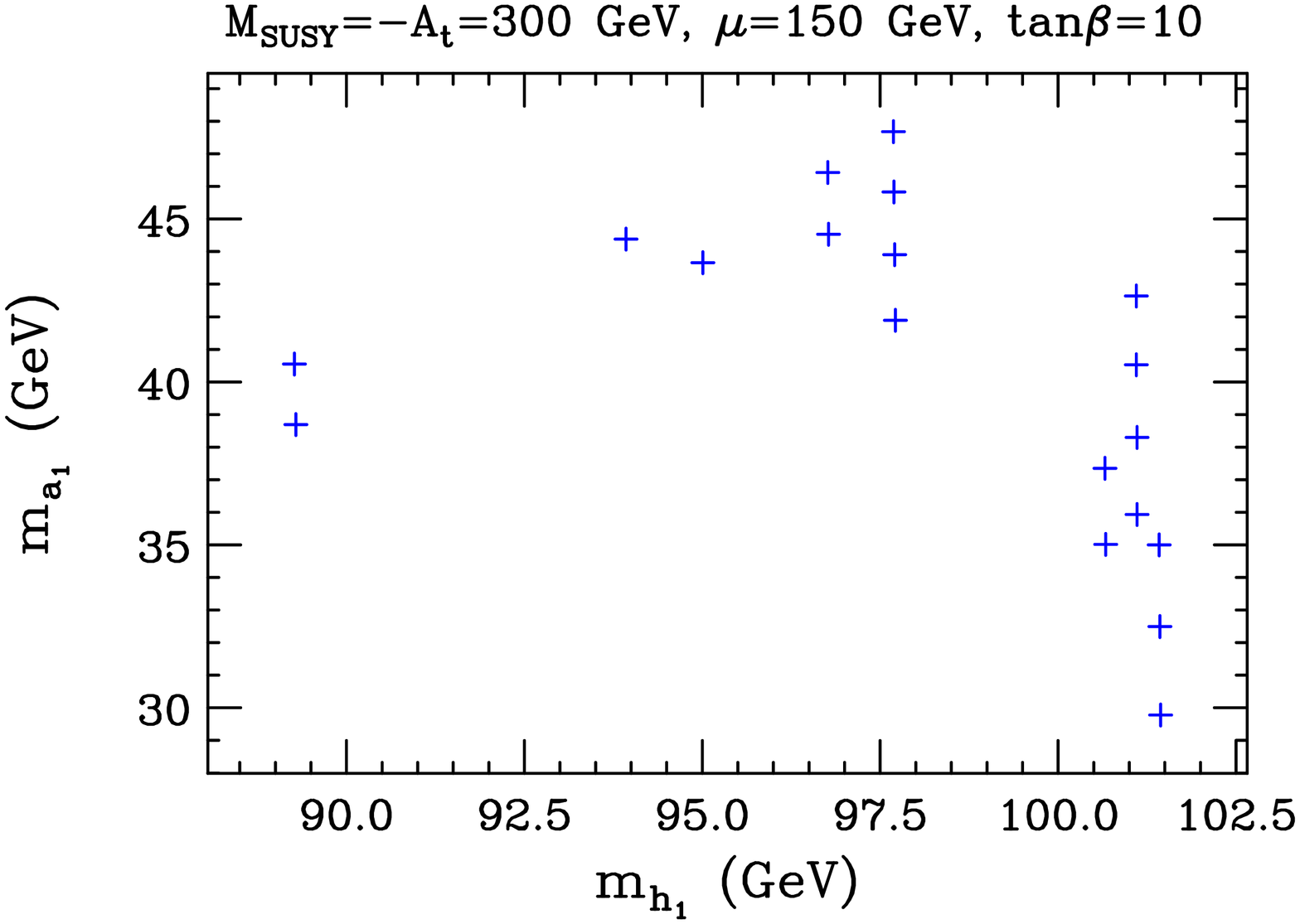}}
\caption{For the points with dominant $\ai\to\gam\gam$ decays and low
  $F$,
we plot: $F$ vs. $\mhi$ (top); and  $\mai$ vs. $\mhi$ (bottom). Note that
there are a number of degeneracies where exactly the same $F$ and
$\mhi$ are predicted for somewhat different parameter choices in the scan.}
\label{4gamplots1}
\end{figure}

\begin{figure}[h!]
  \centerline{\includegraphics[width=2.2in,angle=90]{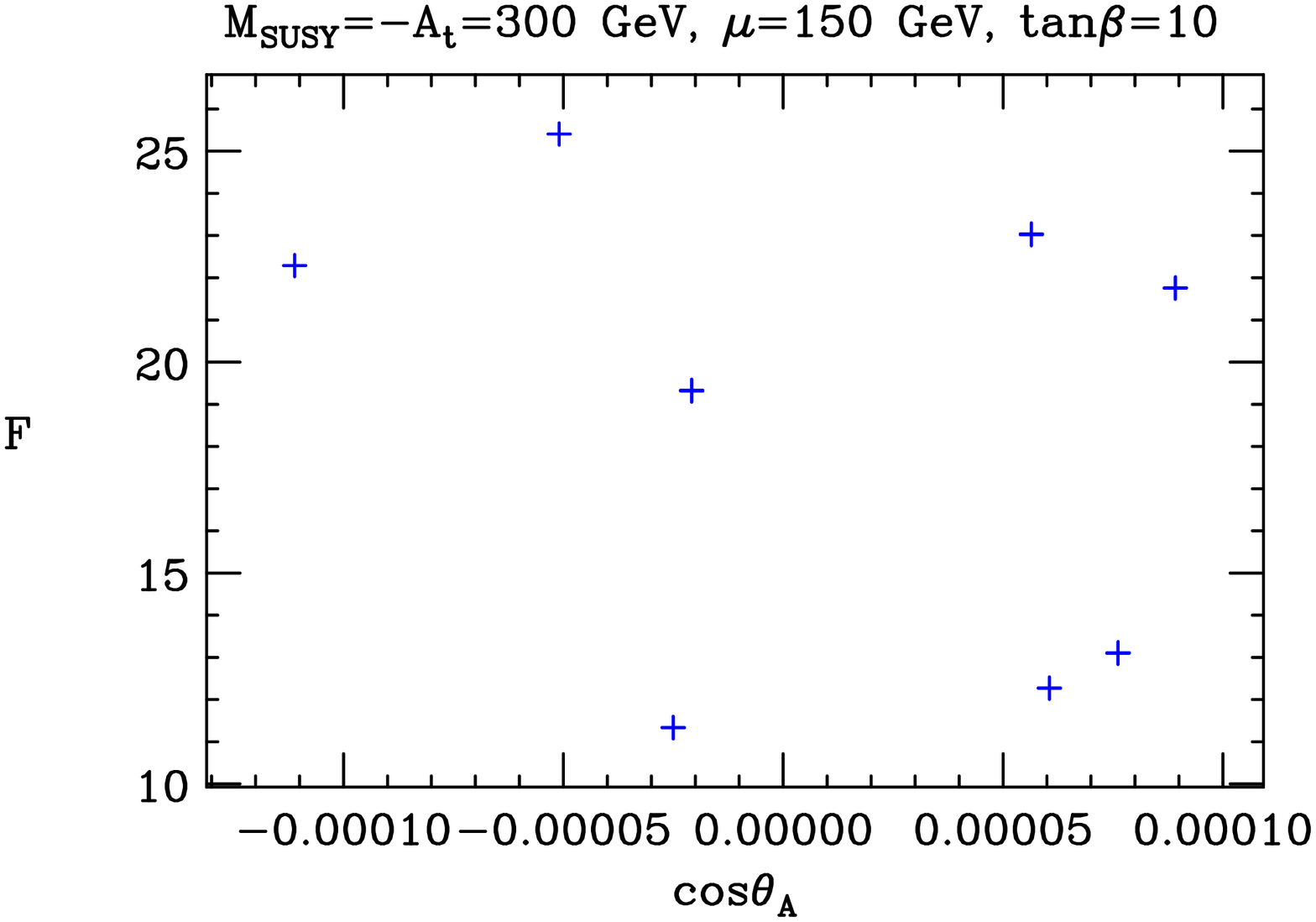}}
  \centerline{\includegraphics[width=2.2in,angle=90]{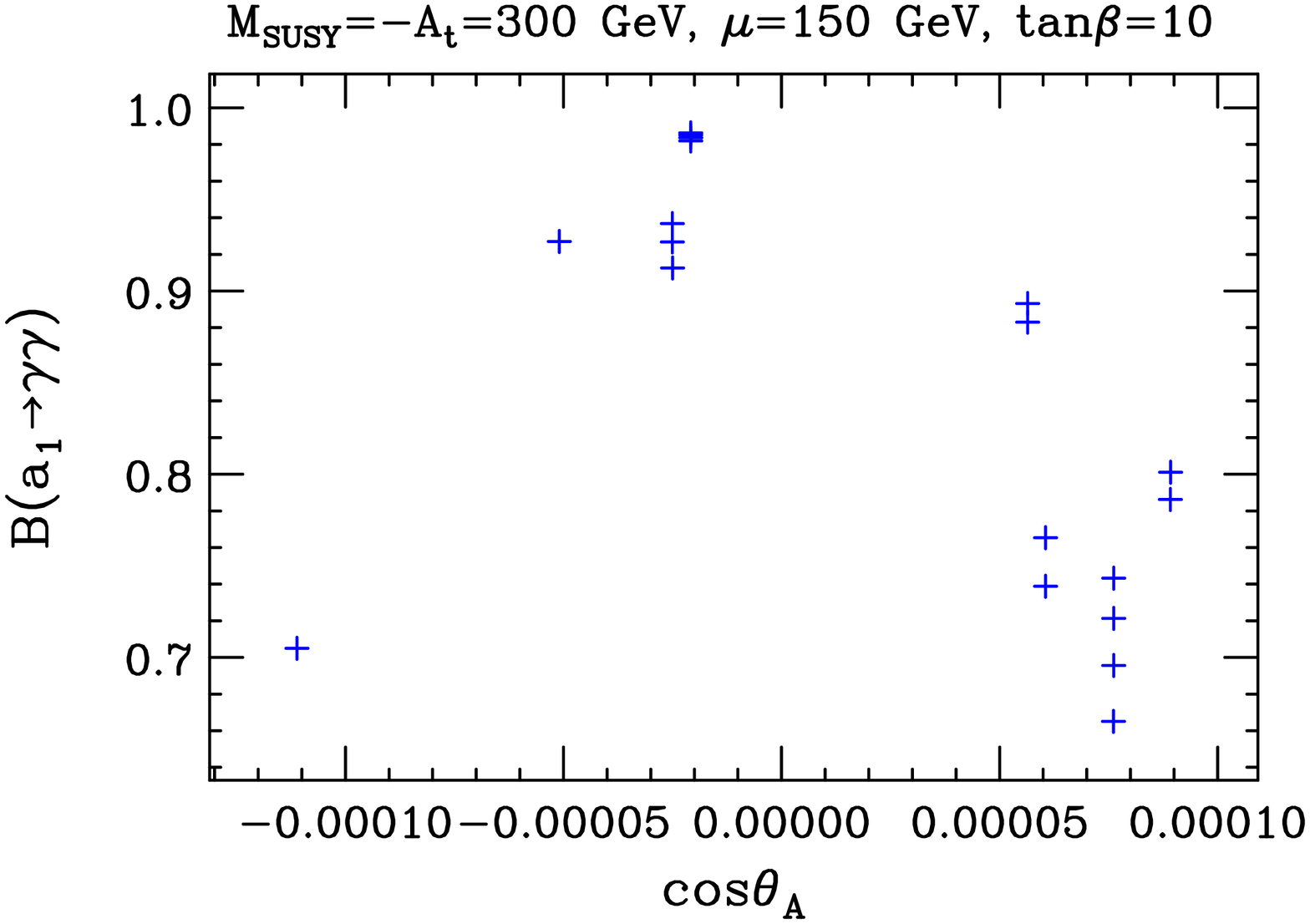}}
  \centerline{\includegraphics[width=2.2in,angle=90]{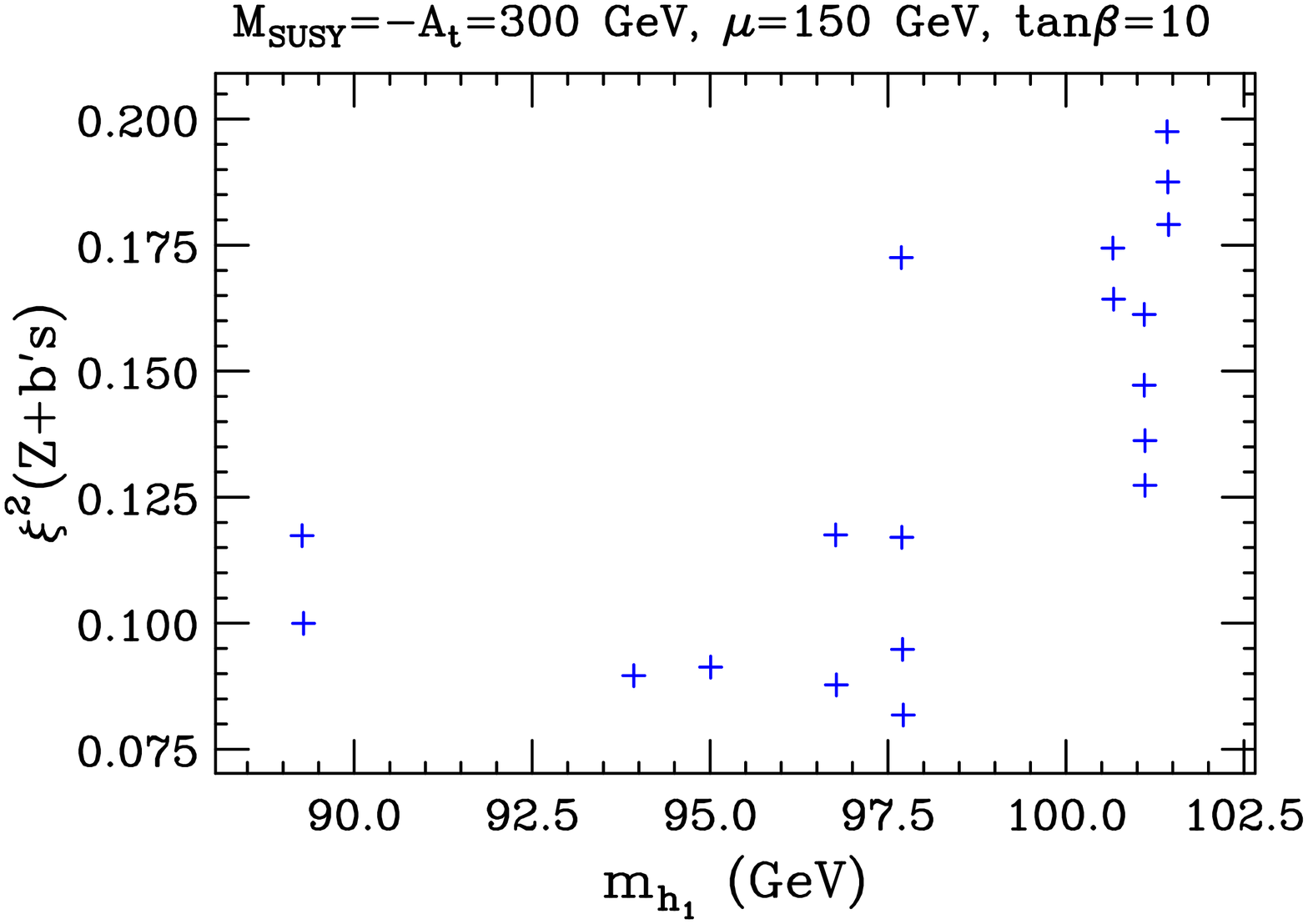}}
\caption{For the points with dominant $\ai\to\gam\gam$ decays and low
  $F$,
we plot: $F$ vs. $\cta$ (top); $\br(\ai\to\gam\gam)$ vs. $\cta$ (middle); 
and $\xi^2(Z+b's)$ vs. $\mhi$ (bottom). Since 
$F$ depends primarily on $\cta$,  there are a number of points in
the $F$ vs. $\cta$ plot that are actually multiple repetitions of exactly
the same $F$ at a given $\cta$ value but with $\br(\ai\to\gam\gam)$
and $\xi^2(Z+b's)$ varying slightly
because of sensitivity to other parameters of the scan.}
\label{4gam_braxisq}
\end{figure}

\begin{figure}[h!]
  \centerline{\includegraphics[width=2.2in,angle=90]{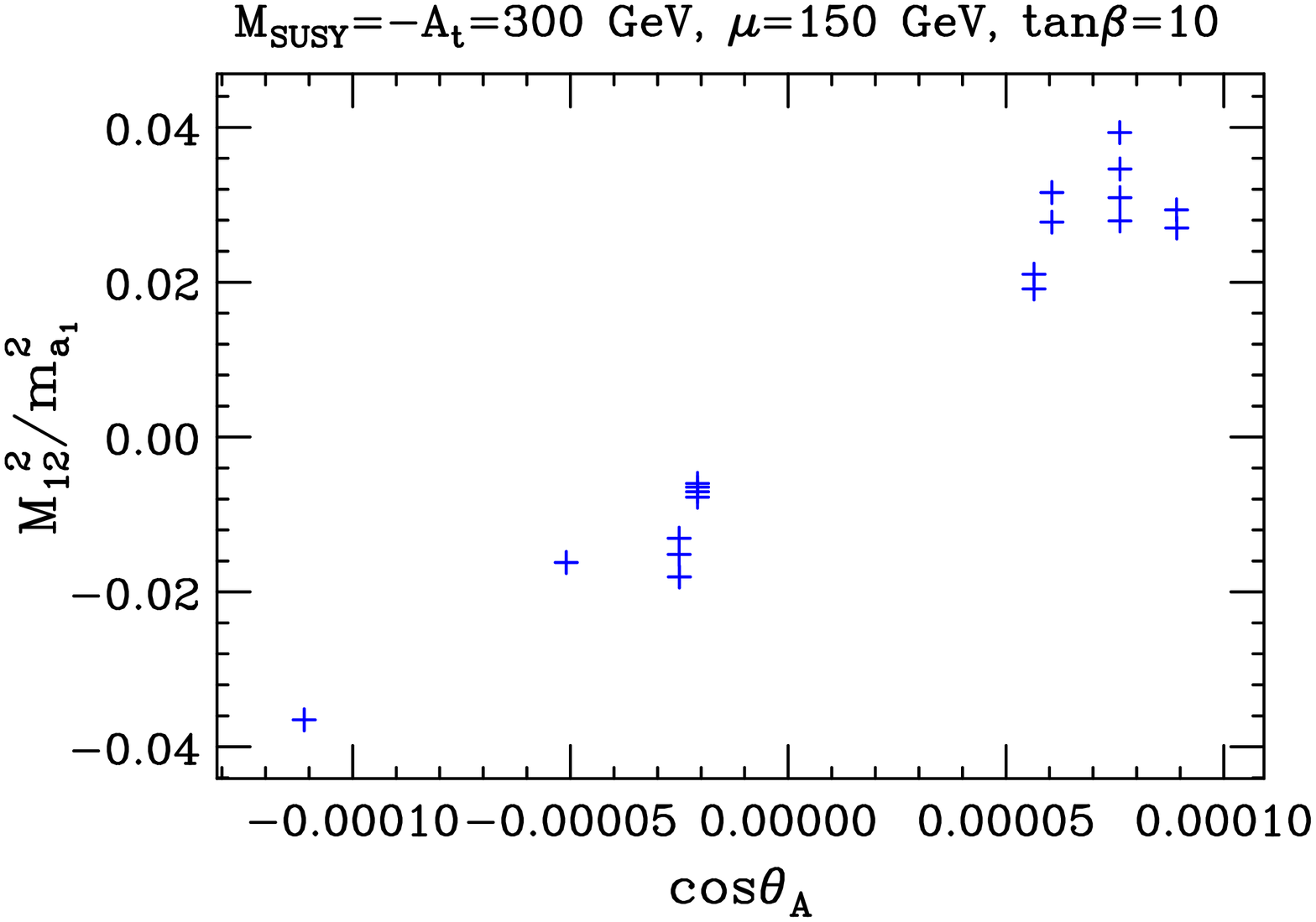}}
  \centerline{\includegraphics[width=2.2in,angle=90]{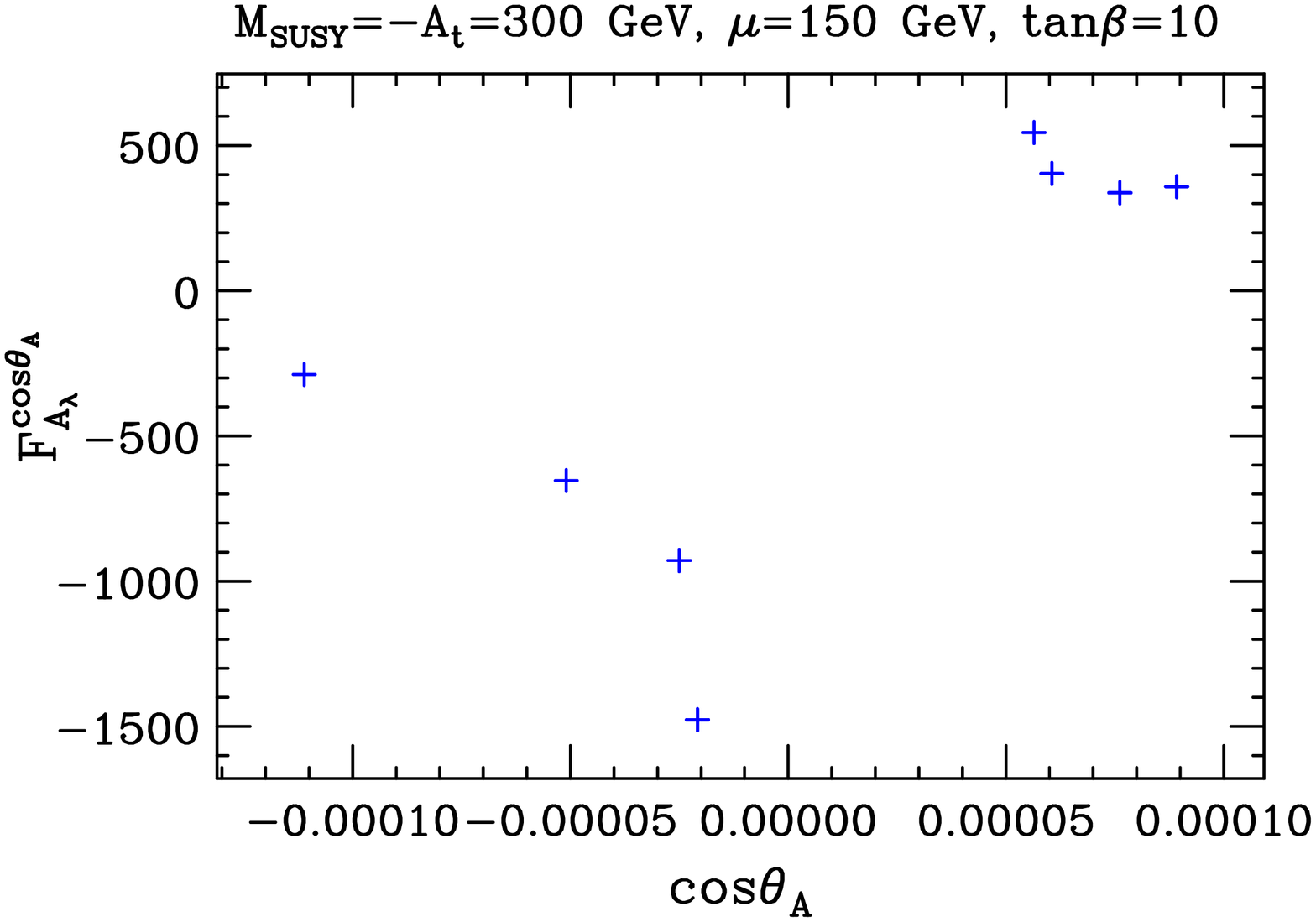}}
\caption{For the points with dominant $\ai\to\gam\gam$ decays and low
  $F$,
we plot:  $M_{12}^2/\mai^2$
vs. $\cta$ (top); and $\fcal$ as a function of
$\cta$ (bottom). (Many of the points have the same $\fcal$ value.)}
\label{4gam_fcalvscta}
\end{figure}

Let us now turn to the special class of points mentioned in
Section~\ref{sec:compare}. These are the low-$F$ points with a SM-like
$\hi$ of mass $\sim 100\gev$ and for which $\mai\gsim 30\gev$. For
these points, $\br(\hi\to\ai\ai)>0.75$ and $\br(\ai\to\gam\gam)\gsim
0.9$. For this to occur, the $\ai$ must be highly singlet in nature so
that the tree-level decays to fermion-antifermion are highly
suppressed, in which case the chargino loop-induced decay to
$\gam\gam$ can be dominant. (The relevant couplings are present even
when the $\ai$ is purely singlet.)  This combination of features
allows consistency with the LEP limits on the $Z+b's$ channel.  We
have not been able to determine if there are relevant limits on the
$Z+4\gam$ channel. This channel would have quite a high rate and most
probably these relatively spectacular events would have been noticed.
These points are also disfavored theoretically since, as detailed
shortly, a very high level of fine tuning of GUT-scale parameters is
required in order to achieve $\cta\lsim 10^{-4}$ as required (for
$\tanb=10$) for $\ai\to 2\gam$ to be the dominant $\ai$ decay channel.
Nonetheless, they should not be entirely discarded as a possible class
and so we give some details regarding them in the following.  These
points were found using an extremely fine grid scanning approach of
the type detailed in \cite{Dermisek:2006wr} with fixed
$\msusy=-A_t=300\gev$, $\mu=150\gev$ and $\tanb=10$.

Basic plots for this scenario appear in Fig.~\ref{4gamplots1}.  The
top plot of this figure gives $F$ as a function of $\mhi$.  We see
again that $\mhi\sim 100\gev$ gives the lowest $F$ value, $F\sim 11$
in this case.  The lower plot shows that $\mai\gsim 30\tev$ is
required for this kind of scenario, with the lowest $F$ obtained for
$\mai\sim 30\gev$.  Fig.~\ref{4gam_braxisq} shows $F$ vs. $\cta$ as
the top plot, $\br(\ai\to\gam\gam)$
as a function of $\cta$ as the middle plot and $\xi^2(Z+b's)$ vs. $\mhi$
as the bottom plot. The top plot is useful for correlating $F$ with
the value of $\cta$. However, note that there is some degeneracy:
essentially the
same values of $F$ and $\cta$ are sometimes obtained even though the basic scan
parameters are different.  The middle plot shows that $\br(\ai\to \gam\gam)\gsim 0.65$ for
these points, with $\br(\ai\to\gam\gam)\sim 0.65$ for the $F\sim 11$
points.  As expected, very small $\cta$ is required in order for the
$\ai\to\gam\gam$ decays to be dominant.  The bottom plot of the figure
shows $\xi^2(Z+b's)$ as a function of $\mhi$. We observe that
$\xi^2(Z+b's)$ is of the right general magnitude to explain the LEP
excess for the lowest $F$ points that have $\mhi\sim 101\gev$. We note
that $\xi^2(Z+b's)$ receives contributions from both the $Z+2b$ final
state from direct $\hi\to b\anti b$ decay and also the $Z+4b$ from
$\hi\to\ai\ai\to 4b$ where $\br(\ai\to b\anti b)<0.35$ due to the
competition from the $\ai\to\gam\gam$ decays.

The careful reader may wonder why it is that we can
have small $\cta$ for these points whereas the $\mai<2\mb$ points have
a lower bound on $\cta$. In fact, it is precisely the {\it
  combination} of the requirements that $\mai<2\mb$ and
$\br(\hi\to\ai\ai)\gsim 0.75$ which forces a lower bound on
$\cta$. Values of $\cta$ small enough to yield large $\br(\ai\to\gam\gam)$
while at the same time $\br(\hi\to\ai\ai)\gsim 0.7$ is maintained are only possible
for relatively large $\mai$.

The fine-tuning required in $\alam$ and $\akap$ to achieve very
small $\cta$ can be quantified via the derivatives
\beq
\fcal\equiv {\partial \cta\over \partial \alam}{\alam\over
  \cta}\,,\quad \fcak\equiv {\partial \cta\over \partial \akap}{\akap\over
  \cta}\,,
\eeq
where all parameters are defined at scale $\mz$. 
Understanding of these quantities can be gleaned from the approximate formula 
\beq
\cos\theta_A\simeq -{M_{12}^2\over M_{11}^2-M_{22}^2}\simeq  -
\frac{\lambda v (A_\lambda - 2 \kappa s) \sin 2 \beta}{2 \lambda s
(A_\lambda + \kappa s) + 3 \kappa A_\kappa s \sin 2 \beta}\,,
\label{ctaform}
\eeq
where we used~\footnote{These mass squared matrix entries receive
  radiative corrections not shown here.}
\begin{eqnarray}
M_{11}^2 &=& \frac{ 2 \lambda s}{\sin 2 \beta} \left(A_\lambda + \kappa s \right), \label{eq:M11sq}\\
M_{12}^2 &=& \lambda v \left( A_\lambda - 2 \kappa s \right), \label{eq:M12sq} \\
M_{22}^2 &=& 2 \lambda \kappa v^2 \sin 2 \beta + \lambda A_\lambda
\frac{v^2 \sin 2 \beta}{2s} - 3 \kappa A_\kappa s.
\label{eq:M22sq}
\end{eqnarray}
Eq.~(\ref{ctaform}) shows that there will be great sensitivity of
$\cta$ to the value of $\alam$ relative to $2\kap s$, and almost no
sensitivity to $\akap$.  Both are confirmed by the numerical results
we now present for fixed $M_{1,2,3}=100,200,300\gev$, $\tanb=10$,
$\mueff=150\gev$, $A_t=-300\gev$, $A_b=A_\tau=0$,
$M_{Q,U,D,L,E}=300\gev$ (for the relevant 3rd generation).  Different
points are obtained by scanning in $\lam,\kap,\alam,\akap$.
(Obviously, many more $4\gam$ points could be found if the
fixed parameters are allowed to vary.  However, large
$A_t<0$ is essential to get small $F$ for such points.)

In Fig.~\ref{4gam_fcalvscta}, we plot $M_{12}^2/\mai^2$ vs. $\cta$ and $\fcal$ vs. $\cta$.  The top
plot shows that $M_{12}^2/\mai^2$
must be small for small $\cta$ (and there is a strong linear
relation).  The bottom plot shows that such small values of $\cta$
imply rather large values of $\fcal$.  Given Eq.~(\ref{alrun}), high
sensitivity to the $\mz$-scale value of $\alam$ implies a high level
of fine-tuning for $\alam$ (at scale $\mz$) with respect to $\alam(\mgut)$, $A_t(\mgut)$ and
$M_3(\mgut)$. One should also note that 
the $\fcal$ tuning measure for $\alam$ is largest ($\sim -1500$) for
the point for which the EWSB fine tuning measure
$F$ is smallest and vice-versa.

\begin{figure}[h!]
  \centerline{\includegraphics[width=2.2in,angle=90]{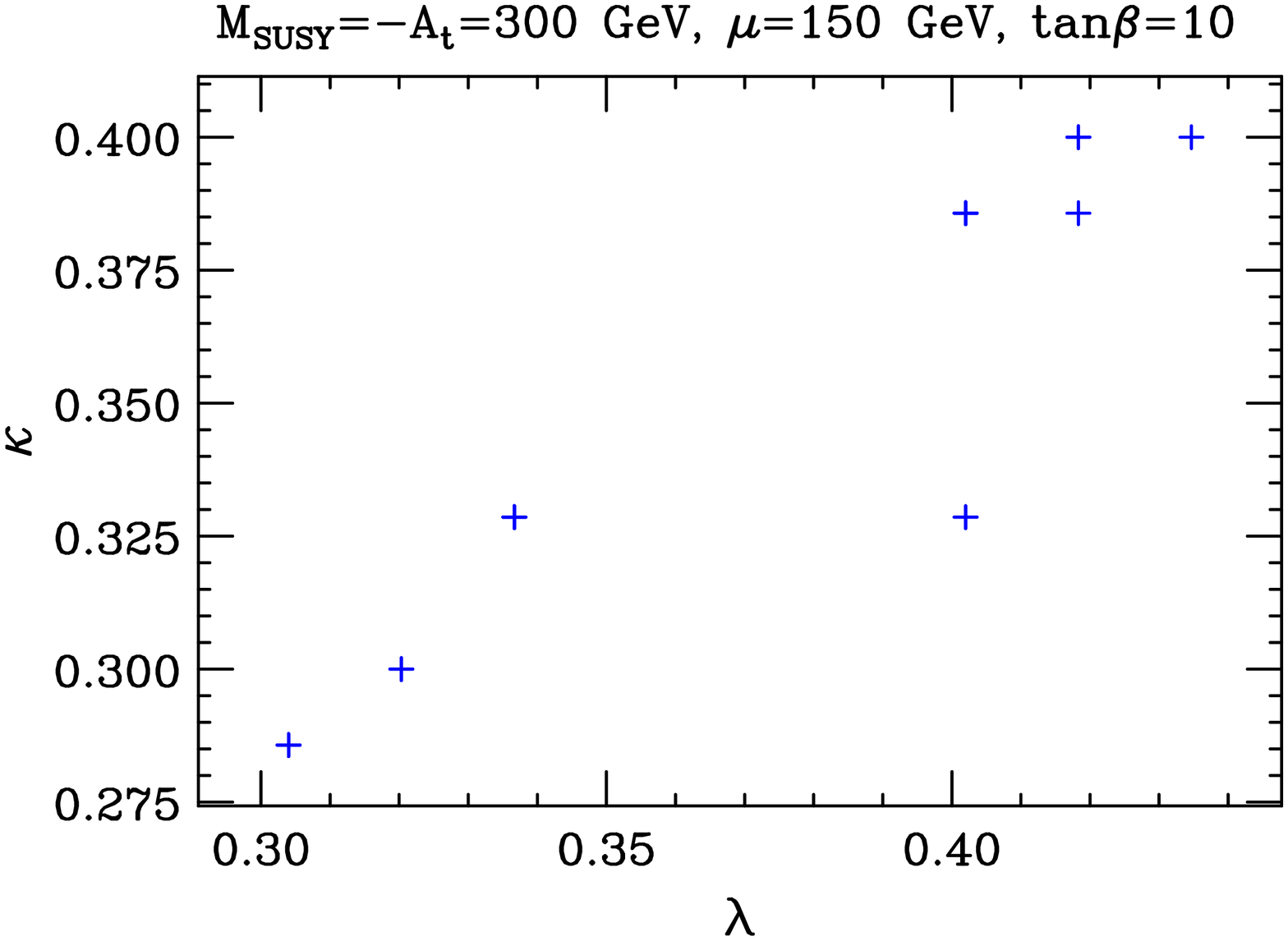}}
  \centerline{\includegraphics[width=2.2in,angle=90]{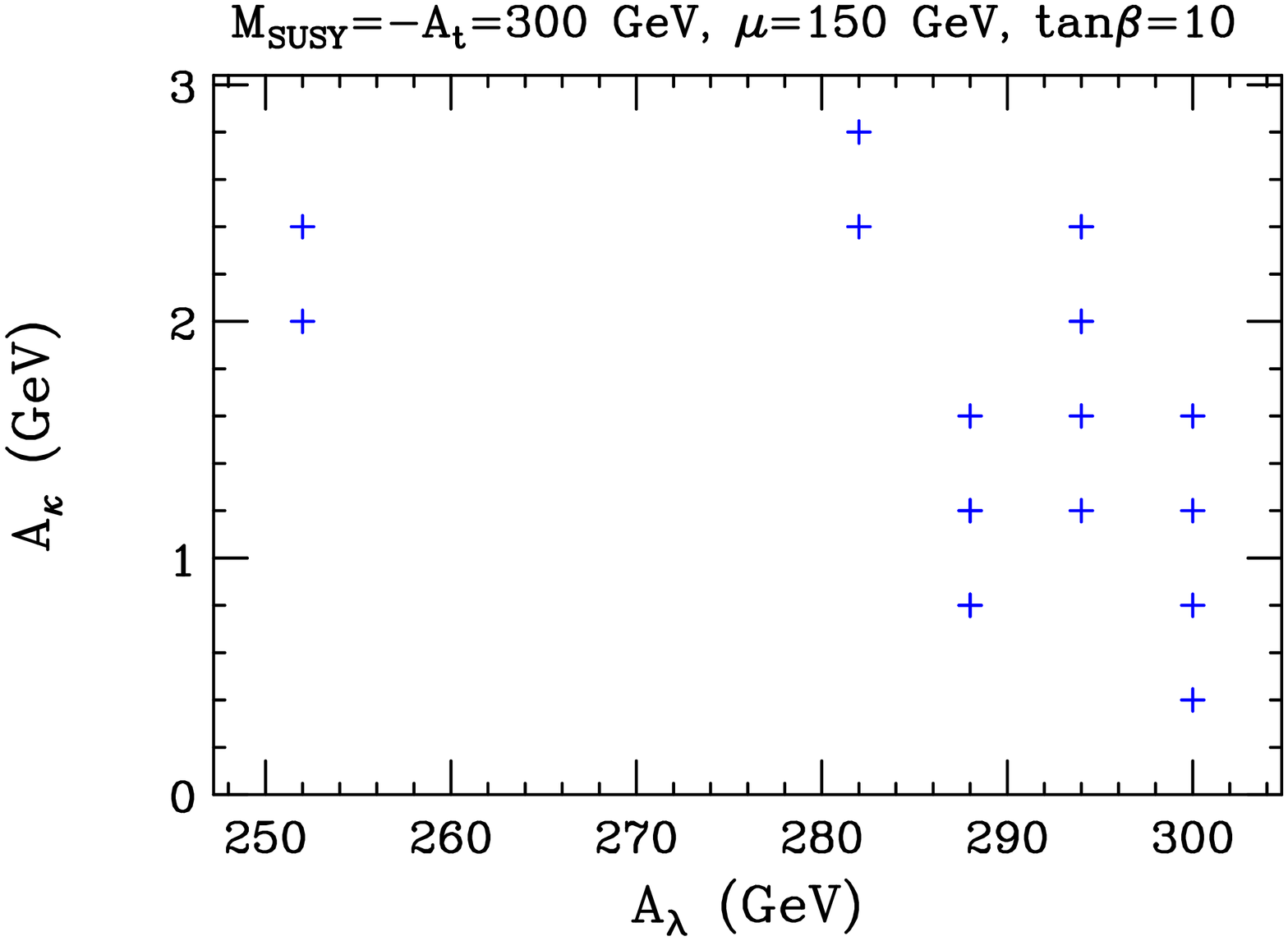}}
\caption{For the points with dominant $\ai\to\gam\gam$ decays and low
  $F$,
we plot: $\kap$ vs. $\lam$ (top); and  $\akap$ vs. $\alam$ (bottom).
}
\label{4gamplots2}
\end{figure}

Finally, the two plots of
Fig.~\ref{4gamplots2} show that these points require largish $\kap$
and $\lam$ that are fairly closely correlated, while $\akap$ must be
quite small.

\section{Conclusions}

There is strong motivation for a supersymmetric model with an extended
Higgs sector containing one or more extra Higgs singlet superfields.
These motivations range from string theory model building, where it is
known that SM-singlets are abundant in string theory
compactifications, to the purely phenomenological, including the fact
that adding anything other than SM singlets to the MSSM will typically
destroy gauge coupling unification.  In this paper, we have studied in detail the
Next-to-Minimal Supersymmetric Model, which contains exactly one
singlet Higgs superfield in addition to the two Higgs doublet
superfields of the MSSM. We have shown that there is a portion of
NMSSM parameter space with an abundance of attractive features, no
outstanding problems and which leads to an important set of
predictions that should be taken quite seriously.  There are many ways
in which the NMSSM is a better benchmark theory than the MSSM, since
it has important flexibilities that are currently leading to
problematical issues for the MSSM.  The attractive features of the
NMSSM include:
\bit
\item a natural explanation for the $\mu$ parameter is provided ---
  since all superpotential couplings are dimensionless in the NMSSM,
  the scale of $\mu$ is given by the scale of soft-SUSY-breaking,
  which (see below) can be well below a TeV;
\item the supersymmetric context provides a highly satisfactory
  solution of the naturalness / hierarchy problem if the squark masses
  (in particular, the stop masses) and the gluino mass are well below
  a TeV (implying possible discovery at the Tevatron and very
  plentiful production at the LHC);
\item in particular, such squark masses imply that fine-tuning with
  respect to GUT-scale parameters is not required in order to obtain
  the observed value of $\mz$ and a light SM-like Higgs boson;
\item low squark masses imply that the lightest Higgs boson, the
  $\hi$, will most naturally be SM-like in its couplings to SM
  particles and have a mass of order $100\gev$, close to the ideal
  value for satisfying precision electroweak constraints;
\item LEP data is fully consistent with such an $\hi$ provided it decays mostly via
  $\hi\to \ai\ai\to \tauptaum\tauptaum$ (requiring $2\mtau<\mai<2\mb$) or
  $\hi\to \ai\ai\to 4j$ (when $\mai<2\mtau$), where the $\ai$ is primarily the
  CP-odd component of the extra complex scalar Higgs singlet field;
\item an appropriately large value of $\br(\hi\to \ai\ai)$ is typically such
  that $\br(\hi\to b\anti b)\sim 0.1$, thereby providing a natural
  explanation for the event excess near $98\gev$ in LEP data for the
  $Z+b\anti b$ channel;
\item  it is quite natural for
  the $\ai$ to be lighter than $2\mb$ while at the same time having sufficient
  $\hi\ai\ai$ coupling for large $\br(\hi\to\ai\ai)$;
\item the optimal scenarios fit nicely with choices for the GUT-scale values
  of the soft-SUSY-breaking Higgs masses squared and  $A$ parameters that
  are quite modest in size, as might be associated with an approximate
  no-scale model for SUSY breaking;
\item in the natural scenarios above, 
the heavier Higgs bosons of the model (two CP-even and one
  CP-odd neutral Higgs bosons and the charged Higgs boson) have
  relatively modest masses that would make them accessible at a hadron
  collider if $\tanb$ is large enough and mostly accessible at a 1 TeV
  linear $\epem$ collider;
\item the light mostly-singlet $\ai$ must have a minimum coupling to the
  SM particles (through mixing with the non-singlet CP-odd state) that
  implies a lower bound, albeit small, on $\br(\Upsilon\to \gam \ai)$;  
\item the $\ai$ could allow for adequate annihilation in the early
  universe of very light neutralinos \cite{Gunion:2005rw}.
\eit
The attractiveness of this scenario suggests that 
the LEP groups should push a re-analysis of the
$Z4\tau$ channel in the hope of either ruling out the $\hi\to \ai\ai\to
4\tau$ scenario, or finding an excess consistent with it for $\mhi$ in
the vicinity of $100\gev$. Either a
positive or negative result would have very important implications for
Higgs searches at the Tevatron and LHC.  We also stress that $B$
factory experiments should attempt to search for a $\Upsilon\to\gam \ai$
signal down to the lowest possible branching ratio (the predicted minimum
in the NMSSM context being of order $10^{-7}$).

We speculate that similar results could emerge in other supersymmetric
models with a Higgs sector that, like the case of the NMSSM, is more
complicated than that of the MSSM. Many such models can be
constructed.  Thus, much of the discussion above regarding Higgs
discovery is quite generic. In general, there might be quite a few
light $a$'s, all of which could appear in the decay of a light SM-like
$h$ and all of which would provide potential signals in reanalyzed LEP
data and in $\Upsilon\to \gam a$ decays.  There is a potential
gold-mine of discovery if one digs deeply enough.

However, whether the $a$ is truly the NMSSM
CP-odd $\ai$ or just a lighter Higgs boson into which
the SM-like $h$ pair-decays, hadron collider detection of the $h$
in its $h\to aa$ decay mode will be very challenging.
Discovery modes that one can hope to demonstrate to be viable include:
\bit
\item $WW$ fusion --- $WW\to h\to aa\to 4\tau$;
\item $t\anti t h$ production with $h\to aa\to 4\tau$;
\item diffractive production \cite{Martin:2005rz}, $pp\to pp h$, with
  $h\to aa\to 4\tau$.
This latter mode looks very promising \cite{hdiff}.
  \eit 
Unfortunately, it seems very doubtful that viable discovery
  signals would be possible for the analogous modes with $h\to aa\to
  4jet $ (that would be the only ones available if $\ma<2\mtau$).
Although $\ma>2\mtau$ is somewhat preferred by naturalness arguments
in the NMSSM case, one should be prepared for the possibility that the
LHC will discover a plethora of supersymmetric particles, and perhaps
some heavy Higgs bosons (if $\tanb$ is large enough) but fail to see
the SM-like light Higgs most closely associated with electroweak
symmetry breaking. The only LHC evidence for its existence would then
be that $WW$ scattering would be found to be fully perturbative, as
predicted if there is a light $h$ with SM-like couplings to $WW$.  

At a linear collider, detection of $\epem\to Zh$ production using the
$\epem\to ZX$ missing mass $\mx$ approach will be completely
straightforward. A $100\gev$ $h$ with SM coupling to $ZZ$ will
result in many events forming a sharp peak in $\mx$, quite
independently of how the $h$ decays.  The decays can then be analyzed
to see what is present and with what branching ratio.  Detection of
an $h$ with unexpected decays at a photon
collider will also be reasonably straightforward \cite{Gunion:2004si}.

\begin{acknowledgments}
  This work was supported by the U.S. Department of Energy under
  grants DE-FG02-90ER40542 and DE-FG03-91ER40674.  JFG thanks the
  Aspen Center for Physics where part of this work was performed. 
\end{acknowledgments}

\end{document}

\bibitem{Belanger:2005kh}
  G.~Belanger, F.~Boudjema, C.~Hugonie, A.~Pukhov and A.~Semenov,
  JCAP {\bf 0509}, 001 (2005)
  [arXiv:hep-ph/0505142].